# Tractable hypergraph properties for constraint satisfaction and conjunctive queries[*]

Dániel Marx[†]

October 28, 2018

**Abstract**

An important question in the study of constraint satisfaction problems (CSP) is understanding how the graph or hypergraph describing the incidence structure of the constraints influences the complexity of the problem. For binary CSP instances (i.e., where each constraint involves only two variables), the situation is well understood: the complexity of the problem essentially depends on the treewidth of the graph of the constraints [27, 43]. However, this is not the correct answer if constraints with unbounded number of variables are allowed, and in particular, for CSP instances arising from query evaluation problems in database theory. Formally, if $\mathcal{H}$ is a class of hypergraphs, then let CSP($\mathcal{H}$) be CSP restricted to instances whose hypergraph is in $\mathcal{H}$. Our goal is to characterize those classes of hypergraphs for which CSP($\mathcal{H}$) is polynomial-time solvable or fixed-parameter tractable, parameterized by the number of variables. Note that in the applications related to database query evaluation, we usually assume that the number of variables is much smaller than the size of the instance, thus parameterization by the number of variables is a meaningful question.

The most general known property of $\mathcal{H}$ that makes CSP($\mathcal{H}$) polynomial-time solvable is bounded fractional hypertree width. Here we introduce a new hypergraph measure called *submodular width*, and show that bounded submodular width of $\mathcal{H}$ (which is a strictly more general property than bounded fractional hypertree width) implies that CSP($\mathcal{H}$) is fixed-parameter tractable. In a matching hardness result, we show that if $\mathcal{H}$ has unbounded submodular width, then CSP($\mathcal{H}$) is not fixed-parameter tractable (and hence not polynomial-time solvable), unless the Exponential Time Hypothesis (ETH) fails. The algorithmic result uses tree decompositions in a novel way: instead of using a single decomposition depending on the hypergraph, the instance is split into a set of instances (all on the same set of variables as the original instance), and then the new instances are solved by choosing a different tree decomposition for each of them. The reason why this strategy works is that the splitting can be done in such a way that the new instances are "uniform" with respect to the number extensions of partial solutions, and therefore the number of partial solutions can be described by a submodular function. For the hardness result, we prove via a series of combinatorial results that if a hypergraph $H$ has large submodular width, then a 3SAT instance can be efficiently simulated by a CSP instance whose hypergraph is $H$. To prove these combinatorial results, we need to develop a theory of (multicommodity) flows on hypergraphs and vertex separators in the case when the function $b(S)$ defining the cost of separator $S$ is submodular, which can be of independent interest.

---

# Contents

# 1  Introduction

There is a long line of research devoted to identifying hypergraph properties that make the evaluation of conjunctive queries tractable (see e.g. [23, 50, 26, 27]). Our main contribution is giving a complete theoretical answer to this question: in a very precise technical sense, we characterize those hypergraph properties that imply tractability for the evaluation of a query. Efficient evaluation of queries is originally a question of database theory; however, it has been noted that the problem can be treated as a constraint satisfaction problem (CSP) and this connection led to a fruitful interaction between the two communities [39, 25, 50]. Most of the literature relevant to the current paper use the language of constraint satisfaction. Therefore, after a brief explanation of the database-theoretic motivation, we switch to the language of CSPs.

**Conjunctive queries.** Evaluation of conjunctive queries (or equivalently, Select-Project-Join queries) is one of the most basic and most studied tasks in relational databases. A relational database consists of a fixed set of relations. A conjunctive query defines a new relation that can be obtained as first taking the join of some relations and then projecting it to a subset of the variables. As an example, consider a relational database that contains three relations: enrolled(Person,Course,Date), teaches(Person,Course,Year), parent(Person1,Person2). The following query $Q$ defines a unary relation ans$(P)$ with the meaning that "$P$ is enrolled in a course taught by her parent."

$$Q : \text{ans}(P) \leftarrow \text{enrolled}(P,C,D) \wedge \text{teaches}(P2,C,Y) \wedge \text{parent}(P2,P).$$

In the *Boolean Conjunctive Query* problem, the task is only to decide if the answer relation is empty or not, that is, if the join of the relations is empty or not. This is usually denoted as the relation "ans" not having any variables. Boolean Conjunctive Query contains most of the combinatorial difficulty of the general problem without complications such that the size of the output being exponentially large. Therefore, the current paper focuses on this decision problem.

In a natural way, we can define the *hypergraph* of a query: its vertices are the variables appearing in the query and for each relation there is a corresponding hyperedge containing the variables appearing in the relation. Intuitively, if the hypergraph has "simple structure," then the query is easy to solve. For example, compare the following two queries:

$$Q_1 : \text{ans} \leftarrow R_1(A,B,C) \wedge R_2(C,D) \wedge R_3(D,E,F) \wedge R_4(E,F,G,H) \wedge R_5(H,I)$$
$$Q_2 : \text{ans} \leftarrow R_1(A,B) \wedge R_2(A,C) \wedge R_3(A,D) \wedge R_4(B,C) \wedge R_5(B,D) \wedge R_6(C,D)$$

Even though more variables appear in $Q_1$, evaluating it seems to be easier: its hypergraph is "path like," thus the query can be answered efficiently by, say, dynamic programming techniques. On the other hand, the hypergraph of $Q_2$ is a clique on 4 vertices and no significant shortcut is apparent compared to trying all possible combinations of values for $(A,B,C,D)$.

What are those hypergraph properties that make Boolean Conjunctive Query tractable? In the early 80s, it has been noted that acyclicity is one such property [9, 19, 53, 8]. Later, more general such properties were identified in the literature: for example, bounded query width [14], bounded hypertree width [23], and bounded fractional hypertree width [42, 28]. Our goal is to find the most general hypergraph property that guarantees an efficient solution for query evaluation.

**Constraint satisfaction.** Constraint satisfaction is a general framework that includes many standard algorithmic problems such as satisfiability, graph coloring, database queries, etc. [26, 20]. A constraint satisfaction problem (CSP) consists of a set $V$ of variables, a domain $D$, and a set $C$ of constraints, where each constraint is a relation on a subset of the variables. The task is to assign a value from $D$ to each variable in such a way that every constraint is satisfied (see Definition 2.1 in Section 2 for the formal definition). For example, 3SAT can be interpreted as a CSP problem where the domain is $D = \{0,1\}$ and the constraints in $C$ correspond to the clauses (thus the arity of each constraint is 3). As another example, let us observe that the $k$-Clique problem (Is there a $k$-clique in a given graph $G$?) can be easily expressed as a CSP instance the following way. Let $D$ be the set of vertices of $G$, let $V$ contain $k$ variables, and let $C$ contain $\binom{k}{2}$ constraints, one constraint on each pair of variables. The binary relation of these constraints require that the two vertices are distinct and adjacent. Therefore, the CSP instance has a solution if and only if $G$ has a $k$-clique.

It is easy to see that Boolean Conjunctive Query can be formulated as the problem of deciding if a CSP instance has a solution: the variables of the CSP instance correspond to the variables appearing in the query and the constraints correspond to the database relations. A distinctive feature of CSP instances obtained this way is that the number of variables is small (as queries are typically small), while the domain of the variables are large (as the database relations usually contain a large number of entries). This has to be contrasted with typical CSP problems from AI, such as 3-colorability and satisfiability, where the domain is small, but the number of variables is large. As our motivation is database-theoretic, in the rest of the paper the reader should keep in mind that we are envisioning scenarios where the number of variables is small and the domain is large.

As the examples above show, solving constraint satisfaction problems is NP-hard in general if there are no additional restrictions on the input instances. The main goal of the research on CSP is to identify tractable special cases of the general problem. The theoretical literature on CSP investigates two main types of restrictions. The first type is to restrict the *constraint language*, that is, the type of constraints that are allowed. This direction includes the classical work of Schaefer [51] and its many generalizations [10, 11, 12, 20, 38]. The second type is to restrict the *structure* induced by the constraints on the variables. The *hypergraph* of a CSP instance is defined to be a hypergraph on the variables of the instance such that for each constraint $c \in C$ there is a hyperedge $e_c$ containing exactly the variables that appear in $c$. If the hypergraph of the CSP instance has very simple structure, then the instance is easy to solve. For example, it is well-known that a CSP instance $I$ with hypergraph $H$ can be solved in time $\|I\|^{O(\text{tw}(H))}$ [22], where $\text{tw}(H)$ denotes the treewidth of $H$ and $\|I\|$ is the size of the representation of $I$ in the input.

Our goal is to characterize the "easy" and "hard" hypergraphs from the viewpoint of constraint satisfaction. However, formally speaking, CSP is polynomial-time solvable for every fixed hypergraph $H$: since $H$ has a constant number $k$ of vertices, every CSP instance with hypergraph $H$ can be solved by trying all $\|I\|^k$ possible combinations on the $k$ variables. It makes more sense to characterize those *classes* of hypergraphs where CSP is easy. Formally, for a class $\mathcal{H}$ of hypergraphs, let CSP($\mathcal{H}$) be the restriction of CSP where the hypergraph of the instance is assumed to be in $\mathcal{H}$. For example, as discussed above, we know that if $\mathcal{H}$ is a class of hypergraphs with bounded treewidth (i.e., there is a constant $w$ such that $\text{tw}(H) \leq w$ for every $H \in \mathcal{H}$), then CSP($\mathcal{H}$) is polynomial-time solvable.

For the characterization of the complexity of CSP($\mathcal{H}$), we can investigate two notions of tractability. CSP($\mathcal{H}$) is *polynomial-time solvable* if there is an algorithm solving every instance of CSP($\mathcal{H}$) in time $(\|I\|)^{O(1)}$, where $\|I\|$ is the length of the representation of $I$ in the input. The following notion interprets tractability in a less restrictive way: CSP($\mathcal{H}$) is *fixed-parameter tractable (FPT)* if there is an algorithm solving every instance $I$ of CSP($\mathcal{H}$) in time $f(H)(\|I\|)^{O(1)}$, where $f$ is an arbitrary computable function of the hypergraph $H$ of the instance. Equivalently, the factor $f(H)$ in the definition can be replaced by a factor $f(k)$ depending only on the number $k$ of vertices of $H$: as the number of hypergraphs on $k$ vertices (without parallel edges) is bounded by a function of $k$, the two definitions result in the same notion. The motivation behind this definition is that if the number of variables is assumed to be much smaller than the the domain size, then we can afford even exponential dependence on the number of variables, as long as the dependence on the size of the instance is polynomial. For a more background on fixed-parameter tractability, the reader is referred to the parameterized complexity literature [18, 21, 45].

**The case of bounded arities.** If the constraints have bounded arity (i.e., the edge size in $\mathcal{H}$ is bounded by a constant $r$), then the complexity of CSP($\mathcal{H}$) is well understood. In this case, bounded treewidth is the only polynomial-time solvable case:

**Theorem 1.1** ([27]). *If $\mathcal{H}$ is a recursively enumerable class of hypergraphs with bounded edge size, then (assuming* FPT $\neq$ W[1]*) the following are equivalent:*

1. *CSP($\mathcal{H}$) is polynomial-time solvable.*

2. *CSP($\mathcal{H}$) is fixed-parameter tractable.*

3. *$\mathcal{H}$ has bounded treewidth.*

The assumption FPT $\neq$ W[1] is a standard hypothesis of parameterized complexity. Thus in the bounded arity case bounded treewidth is the only property of the hypergraph that can make the problem polynomial-time solvable. By definition, polynomial-time solvability implies fixed-parameter tractability, but Theorem 1.1 proves the surprising result that whenever CSP($\mathcal{H}$) is fixed-parameter tractable, it is polynomial-time solvable as well.

The following sharpening of Theorem 1.1 shows that there is no algorithm whose running time is significantly better than the $\|I\|^{O(\text{tw}(H))}$ bound of the treewidth based algorithm, and this is true if we restrict the problem to *any* class $\mathcal{H}$ of hypergraphs. The result is proved under the Exponential Time Hypothesis (ETH) [35] stating that there is no $2^{o(n)}$ time algorithm for *n*-variable 3SAT, which is a somewhat stronger assumption than FPT $\neq$ W[1].

**Theorem 1.2** ([43]). *If there is a function f and a recursively enumerable class $\mathcal{H}$ of hypergraphs with bounded edge size and unbounded treewidth such that the problem CSP($\mathcal{H}$) can be solved in time $f(H)\|I\|^{o(\text{tw}(H)/\log \text{tw}(H))}$ for instances I with hypergraph $H \in \mathcal{H}$, then ETH fails.*

This means that the treewidth-based algorithm is almost optimal on every class of hypergraphs: in the exponent only an $O(\log \text{tw}(H))$ factor improvement is possible. It is conjectured in [43] that Theorem 1.2 can be made tight, i.e., the lower bound holds even if the logarithmic factor is removed from the exponent.

**Conjecture 1.3** ([43]). *If $\mathcal{H}$ is a class of hypergraphs with bounded edge size, then there is no algorithm that solves CSP($\mathcal{H}$) in time $f(H)\|I\|^{o(\text{tw}(H))}$ for instances I with hypergraph $H \in \mathcal{H}$, where f is an arbitrary function.*

**Unbounded arities.** The situation is less understood in the unbounded arity case, i.e., when there is no bound on the maximum edge size in $\mathcal{H}$. First, the complexity in the unbounded-arity case depends on how the constraints are represented. In the bounded-arity case, if each constraint contains at most *r* variables (*r* being a fixed constant), then every reasonable representation of a constraint has size $|D|^{O(r)}$. Therefore, the size of the different representations can differ only by a polynomial factor. On the other hand, if there is no bound on the arity, then there can be exponential difference between the size of succinct representations (e.g., formulas [15]) and verbose representations (e.g., truth tables [44]). The running time of an algorithm is expressed as a function of the input size, hence the complexity of the problem can depend on how the input is represented: longer representation means that it is potentially easier to obtain a polynomial-time algorithm.

The most well-studied representation of constraints is listing all the tuples that satisfy the constraint. This representation is perfectly compatible with our database-theoretic motivation: the constraints are relations of the database, and a relation is physically stored as a table containing all the tuples in the relation. For this representation, there are classes $\mathcal{H}$ with unbounded treewidth such that CSP restricted to this class is polynomial-time solvable. A trivial example is the class $\mathcal{H}$ of all hypergraphs having only a single hyperedge of arbitrary size. The treewidth of such hypergraphs can be arbitrarily large (as the treewidth of a hypergraph consisting of a single edge $e$ is exactly $|e| - 1$), but CSP($\mathcal{H}$) is trivial to solve: we can pick any tuple from the constraint corresponding to the single edge. There are other, nontrivial, classes of hypergraphs with unbounded treewidth such that CSP($\mathcal{H}$) is solvable in polynomial time: for example, classes with bounded *(generalized) hypertree width* [24], bounded *fractional edge cover number* [28], and bounded *fractional hypertree width* [28, 42]. Thus, unlike in the bounded-arity case, treewidth is not the right measure for characterizing the complexity of the problem.

**Our results.** We introduce a new hypergraph width measure that we call *submodular width*. Small submodular width means that for every monotone submodular function $b$ on the vertices of the hypergraph $H$, there is a tree decomposition where $b(B)$ is small for every bag $B$ of the decomposition. (This definition makes sense only if we normalize the considered functions: for this reason, we require that $b(e) \leq 1$ for every edge $e$ of $H$.) The main result of the paper is showing that bounded submodular width is the property that precisely characterizes the complexity of CSP($\mathcal{H}$):

**Theorem 1.4** (Main). *Let $\mathcal{H}$ be a recursively enumerable class of hypergraphs. Assuming the Exponential Time Hypothesis, CSP($\mathcal{H}$) parameterized by H is fixed-parameter tractable if and only if $\mathcal{H}$ has bounded submodular width.*

Theorem 1.4 has an algorithmic side (algorithm for bounded submodular width) and a complexity side (hardness result for unbounded submodular width). Unlike previous width measures in the literature, where small value of the measure suggests a way of solving CSP($\mathcal{H}$) it is not at all clear how bounded submodular width is of any help. In particular, it is not obvious what submodular functions have to do with CSP instances. The main idea of our algorithm is that a CSP instance can be "split" into a small number of "uniform" CSP instances; for this purpose, we use a partitioning procedure inspired by a result of Alon et al. [4]. More precisely, splitting means that we partition the set of tuples appearing in the constraint relations in a certain way and each new instance inherits only one class of the partition (thus each new instance has the same set of variables as the original). Uniformity means that for any subset $B \subseteq A$ of variables, every solution for the problem restricted to $B$ has roughly the same number of extensions to $A$. The property of uniformity allows us to bound the logarithm of the number of solutions on the different subsets by a submodular function. Therefore, bounded submodular width guarantees that each uniform instance has a tree decomposition where only a polynomially bounded number of solutions has to be considered in each bag.

Conceptually, our algorithm goes beyond previous decomposition techniques in two ways. First, the tree decomposition that we use depends not only on the hypergraph, but on the actual constraint relations in the instance (we remark that this idea first appeared in [44] in a different context that does not directly apply to our problem). Second, we are not only decomposing the set of variables, but we also split the constraint relations. This way, we can apply different decompositions to different parts of the solution space.

The proof of the complexity side of Theorem 1.4 follows the same high-level strategy as the proof of Theorem 1.2 in [43]. In a nutshell, the argument of [43] is the following: if treewidth is large, then there is subset of vertices which is highly connected in the sense that the set does not have a small balanced separator; such a highly connected set implies that there is uniform concurrent flow (i.e., a compatible set of flows connecting every pair of vertices in the set); the paths in the flows can be used to embed the graph of a 3SAT formula; and finally this embedding can be used to reduce 3SAT to CSP. These arguments build heavily on well-known characterizations of treewidth and results from combinatorial optimization (such as the $O(\log k)$ integrality gap of sparsest cut). The proof of Theorem 1.4 follows this outline, but now no such well-known tools are available: we are dealing with hypergraphs and submodular functions in a way that was not explored before in the literature. Thus we have to build from scratch all the necessary tools. One of the main difficulties of obtaining Theorem 1.4 is that we have to work in three different domains:

- **CSP instances.** As our goal is to investigate the existence of algorithms solving CSP, the most obvious domain is CSP instances. In light of previous results, we are especially interested in algorithms based on tree decompositions. For such algorithms, what matters is the existence of subsets of vertices such that restricting the instance to any of these subsets gives an instance with "small" number of solutions. In order to solve the instance, we would like to find a tree decomposition where every bag is such a small set.

- **Submodular functions.** Submodular width is defined in terms of submodular functions, thus submodular functions defined on hypergraphs is our second natural domain. We need to understand what large submodular width means, that is, what property of the submodular function and the hypergraph makes it impossible to obtain a tree decomposition where every bag has small value.

- **Flows and embeddings in hypergraphs.** In the hardness proof, our goal is to embed the graph of a 3SAT formula into a hypergraph. Thus we need to define an appropriate notion of embedding and study what guarantees the existence of embeddings with suitable properties. As in [43], we use the paths appearing in flows to construct embeddings. For our purposes, the right notion of flow is a collection of weighted paths where the total weight of the paths intersecting each hyperedge is at most 1. This notion of flows has not been studied in the literature before, thus we need to obtain basic results on such flows, such as exploring the duality between flows and separators.

A key question is how to find connections between these domains. As mentioned above and detailed in Section 4, we have a procedure that reduces a CSP instance into a set of uniform CSP instances, and the number of solutions on the different subsets of variables in a uniform CSP instance can be described by a submodular function. This method allows us to move from the domain of CSP instances to the domain of submodular functions. Section 5 is devoted to showing that if submodular width of a hypergraph is large, then there is a certain "highly connected" set in the hypergraph. Highly connected set is defined as a property of the hypergraph and has no longer anything to do with submodular functions. Thus this connection allows us to move from the domain of submodular functions to the study of hypergraphs. In Section 6, we show that a highly connected set in a hypergraph means that graphs can be efficiently embedded into the hypergraph. In particular, the graph of a 3SAT formula can be embedded into the hypergraph, which gives us (as shown in Section 7) a reduction from 3SAT to CSP($\mathcal{H}$). This connection allows us to move from the domain of embeddings back to the domain of CSP instances. We remark that Sections 4–7 are written in a self-contained way: only the first theorem of each section is used outside the section.

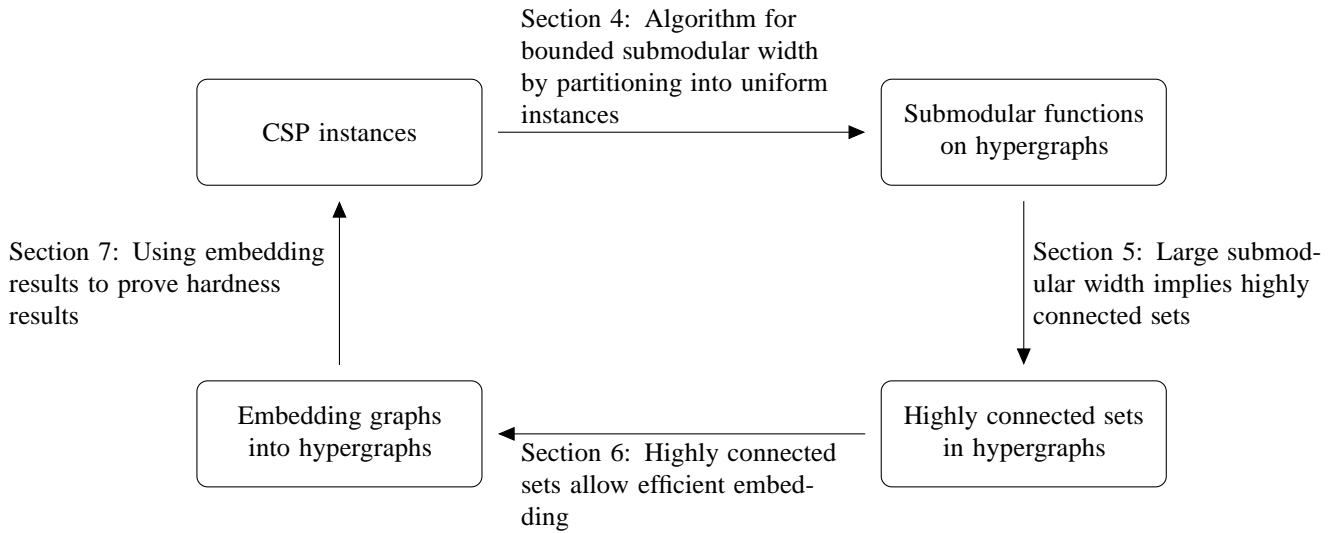

Figure 1: Connections between different domains.

As a consequence of our characterization of submodular width, we obtain the surprising result that bounded submodular width equals bounded adaptive width (defined in [44]):

**Theorem 1.5.** *A class of hypergraphs has bounded submodular width if and only if it has bounded adaptive width.*

It is proved in [44] that there are classes of hypergraphs having bounded adaptive width (and hence bounded submodular width), but unbounded fractional hypertree width. Previously, bounded fractional hypertree width was the most general property that was known to guarantee fixed-parameter tractability [28]. Thus Theorem 1.4 not only gives a complete characterization of the parameterized complexity of CSP($\mathcal{H}$), but its algorithmic side proves fixed-parameter tractability in a strictly more general case than what was known before.

**Why fixed-parameter tractability?** We argue that investigating the fixed-parameter tractability of CSP($\mathcal{H}$) is at least as interesting as investigating polynomial-time solvability. In problems coming from our database-theoretic motivation, the size of the hypergraph (that is, the size of the query) is assumed to be much smaller than the input size (which is usually dominated by the size of the database), hence a constant factor in the running time depending only on the number of variables (or on the hypergraph) is acceptable[1]. Even the STOC 1977 landmark paper of Chandra and Merlin [13], which started the complexity research on conjunctive queries, suggests spending exponential time (in the size of the query) on finding the best possible evaluation order. Furthermore, the notion of fixed-parameter tractability formalizes the usual viewpoint of the literature on conjunctive queries: in the complexity analysis, we should analyze separately the contribution of the query size and the contribution of the database size.

---
[1]This assumption is valid only for evaluation problems (where the problem instance includes a large database) and not for problems that involves only queries, such as the Conjunctive Query Containment problem.

By aiming for fixed-parameter tractability, we can focus more on the core algorithmic question: is there some method for decomposing the space of all solutions in a way that allows efficient evaluation of the query? Some of the progress in this area was made by introducing new decomposition techniques, without showing how to actually find such decompositions. For example, this was the case for the papers introducing query width [14] and fractional hypertree width [28]: it was shown that if a certain type of decomposition is given, then the problem can be solved in polynomial time. In our terminology, these results already show the fixed-parameter tractability of CSP($\mathcal{H}$) for the classes $\mathcal{H}$ where such decompositions exist (since the time required to find an appropriate decomposition can be bounded by a function of the hypergraph $H$ only), but do not give polynomial-time algorithms. It took some more time and effort to come up with polynomial-time (approximation) algorithms for finding such decompositions [23, 42]. While investigating algorithms for finding decompositions give rise to interesting and important problems, they are purely combinatorial problems on graphs and hypergraphs, and no longer has anything to do with query evaluation, constraints, or databases. Thus fixed-parameter tractability gives us a formal way of ignoring these issues and focusing exclusively on the evaluation problem.

On the complexity side, fixed-parameter tractability of CSP($\mathcal{H}$) seems to be a more robust question than polynomial-time solvability. For example, any polynomial-time reduction to CSP($\mathcal{H}$) should be able to pick a member of $\mathcal{H}$, thus it seems that polynomial-time reduction to CSP($\mathcal{H}$) is only possible if certain artificial technical conditions are imposed on $\mathcal{H}$ (such as there is an algorithm efficiently generating appropriate members of $\mathcal{H}$). Furthermore, there are classes $\mathcal{H}$ for which CSP($\mathcal{H}$) is polynomial-time equivalent to LOG CLIQUE [27], thus we cannot hope to classify CSP($\mathcal{H}$) into polynomial-time solvable and NP-hard cases. Another difficulty in understanding polynomial-time solvability is that it can depend on the "irrelevant" parts of the hypergraph. Suppose for example that there is class $\mathcal{H}$ for which CSP($\mathcal{H}$) is not polynomial-time solvable, but it is fixed-parameter tractable: it can be solved in time $f(H) \cdot (\|I\|)^{O(1)}$. Let $\mathcal{H}'$ be constructed the following way: for every $H \in \mathcal{H}$, class $\mathcal{H}'$ contains a hypergraph $H'$ that is obtained from $H$ by adding a new component that is a path of length $f(H)$. This new path is trivial with respect to the CSP problem, thus any algorithm for CSP($\mathcal{H}$) can be used for CSP($\mathcal{H}'$) as well. Consider an instance $I$ of CSP($\mathcal{H}'$) having hypergraph $H'$, which was obtained from hypergraph $H$. After taking care of the path, the assumed algorithm for CSP($\mathcal{H}$) can solve this instance in time $f(H) \cdot (\|I\|)^{O(1)}$, which is polynomial in $\|I\|$: instance $I$ contains a representation of $H'$, which has at least $f(H)$ vertices, thus $\|I\|$ is at least $f(H)$. Therefore, CSP($\mathcal{H}'$) is polynomial-time solvable. This example shows that aiming for polynomial-time solvability instead of fixed-parameter tractability might require understanding such subtle, but mostly irrelevant phenomena.

In the hardness results obtained so far, evidence for the non-existence of polynomial-time algorithms is given not in the form of NP-hardness, but by giving evidence that the problem is not even fixed-parameter tractable. In Theorem 1.1, it is a remarkable coincidence that polynomial-time solvability and fixed-parameter tractability are equivalent. However, there is no reason to expect this to remain true in more general cases. Therefore, as discussed above, it makes sense to focus first on understanding the fixed-parameter tractability of the problem.

**Organization.** For convenience, Section 2 collects many of the definitions appearing in the papers. The reader might want to skim through this at first and refer to appropriate parts of it later. Submodular width and other width measures are defined in Section 3. Section 4 contains the algorithmic part of the paper: the algorithm for classes with bounded submodular width. Section 5 characterizes large submodular width with highly connected sets, while Section 6 uses highly connected sets to find good embeddings in hypergraph. The main hardness result of the paper is proved in Section 7.

## 2 Preliminaries

**Constraint satisfaction problems.** We briefly recall the most important notions related to CSP. For more background, see e.g., [26, 20].

**Definition 2.1.** An instance of a *constraint satisfaction problem* is a triple $(V, D, C)$, where:

- $V$ is a set of variables,

- $D$ is a domain of values,
- $C$ is a set of constraints, $\{c_1, c_2, \ldots, c_q\}$. Each constraint $c_i \in C$ is a pair $\langle s_i, R_i \rangle$, where:
    - $s_i$ is a tuple of variables of length $m_i$, called the *constraint scope,* and
    - $R_i$ is an $m_i$-ary relation over $D$, called the *constraint relation.*

For each constraint $\langle s_i, R_i \rangle$ the tuples of $R_i$ indicate the allowed combinations of simultaneous values for the variables in $s_i$. The length $m_i$ of the tuple $s_i$ is called the *arity* of the constraint. A *solution* to a constraint satisfaction problem instance is a function $f$ from the set of variables $V$ to the domain of values $D$ such that for each constraint $\langle s_i, R_i \rangle$ with $s_i = \langle v_{i_1}, v_{i_2}, \ldots, v_{i_m} \rangle$, the tuple $\langle f(v_{i_1}), f(v_{i_2}), \ldots, f(v_{i_m}) \rangle$ is a member of $R_i$. We say that an instance is *binary* if each constraint relation is binary, i.e., $m_i = 2$ for each constraint.[2] It can be assumed that the instance does not contain two constraints $\langle s_i, R_i \rangle$, $\langle s_j, R_j \rangle$ with $s_i = s_j$, since in this case the two constraints can be replaced by the constraint $\langle s_i, R_i \cap R_j \rangle$.

In the input, the relation appearing in a constraint is represented by listing all the tuples of the constraint. We denote by $\|I\|$ the size of the representation of the instance $I = (V, D, C)$. It can be assumed that $|D| \le \|I\|$: elements of $D$ that do not appear in any relation can be safely removed.

Let $I = (V, D, C)$ be a CSP instance and let $V' \subseteq V$ be a nonempty subset of variables. If $f$ is a solution of $I$, then $\mathrm{pr}_{V'} f$ is the *projection* of $f$ to $V'$, which is simply the restriction of the function $f : V \to D$ to $V' \subseteq V$. If $R$ is a set of solutions for $I$, then we let $\mathrm{pr}_{V'} R = \{\mathrm{pr}_{V'} f \mid f \in R\}$.

The *projection* $\mathrm{pr}_{V'} I$ of $I$ to $V'$ is a CSP $I' = (V', D, C')$, where $C'$ is defined the following way: For each constraint $c = \langle (v_1, \ldots, v_k), R \rangle$ having at least one variable in $V'$, there is a corresponding constraint $c'$ in $C'$. Suppose that $v_{i_1}, \ldots, v_{i_\ell}$ are the variables among $v_1, \ldots, v_k$ that are in $V'$. Then the constraint $c'$ is defined as $\langle (v_{i_1}, \ldots, v_{i_\ell}), R' \rangle$, where the relation $R'$ is the projection of $R$ to the coordinates $i_1, \ldots, i_\ell$, that is, $R'$ contains an $\ell$-tuple $(d'_1, \ldots, d'_\ell) \in D^\ell$ if and only if there is a $k$-tuple $(d_1, \ldots, d_k) \in R$ such that $d'_j = d_{i_j}$ for $1 \le j \le \ell$. Clearly, if $f$ is a solution of $I$, then $\mathrm{pr}_{V'} f$ is a solution of $\mathrm{pr}_{V'} I$ (but the converse is not true). For a subset $V' \subseteq V$, we denote by $\mathrm{sol}_I(V')$ the set of all solutions of $\mathrm{pr}_{V'} I$ (which can contain a solution which is not the projection of any solution of $I$). If the instance $I$ is clear from the context, we drop the subscript.

The *primal graph* (or *Gaifman graph*) of a CSP instance $I = (V, D, C)$ is a graph with vertex set $V$ such that $u, v \in V$ are adjacent if and only if there is a constraint whose scope contains both $u$ and $v$. The *hypergraph* of a CSP instance $I = (V, D, C)$ is a hypergraph $H$ with vertex set $V$, where $e \subseteq V$ is an edge of $H$ if and only if there is a constraint whose scope is $e$ (more precisely, where the scope is an $|e|$-tuple $s$, whose coordinates form a permutation of the elements of $e$). For a class $\mathcal{H}$ of graphs, we denote by $\mathrm{CSP}(\mathcal{H})$ the problem restricted to instances whose hypergraph is in $\mathcal{H}$.

**Graphs and hypergraphs.** If $G$ is a graph or hypergraph, then we denote by $V(G)$ and $E(G)$ the set of vertices and the set of edges of $G$, respectively. Vertices $u, v \in V(G)$ are *adjacent* if there is an edge $e \in E(G)$ with $u, v \in e$. A set $K \subseteq V(G)$ is a *clique* if the vertices in $K$ are pairwise adjacent. If $H$ is a hypergraph and $V' \subseteq V(H)$, then the *subhypergraph induced by $V'$* is a hypergraph $H'$ with vertex set $S$ and $\emptyset \subset e' \subseteq V'$ is an edge of $H'$ if and only if there is an edge $e \in E(H)$ with $e \cap V' = e'$. We denote by $H \setminus S$ the subhypergraph of $H$ induced by $V(H) \setminus S$.

**Paths, separators, and flows in hypergraphs.** A *path* $P$ in hypergraph $H$ is an ordered sequence $v_0$, $v_1$, $\ldots$, $v_r$ of vertices such that $v_i$ and $v_{i-1}$ are adjacent for every $1 \le i < r$. We distinguish the endpoints of a path: vertex $v_0$ is the *first endpoint* of $P$ and $v_r$ is the *second endpoint* of $P$. For a path of length zero, the first and second endpoints coincide. A path is an $X - Y$ *path* if its first endpoint is in $X$ and its second endpoint is in $Y$. A path $P = v_1 v_2 \ldots v_t$ is *minimal* if there are no shortcuts, i.e., $v_i$ and $v_j$ are not adjacent if $|i - j| > 1$. Note that a minimal path intersects each edge at most twice.

Let $H$ be a hypergraph and $X, Y \subseteq V(H)$ be two (not necessarily disjoint) sets of vertices. An $(X, Y)$-*separator* is a set $S \subseteq V(H)$ of vertices such that there is no $(X \setminus S) - (Y \setminus S)$ path in $H \setminus S$, or in other words, every $X - Y$ path of $H$ contains at least one vertex of $S$. In particular, this means that $X \cap Y \subseteq S$.

---

[2]It is unfortunate that some communities use the notion "binary CSP" in the sense that each constraint is binary (as this paper), while other communities use it in the sense that the variables are 0-1, i.e., the domain size is 2.

An assignment $s : E(H) \to \mathbb{R}^+$ is a *fractional $(X,Y)$-separator* if every $X - Y$ path $P$ is *covered* by $s$, that is, $\sum_{e \in E(H), e \cap P \neq \emptyset} s(e) \geq 1$. The *weight* of the fractional separator $s$ is $\sum_{e \in E(H)} s(e)$.

Let $H$ be a hypergraph and let $\mathcal{P}$ be the set of all paths in $H$. A *flow* of $H$ is an assignment $f : \mathcal{P} \to \mathbb{R}^+$ such that $\sum_{P \in \mathcal{P}, P \cap e \neq \emptyset} f(P) \leq 1$ for every $e \in E(H)$. The *value* of the flow $f$ is $\sum_{P \in \mathcal{P}} f(P)$. We say that a path $P$ *appears* in flow $f$, or simply $P$ is a *path of $f$* if $f(P) > 0$. For some $X, Y \subseteq V(H)$, an $(X,Y)$-*flow* is a flow $f$ such that only $X - Y$ paths appear in $f$. A standard LP duality argument shows that the minimum weight of a fractional $(X,Y)$-separator is equal to the maximum value of an $(X,Y)$-flow.

If $f, f'$ are flows such that $f'(P) \leq f(P)$ for every path $P$, then $f'$ is a *subflow* of $f$. The *sum* of the flows $f_1$, ..., $f_r$ is a mapping that assigns weight $\sum_{i=1}^r f_i(P)$ to each path $P$. Note that the sum of flows is not necessarily a flow itself. If the sum of $f_1$, ..., $f_r$ happens to be a flow, then we say that $f_1$, ..., $f_r$ are *compatible*.

**Highly connected sets.** An important step in understanding various width measures is showing that if the measure is large, then the (hyper)graph contains a highly connected set (in a certain sense). We define here the notion of highly connectedness that will be used in the paper. First, recall that a *fractional independent set* of a hypergraph $H$ is a mapping $\mu : V(H) \to [0,1]$ such that $\sum_{v \in e} \mu(v) \leq 1$ for every $e \in E(H)$. We extend functions on the vertices of $H$ to subsets of vertices of $H$ the natural way by setting $\mu(X) := \sum_{v \in X} \mu(v)$, thus $\mu$ is a fractional independent set if and only if $\mu(e) \leq 1$ for every $e \in E(H)$.

Let $\mu$ be a fractional independent set of hypergraph $H$ and let $\lambda > 0$ be a constant. We say that a set $W \subseteq V(H)$ is $(\mu, \lambda)$-*connected* if for any two disjoint sets $A, B \subseteq W$, the minimum weight of a fractional $(A, B)$-separator is at least $\lambda \cdot \min\{\mu(A), \mu(B)\}$. Note that if $W$ is $(\mu, \lambda)$-connected, then $W$ is $(\mu, \lambda')$-connected for every $\lambda' < \lambda$ and every $W' \subseteq W$ is also $(\mu, \lambda)$-connected. Informally, if $W$ is $(\mu, \lambda)$-lambda connected for some fractional independent set $\mu$ such that $\mu(W)$ is "large", then we call $W$ a highly connected set. For $\lambda > 0$, we denote by $\text{con}_\lambda(H)$ the maximum of $\mu(W)$, taken over every $(\mu, \lambda)$-connected set $W$ of $H$. Note that if $\lambda' < \lambda$, then $\text{con}_{\lambda'}(H) > \text{con}_\lambda(H)$. Throughout the paper, $\lambda$ can be thought of as a sufficiently small universal constant, say, 0.001.

**Embeddings.** The hardness result presented in the paper and earlier hardness results for CSP($\mathcal{H}$) [27, 44, 43] are based on embedding some other problem (with a certain graph structure) in a CSP instance whose hypergraph is a member of $\mathcal{H}$. Thus we need appropriate notions of embedding a graph in a (hyper)graph. Let us first recall the definition of minors in graphs. A graph $H$ is a *minor* of $G$ if $H$ can be obtained from $G$ by a sequence of vertex deletions, edge deletions, and edge contractions. The following alternative definition is more relevant from the viewpoint of embeddings: a graph $F$ is a minor of $G$ if there is a mapping $\psi$ that maps each vertex of $F$ to a connected subset of $V(G)$ such that $\psi(u) \cap \psi(v) = \emptyset$ for $u \neq v$, and if $u, v \in V(F)$ are adjacent in $F$, then there is an edge in $E(G)$ connecting $\psi(u)$ and $\psi(v)$.

A crucial difference between the proof of Theorem 1.1 in [27] and the proof of Theorem 1.2 in [43] is that the former result is a based on finding a minor embedding of a grid, while the latter result uses a more general notion of embedding where the images of distinct vertices are not necessarily disjoint, but can overlap in a controlled way. We define such embeddings the following way. We say that two sets of vertices $X, Y \subseteq V(H)$ *touch* if either $X \cap Y \neq \emptyset$, or there is an edge $e \in E(H)$ intersecting both $X$ and $Y$. An *embedding* of graph $G$ into hypergraph $H$ is a mapping $\psi$ that maps each vertex of $H$ to a connected subset of $V(G)$ such that if $u$ and $v$ are adjacent in $G$, then $\psi(u)$ and $\psi(v)$ touch. The *depth* of a vertex $v \in V(H)$ in embedding $\psi$ is $d_\psi(v) := |\{u \in V(G) \mid v \in \psi(u)\}|$, the number of vertices of $G$ whose images contain $v$. The *vertex depth* of the embedding is $\max_{v \in V(H)} d_\psi(v)$. Observe that $\psi$ is a minor mapping if and only if it has vertex depth 1. Because in our case we want to control the size of the constraint relations, we need a notion of depth that is sensitive to "what the edges see." We define the depth $d_\psi(e)$ of an edge as $d_\psi(e) = \sum_{v \in e} d_\psi(e)$ and the *edge depth* to be the maximum of $e$ taken over all edges $e \in E(H)$. Equivalently, we can define the depth of an edge as $d_\psi(E) = \sum_{v \in V(G)} |\psi(v) \cap e|$, that is, each vertex $v$ contributes $|\psi(v) \cap e|$ to the depth. (A different, perhaps more natural, definition of edge depth would be to define it simply as a maximum number of sets $\psi(v)$ that intersect an edge. Somewhat unexpectedly, most results of the paper remain true with both notions; see Remarks 7.6–7.7.)

Trivially, for any graph $G$ and hypergraph $H$, there is an embedding of $G$ into $H$ having vertex depth and edge depth at most $|V(G)|$. If $G$ has $m$ edges and no isolated vertices, then $|V(G)|$ is at most $2m$. We are interested in how much we can gain compared to this trivial solution of depth $O(m)$. We define the *embedding* *power* emb($H$) to be the maximum (supremum) value of $\alpha$ for which there is an integer $m_\alpha$ such that every graph $G$ with $m \geq m_\alpha$ edges has an embedding into $H$ with edge depth $m/\alpha$. It might look unmotivated that we define embedding power in terms of the number of edges of $G$: defining it in terms of the number of vertices might look more natural. However, if we replace the number $m$ of edges with the number $n$ of vertices in the definition, then the worst case occurs if $H$ is a clique on $n$ vertices. Such a definition would describe how well cliques can be embedded, and would give us no information about how sparse graphs can be embedded.

## 3 Width parameters

Treewidth and its various generalizations are defined in this section. We follow the framework of width functions introduced by Adler [1]. A *tree decomposition* of a hypergraph $H$ is a tuple $(T, (B_t)_{t \in V(T)})$, where $T$ is a tree and $(B_t)_{t \in V(T)}$ is a family of subsets of $V(H)$ satisfying the following two conditions: (1) for each $e \in E(H)$ there is a node $t \in V(T)$ such that $e \subseteq B_t$, and (2) for each $v \in V(H)$ the set $\{t \in V(T) \mid v \in B_t\}$ is connected in $T$. The sets $B_t$ are called the *bags* of the decomposition. Let $f : 2^{V(H)} \to \mathbb{R}^+$ be a function that assigns a nonnegative real number to each nonempty subset of vertices. The *$f$-width* of a tree-decomposition $(T, (B_t)_{t \in V(T)})$ is $\max\{f(B_t) \mid t \in V(T)\}$. The *$f$-width* of a hypergraph $H$ is the minimum of the $f$-widths of all its tree decompositions. In other words, $f$-width($H$) $\leq w$ if and only if there is a tree decomposition of $H$ where $f(B) \leq w$ for every bag $B$.

The main idea of tree decomposition based algorithms is that if we have a tree decomposition for instance $I$ such that at most $C$ assignments on $B_t$ have to be considered for each bag $B_t$, then the problem can be solved by dynamic programming in time polynomial in $C$ and $\|I\|$. The various width notions try to guarantee the existence of such decompositions. The simplest such notion, treewidth, can be defined as follows:

**Definition 3.1.** Let $s(B) = |B| - 1$. The *treewidth* of $H$ is tw($H$) := $s$-width($H$).

Further width notions defined in the literature can also be conveniently defined using this setup. A subset $E' \subseteq E(H)$ is an *edge cover* if $\bigcup E' = V(H)$. The *edge cover number* $\rho(H)$ is the size of the smallest edge cover (here we assume that $H$ has no isolated vertices). For $X \subseteq V(H)$, let $\rho_H(X)$ be the size of the smallest set of edges covering $X$.

**Definition 3.2.** The *generalized hypertree width* of $H$ is hw($H$) := $\rho_H$-width($H$).

The original (nongeneralized) definition [23] of hypertree width includes an additional requirement on the decomposition (we omit the details), thus it cannot be less than generalized hypertree. However, it is known that hypertree width and generalized hypertree width can differ by at most a constant factor [2].

Grohe and Marx [28] further generalized hypertree width by considering linear relaxations of edge covers. A function $\gamma : E(H) \to [0, 1]$ is a *fractional edge cover* of $H$ if $\sum_{e \in E(H): v \in e} \gamma(e) \geq 1$ for every $v \in V(H)$. The *fractional cover number* $\rho^*(H)$ of $H$ is the minimum of $\sum_{e \in e(H)} \gamma(e)$ taken over all fractional edge covers of $H$ (it is well known that this minimum is achieved by some rational $\gamma$). We define $\rho_H^*(X)$ analogously to $\rho_H(X)$: the requirement $\sum_{e: v \in e} \gamma(e) \geq 1$ is restricted to vertices $v \in X$.

**Definition 3.3.** The *fractional hypertree width* of $H$ is fhw($H$) := $\rho_H^*$-width($H$).

A crucial idea in [44] is to make the choice of tree decomposition adaptive: instead of assigning a single decomposition to each hypergraph, we choose the best decomposition based on additional properties of the current instance. Motivated by this idea, we generalize the notion of $f$-width from a single function $f$ to a class of functions $\mathcal{F}$. Let $H$ be a hypergraph and let $\mathcal{F}$ be an arbitrary (possibly infinite) class of functions that assign nonnegative real numbers to nonempty subsets of vertices of $H$. The *$\mathcal{F}$-width* of $H$ is $\mathcal{F}$-width($H$) := $\sup\{f$-width($H$) $\mid f \in \mathcal{F}\}$. Thus if $\mathcal{F}$-width($H$) $\leq k$, then for every $f \in \mathcal{F}$, hypergraph $H$ has a tree decomposition with $f$-width at most $k$. Note that this tree decomposition can be different for the different functions $f$. For normalization purposes, we consider only functions $f$ on $V(H)$ that satisfy $f(\emptyset) = 0$ and that are *edge-dominated*, that is, $f(e) \leq 1$ holds for every $e \in E(H)$.

Using these definitions, we can define adaptive width, introduced in [44], as follows. Recall that in Section 2, we stated that if $\mu$ is a fractional independent set, then $\mu$ is extended to subsets of vertices by defining $\mu(X) := \sum_{v \in X} \mu(v)$ for every $X \subseteq V(H)$.

**Definition 3.4.** The *adaptive width* adw($H$) of a hypergraph $H$ is $\mathcal{F}$-width($H$), where $\mathcal{F}$ is the set of all fractional independent sets of $H$.

A function $f : 2^{V(H)} \to \mathbb{R}$ is *modular* if $f(X) = \sum_{v \in X} c_v$ for some constants $c_v$ ($v \in V(H)$). The function $\mu(X)$ arising from a fractional independent set is clearly a modular and edge dominated function, in fact, in Definition 3.4 we can define $\mathcal{F}$ as the set of all nonnegative modular edge-dominated functions on $V(H)$. The main new definition of the paper is a new width measure, which is obtained by imposing a requirement weaker than modularity on the functions in $\mathcal{F}$ (hence the considered set $\mathcal{F}$ of functions is larger):

**Definition 3.5.** A function $b : 2^{V(H)} \to \mathbb{R}^+$ is *submodular* if $b(X) + b(Y) \geq b(X \cap Y) + b(X \cup Y)$ holds for every $X, Y \subseteq V(H)$. Given a hypergraph $H$, let $\mathcal{F}$ contain every edge-dominated monotone submodular function $b$ on $V(H)$ with $b(\emptyset) = 0$. The *submodular width* of hypergraph $H$ is subw($H$) := $\mathcal{F}$-width($H$).

It is well-known that submodular functions can be equivalently characterized by the property that $b(X \cup v) - b(X)$, the *marginal value* of $v$ with respect to $X$, is a nonincreasing function of $X$. That is, for every $v$ and $X \subseteq Y$,
$$b(X \cup v) - b(X) \geq b(Y \cup v) - b(Y). \tag{1}$$

It is clear that subw($H$) $\geq$ adw($H$): Definition 3.5 considers a larger set of functions than Definition 3.4. Furthermore, we show that subw($H$) is at most the fractional hypertree width of $H$. This is a straightforward consequence of the fact that an edge-dominated submodular function is always bounded by the fractional cover number:

**Lemma 3.6.** *Let $H$ be a hypergraph and $b$ be a monotone edge-dominated submodular function with $b(\emptyset) = 0$. Then $b(S) \leq \rho_H^*(S)$ for every $S \subseteq V(H)$.*

*Proof.* The statement can be proved along the same lines as the proof of Shearer's Lemma [16] attributed to Radhakrishnan goes. It is sufficient to prove the statement for the case $S = V(H)$: otherwise, we can consider the subhypergraph of $H$ induced by $S$ and the function $b$ restricted to $S$. Let $\gamma : E(H) \to \mathbb{R}^+$ be a minimum fractional edge cover of $S$. Let $v_1, \ldots, v_n$ be an arbitrary ordering of $V(H)$ and let $V_i = \{v_1, \ldots, v_i\}$, $V_0 = \emptyset$. For every $e \in E(H)$, we have $b(e) = \sum_{v_i \in e}(b(e \cap V_i) - b(e \cap V_{i-1})) \geq \sum_{v_i \in e}(b(V_i) - b(V_{i-1}))$ (the equality is a simple telescopic sum; the inequality uses (1), i.e., the marginal value of $v_i$ with respect to $V_{i-1}$ is not greater than with respect to $e \cap V_{i-1}$).

$$\rho_H^*(V(H)) = \sum_{e \in E(H)} \gamma(e) \geq \sum_{e \in E(H)} \gamma(e) b(e) \geq \sum_{e \in E(H)} \gamma(e) \sum_{v_i \in e} (b(V_i) - b(V_{i-1}))$$
$$= \sum_{i=1}^n \left( (b(V_i) - b(V_{i-1})) \sum_{e \in E(H), v_i \in e} \gamma(e) \right) \geq \sum_{i=1}^n (b(V_i) - b(V_{i-1})) = b(V(H))$$

(in the first inequality, we use that $f$ is edge dominated; in the last inequality, we use that $\gamma$ is a fractional edge cover). □

**Proposition 3.7.** *For every hypergraph $H$, subw($H$) $\leq$ fhw($H$).*

*Proof.* Let $(T, B_{t \in V(T)})$ be a tree decomposition of $H$ whose $\rho_H^*$-width is fhw($H$). If $b$ is an edge-bounded monotone submodular function with $b(\emptyset) = 0$, then by Lemma 3.6, $b(B_t) \leq \rho_H^*(B_t) \leq$ fhw($H$) for every bag $B_t$ of the decomposition, i.e., $b$-width($H$) $\leq$ fhw($H$). This is true for every such function $b$, hence subw($H$) $\leq$ fhw($H$). □

Since adw($H$) $\leq$ subw($H$) $\leq$ fhw($H$), if a class $\mathcal{H}$ of hypergraphs has bounded fractional hypertree width, then it has bound submodular width, and if a class $\mathcal{H}$ has bounded submodular width, then it has bounded adaptive width. Surprisingly, it turns out that the latter implication is actually an equivalence: Corollary 6.10 shows that subw($H$) is at most $O(\text{adw}(H)^4)$, thus a class of hypergraphs has bounded submodular width if and only if it has bounded adaptive width. In other words, large submodular width can be certified already by modular functions: if submodular width is unbounded in $\mathcal{H}$ and we want to choose an $H \in \mathcal{H}$ and a submodular function $b$ such that the $b$-width of $H$ is larger than some constant $k$, then we can choose $H$ and $b$ such that $b$ is actually modular. There is no intuitive reason why this is true: submodular functions seem to be much more powerful than modular functions. Still, we obtain this result as a byproduct of our characterization of submodular width.

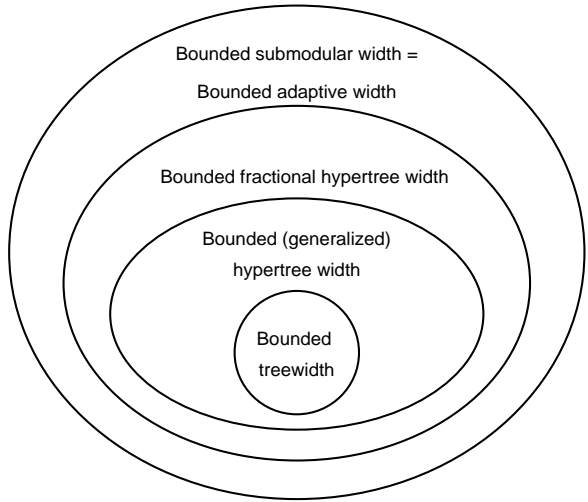

Figure 2: Hypergraph properties that make CSP fixed-parameter tractable.

There is no such connection between adaptive width and fractional hypertree width: it is shown in [44] that there is a class of hypergraphs with bound adaptive width and unbounded fractional hypertree width. Thus the property bounded fractional hypertree width is a strictly weaker property than bounded adaptive/submodular width.

Figure 2 shows the relations of the hypergraph properties defined in this section (note that the elements of this Venn diagram are sets of hypergraphs; e.g., the set "bounded treewidth" contains every set $\mathcal{H}$ of hypergraphs with bounded treewidth). As discussed above, all the inclusions in the figure are proper.

Finally, let us remark that there have been investigations of tree decompositions and branch decompositions of submodular functions and matroids in the literature [33, 47, 34, 32, 5]. However, in those results the submodular function is a connectivity function: $b(S)$ describes the boundary of $S$, that is, the cost of separating $S$ from its complement. In our case, $b(S)$ describes the cost of the separator $S$ itself. Therefore, we are in a completely different setting and the previous results cannot be used.

## 4 From CSP instances to submodular functions

In this section, we prove the main algorithmic result of the paper: $\mathrm{CSP}(\mathcal{H})$ is fixed-parameter tractable if $\mathcal{H}$ has bounded submodular width.

**Theorem 4.1.** *Let $\mathcal{H}$ be a class of hypergraphs such that $\mathrm{subw}(H) \le c_0$ for every $H \in \mathcal{H}$. Then $\mathrm{CSP}(\mathcal{H})$ can be solved in time $2^{c_0 \cdot 2^{O(|V(H)|)}} \cdot \|I\|^{O(c_0)}$.*

The proof of Theorem 4.1 is based on two main ideas:

1. A CSP instance $I$ can be decomposed into a bounded number of "uniform" CSP instances $I_1, \ldots, I_t$ (Lemma 4.11). Here uniform means that if $B \subseteq A$ are two sets of variables, then every solution of $\mathrm{pr}_B I_j$ has roughly the same number of extensions to $\mathrm{pr}_A I_j$.

2. If $I$ is a uniform CSP instance, then (the logarithm of) the number of solutions on the different projections of $I$ can be described by an edge-dominated monotone submodular function $b$ (Lemma 4.12). Therefore, if the hypergraph $H$ of $I$ has bounded submodular width, then it follows that there is a tree decomposition where every bag has a bounded number of solutions. This means that the existence of a solution can be tested by standard techniques.

While our algorithm is based on these two ideas, the technical implementation is slightly different. First, we can achieve uniformity only on "small sets" of variables. For technical reasons, we have to ensure a certain consistency condition (for example, to ensure that the submodular function $b$ is monotone). It follows from the consistency condition that when find a tree decomposition for a uniform instance such that every bag has a small number of solutions, then this automatically implies that the instance has a solution; we do not even have to use the tree decomposition (see Lemma 4.7).

In Section 4.1 we define the notion of consistence that we use and discuss how it can be reached. Section 4.2 describes how the instance can be partitioned into uniform instances. Section 4.3 shows how a submodular function can be defined based on a uniform instance, connecting our algorithm to submodular width.

## 4.1 Consistency

Recall from Section 2 that $\text{pr}_A I$ is instance $I$ projected to a set $A$ of variables and $\text{sol}_I(A)$ is the set of all solution of $\text{pr}_A I$. In the implementation of the first idea (Lemma 4.11), we guarantee uniformity only to subsets of variables that are "small" in the following hereditary sense (note that in general it is possible that $|\text{sol}_I(S')| > |\text{sol}_I(S)|$ for some $S' \subset S$):

**Definition 4.2.** Let $I$ be a CSP instance and $M \geq 1$ an integer. We say that $S \subseteq V$ is *M-small* if $|\text{sol}_I(S')| \leq M$ for every $S' \subseteq S$.

It is not difficult to find all the $M$-small sets, and every solution of the instances projected onto these sets:

**Lemma 4.3.** *Let $I = (V,D,C)$ be a CSP instance and $M \geq 1$ an integer. There is an algorithm with running time $2^{O(|V|)} \cdot \text{poly}(\|I\|,M)$ that finds the set $\mathcal{S}$ of all M-small sets $S \subseteq V$ and constructs $\text{sol}_I(S)$ for each such $S \in \mathcal{S}$.*

*Proof.* For $i = 1, 2, \ldots, |V|$, we find every $M$-small set $S$ of size $i$ and construct $\text{sol}_I(S)$. This is trivial to do for $i = 1$. Suppose that we have already found the collection $\mathcal{S}_i$ of all $M$-small sets of size exactly $i$. By definition, every size $i$ subset $S$ of an $M$-small set $S$ of size $i+1$ is an $M$-small set. Thus we can find every $M$-small set of size $i+1$ by enumerating every $S \in \mathcal{S}_i$ and checking for every $v \in V \setminus S$ whether $S' := S \cup \{v\}$ is $M$-small. To check whether $S'$ is $M$-small, we first check whether every subset of size $i$ is $M$-small, which is easy to do using the set $\mathcal{S}_i$. Then we construct $\text{sol}_I(S')$: this can be done by enumerating every tuple $s \in \text{sol}_I(S)$ and every extension of $s$ by a new value from $D$. Thus we need to consider at most $|\text{sol}_I(S)| \cdot |D| \leq M \cdot |D|$ tuples as possible members in $\text{sol}_I(S')$, which means that $\text{sol}_I(S')$ can be constructed in time polynomial in $M$ and $\|I\|$. If $|\text{sol}_I(S')| \leq M$, then we put $S'$ into $\mathcal{S}_{i+1}$. As the size of each set $\mathcal{S}_i$ is at most $2^{|V|}$ and every operation is polynomial in $M$ and $\|I\|$, the total running time is $2^{O(|V|)} \cdot \text{poly}(\|I\|,M)$. □

We want to avoid dealing with assignments $b \in \text{sol}(B)$ that cannot be extended to a member of $\text{sol}(A)$ for some $A \supseteq B$. Of course, there is no easy way to avoid this in general (or even to detect if there is such a $b$): for example, if $A$ is the set of all variables, then we would need to check if $b$ can be extended to a solution. Therefore, we require only that there is no such unextendable $b$ if $A$ and $B$ are $M$-small:

**Definition 4.4.** A CSP instance is *M-consistent* if $\text{sol}(B) = \text{pr}_B \text{sol}(A)$ for all $M$-small sets $B \subseteq A$.

The notion of $M$-consistency is very similar to $k$-consistency, a standard notion in the constraint satisfaction literature [7, 17, 40]. However, we restrict the considered subsets not by the number of variables, but by the number of solutions (more precisely, by considering only $M$-small sets). Similarly to usual $k$-consistency, we can achieve $M$-consistency by throwing away partial solutions that violate the requirements: if we use the algorithm of Lemma 4.3 to find all possible assignments of the $M$-small sets, then we can check if there is such an unextendable $b$ for some $M$-small sets $A$ and $B$. If there is such a $b$, then we can exclude it from consideration (without losing any solution of the instance) by introducing a new constraint on $B$. By repeatedly excluding the unextendable assignments, we can avoid all such problems. We say that $I' = (V,D,C')$ is a *refinement* of $I = (V,D,C)$ if for every constraint $\langle s,R \rangle \in C$, there is a constraint $\langle s,R' \rangle \in C'$ such that $R' \subseteq R$.

**Lemma 4.5.** *Let $I = (V, D, C)$ be a CSP instance and $M \geq 1$ an integer. There is an algorithm with running time $2^{O(|V|)} \cdot \text{poly}(\|I\|, M)$ that produces an M-consistent CSP instance $I'$ that is a refinement of $I$ with $\text{sol}(I) = \text{sol}(I')$.*

*Proof.* Using the algorithm of Lemma 4.3, we can find all the $M$-small sets and then we can easily check if there are two $M$-small sets $S \subseteq S'$ violating consistency, i.e., $\text{sol}(S) \not\subseteq \text{pr}_S \text{sol}(S')$. In this case, let $s$ be a $|S|$-tuple whose coordinates contain $S$ in an arbitrary order and let us add the constraint $\langle s, \text{pr}_S \text{sol}(S') \rangle$; it is clear that $\text{sol}(V)$ does not change but $|\text{sol}(S)|$ strictly decreases. We repeat this step until the instance becomes $M$-consistent. Note that adding the new constraint can make a set $M$-small that was not $M$-small before, thus we need to rerun the algorithm of Lemma 4.3. To bound the number of iterations before $M$-consistency is reached, observe that adding a new constraint does not increase $|\text{sol}(A)|$ for any $A$ and strictly decreases $|\text{sol}(S)|$ for some $M$-small set $S$. As there are at most $2^{|V|}$ sets $S$ and $|\text{sol}(S)| \leq M$ for every $M$-small set $S$, it follows that this step can be repeated at most $2^{|V|} \cdot M$ times. The size of the instance increases in each step by adding a new constraint with at most $M$ tuples, thus the size of the instance at the end of the process can be still bounded by $2^{O(|V|)} \cdot \text{poly}(\|I\|, M)$. Thus the total time required to ensure that instance $I$ is $M$-consistent can be bounded by $2^{O(|V|)} \cdot \text{poly}(\|I\|, M)$. □

We want to avoid degenerate cases where there is no solution even for some $M$-small sets. Consistency implies that it is sufficient to require this for sets of size 1. We say that a CSP instance is *nontrivial* if $\text{sol}(\{v\}) \neq \emptyset$ for every $v \in V$. The following is immediate:

**Proposition 4.6.** *If $I$ is an M-consistent nontrivial CSP instance, then $\text{sol}(S) \neq \emptyset$ for every M-small set S.*

It is well known that by achieving $k$-consistency, we can solve CSP instances with treewidth $k$: the key observation is that if an instance $I$ with treewidth at most $k$ has a $k$-consistent nontrivial refinement $I'$, then $I$ has a solution. The following lemma adapts this statement to our setting.

**Lemma 4.7.** *Let $I = (V, D, C)$ be a CSP instance and $M \geq 1$ an integer. Let $I'$ be an M-consistent nontrivial refinement of $I$. If the hypergraph H of I has a tree decomposition where every bag B is M-small in $I'$, then I has a solution.*

*Proof.* Suppose that there is such a tree decomposition $(T, (B_t)_{t \in V(T)})$. Assume that $T$ is rooted and for every node $t \in V(T)$, let $V_t$ be the union of the bags that are descendants of $t$ (including $B_t$). We claim that every assignment in $\text{sol}_{I'}(B_t)$ can be extended to an assignment of $V_t$ that satisfies every constraint of $I$ whose scope is fully contained in $V_t$. Applying this statement to the root of $T$ proves that there exists a solution for $I$. (Recall that every edge of the hypergraph $H$, and hence the scope of every constraint, is fully contained in one of the bags.)

We prove the claim for every node of $T$ in a bottom up order. The statement is trivial for the leaves. Let $t_1$, ..., $t_\ell$ be the children of $t$ and suppose the claim is true for these nodes. Consider an assignment $g \in \text{sol}_{I'}(B_t)$. Since $I'$ is $M$-consistent and $B_{t_i}$ is $M$-small, assignment $g_{|B_t \cap B_{t_i}}$ can be extended to an assignment $g_i \in \text{sol}_{I'}(B_{t_i})$. As the claim is true for node $t_i$, assignment $g_i$ can be extended to an assignment $g'_i$ of $V_{t_i}$. The assignments $g, g'_1, \ldots, g'_\ell$ can be combined to obtain an assignment $g'$ on $V_t$ (note that this is well defined: the intersection of $V_{t_i}$ and $V_{t_j}$ is in $V_t$, which means that a variable appearing in both $V_{t_i}$ and $V_{t_j}$ has the same value in $g$, $g'_i$, and $g'_j$). Furthermore, every edge $e$ of $H$ that is fully contained in $V_t$ is fully contained in at least one of $B_t, V_{t_1}, \ldots, V_{t_\ell}$, and the corresponding assignment among $g, g'_1, \ldots, g'_\ell$ shows that $g'$ satisfies the constraint corresponding to $e$. □

Note the subtle detail that Lemma 4.7 does not claim that $I'$ has a solution. Furthermore, when Lemma 4.5 creates an $M$-consistent instance, then it possibly adds many new constraints and the hypergraph of $I'$ can be very dense even if the hypergraph of $I$ has nice structure. However, this is not a problem, as Lemma 4.7 does not require any property on the hypergraph of $I'$.

## 4.2 Decomposition into uniform CSP instances

Our algorithm for decomposing a CSP instance into uniform CSP instances is inspired by a combinatorial result of Alon et al. [4], which shows that, for every fixed $n$, an $n$-dimensional point set $S$ can be partitioned into $\text{polylog}(|S|)$ classes such that each class is $O(1)$-uniform. We follow the same proof idea: the instance is split into two instances if uniformity is violated somewhere, and we analyze the change of an appropriately defined weight function to bound the number of splits performed. However, the parameter setting is different in our proof: we want to partition into $f(|V|)$ classes, but we are satisfied with somewhat weaker uniformity. Another minor technical difference is that we require uniformity only on the $N^c$-small sets.

The following definitions gives the precise notion of uniformity that we use:

**Definition 4.8.** Let $I = (V, D, C)$ be a CSP instance. For $B \subseteq A \subseteq V$ and an assignment $b : B \to D$, let $\text{sol}_I(A|B = b) := \{a \in \text{sol}_I(A) \mid \text{pr}_B a = \text{pr}_B b\}$, the set of all extensions of $b$ to a solution of $\text{pr}_A I$. Let $\max_I(A|B) = \max_{b \in \text{sol}_I(B)} |\text{sol}_I(A|B = b)|$ (if $\text{sol}_I(B) = \emptyset$, then $\max_I(A|B) = 0$). We define $\max_I(A|\emptyset) = |\text{sol}_I(A)|$ and $\max_I(\emptyset|\emptyset) = 1$. We will drop $I$ from the subscript of max if it is clear from the context.

Let us prove two straightforward properties of the function $\max(A|B)$:

**Proposition 4.9.** For every $B \subseteq A \subseteq V$ and $C \subseteq V$, we have

1. $\max(A|B) \geq |\text{sol}(A)|/|\text{sol}(B)|$,

2. $\max(A|B) \geq \max(A \cup C|B \cup C)$.

*Proof.* If every $b \in \text{sol}(B)$ has at most $\max(A|B)$ extensions to $A$, then clearly $|\text{sol}(A)|$ is at most $|\text{sol}(B)| \cdot \max(A|B)$, proving the first statement. To show the second statement, consider an $x \in \text{sol}(B \cup C)$ with $\max(A \cup C|B \cup C)$ extensions to $A \cup C$. For any two $y_1, y_2 \in \text{sol}(A \cup C|B \cup C = x)$ with $y_1 \neq y_2$, we have $\text{pr}_C y_1 = \text{pr}_C y_2 = \text{pr}_C x$, hence $y_1$ and $y_2$ can be different only if $\text{pr}_A y_1 \neq \text{pr}_A y_2$. This means that $\text{pr}_A y_1$ and $\text{pr}_A y_2$ are two different extensions of $\text{pr}_B x$ to $A$. Therefore,

$$\max(A|B) \geq |\text{sol}(A|B = \text{pr}_B x)| \geq |\text{sol}(A \cup C|B \cup C = x)| = \max(A \cup C|B \cup C),$$

what we had to show. □

Notice that (2) in Prop. 4.9 gives a hint that submodularity will be relevant: it is analogous to inequality (1) expressing that marginal value is larger with respect to a smaller set.

**Definition 4.10.** We say that $A \subseteq V$ is *c-uniform* (for some integer $c$) if, for every $B \subseteq A$,

$$\max_I(A|B) \leq c|\text{sol}_I(A)|/|\text{sol}_I(B)|.$$

A CSP instance is $(N, c, \varepsilon)$-*uniform* if every $N^c$-small set is $N^\varepsilon$-uniform.

That is, $A$ is $c$-uniform if every solution on of $\text{sol}_I(B)$ has at most $c$ times as many extensions as the average number of extensions.

**Lemma 4.11.** *Let $I = (V, D, C)$ be a CSP instance, let $N$ an be an integer, and let $c \geq 1$, $\varepsilon > 0$ real numbers. There is an algorithm with running time $2^{2^{O(|V|)} \cdot c/\varepsilon} \cdot \text{poly}(\|I\|, N^c)$ that produces a set of $(N, c, \varepsilon)$-uniform $N^c$-consistent nontrivial instances $I_1, \ldots, I_t$ with $0 \leq t \leq 2^{2^{O(|V|)} \cdot c/\varepsilon}$, all on the set $V$ of variables, such that*

1. *every solution of $I$ is a solution of exactly one instance $I_i$,*

2. *for every $1 \leq i \leq t$, instance $I_i$ is a refinement of $I$.*

*Proof.* The main step of the algorithm takes a CSP instance $I$ and either makes it $(N,c,\varepsilon)$-uniform and $N^c$-consistent without losing any solutions, or splits it into two instances $I_{\text{small}}$, $I_{\text{large}}$. By applying the main step recursively on $I_{\text{small}}$ and $I_{\text{large}}$, we eventually arrive to a set of $(N,c,\varepsilon)$-uniform $N^c$-consistent instances. We will argue that the number of constructed instances is $2^{2^{O(|V|)}\cdot c/\varepsilon}$.

In the main step, we first check if the instance is trivial; in this case we can stop with $t = 0$. Otherwise, we invoke the algorithm of Lemma 4.5 to obtain an $N^c$-consistent refinement of the instance, without losing any solution. Next we check if this $N^c$-consistent instance $I$ is $(N,c,\varepsilon)$-uniform. This can be tested in time $2^{O(|V|)} \cdot \text{poly}(\|I\|, N^c)$ if we use Lemma 4.3 to find all the $N^c$-small sets and the corresponding sets of solutions. Suppose that $N^c$-small sets $B \subseteq A$ violate uniformity, that is,

$$\max(A|B) > N^\varepsilon |\text{sol}(A)|/|\text{sol}(B)|.$$

Let $\text{sol}_{\text{small}}(B)$ contain those tuples $b$ for which $|\text{sol}(A|B=b)| \leq \sqrt{N^\varepsilon}|\text{sol}(A)|/|\text{sol}(B)|$ and let $\text{sol}_{\text{large}}(B) = \text{sol}(B) \setminus \text{sol}_{\text{small}}(B)$. Note that $|\text{sol}(A)| \geq |\text{sol}_{\text{large}}(B)| \cdot (\sqrt{N^\varepsilon}|\text{sol}(A)|/|\text{sol}(B)|)$ (as every tuple $b \in \text{sol}_{\text{large}}(B)$ has at least $\sqrt{N^\varepsilon}|\text{sol}(A)|/|\text{sol}(B)|$ extensions to $A$), hence

$$|\text{sol}_{\text{large}}(B)| \leq |\text{sol}(B)|/\sqrt{N^\varepsilon}. \tag{2}$$

Let instance $I_{\text{small}}$ (resp., $I_{\text{large}}$) be obtained from $I$ by adding the constraint $\langle B, \text{sol}_{\text{small}}(B)\rangle$ (resp., $\langle B, \text{sol}_{\text{large}}(B)\rangle$). Clearly, the set of solutions of $I$ is the disjoint union of the sets of solutions of $I_{\text{small}}$ and $I_{\text{large}}$. This completes the description of the main step.

It is clear that if the recursive procedure stops, then the instances at the leaves of the recursion satisfy the two requirements: the application of Lemma 4.5 does not lose any solution and each resulting instance is $N^c$-consistent and $(N,c,\varepsilon)$-uniform. We show that the height of the recursion tree can be bounded from above by a function $h(|V|,c,\varepsilon) = 2^{O(|V|)} \cdot c/\varepsilon$ depending only on $|V|$, $c$, and $\varepsilon$; in particular, this shows that the recursive algorithm eventually stops and produces at most $t = 2^{h(|V|,c,\varepsilon)} = 2^{2^{O(|V|)}\cdot c/\varepsilon}$ instances.

Let us consider a path in the recursion tree starting at the root, and let $I^1, I^2, \ldots, I^p$ be the corresponding $N^c$-consistent instances. If a set $S$ is $N^c$-small in $I^j$, then it is $N^c$-small in $I^{j'}$ for every $j' > j$: the main step cannot increase $|\text{sol}(S)|$ for any $S$. Thus, with the exception of at most $2^{|V|}$ values of $j$, instances $I^j$ and $I^{j+1}$ have the same $N^c$-small sets. Let us consider a subpath $I^x, \ldots, I^y$ such that all these instances have the same $N^c$-small sets. We show that the length of this subpath is $O(3^{|V|} \cdot c/\varepsilon)$, hence $p = O(2^{|V|} \cdot 3^{|V|} \cdot c/\varepsilon)$. As this holds for any path starting at the root, we obtain that the height of the recursion tree is $2^{O(|V|)} \cdot c/\varepsilon$ and hence $t = 2^{2^{O(|V|)}\cdot c/\varepsilon}$.

For the instance $I^j$, let us define the following weight:

$$W^j = \sum_{\substack{\emptyset \subseteq B \subseteq A \subseteq V \\ A, B \text{ are } N^c\text{-small in } I^j}} \log \max_{I^j}(A|B).$$

We bound the length of the subpath $I^x, \ldots, I^y$ by analyzing how this weight changes in each step. Observe first that when invoking the algorithm of Lemma 4.5 to find an $N^c$-consistent refinement, then the weight does not increase: adding new constraints cannot increase $\max(A|B)$ for any $A, B \subseteq V$ and cannot create new $N^c$-small sets by the assumption on the subpath $I^x$ and $I^y$. Thus it is sufficient to analyze how the weight decreases in $I_{\text{large}}$ and $I_{\text{small}}$ compared to $I$. Note that $0 \leq W^j \leq 3^{|V|} \log N^c = 3^{|V|} \cdot c \log N$: the sum consists of at most $3^{|V|}$ terms and (as $A$ is $N^c$-small and the instance $I^j$ is $N^c$-consistent and nontrivial) $\max_{I^j}(A|B)$ is between 1 and $N^c$. We show that $W^{j+1} \leq W^j - (\varepsilon/2)\log N$, which immediately implies that the length of the subpath is $O(3^{|V|} \cdot c/\varepsilon)$. Let us inspect how $W^{j+1}$ changes compared to $W^j$. Since $I^j$ and $I^{j+1}$ have the same $N^c$-small sets by assumption, no new term can appear in $W^{j+1}$. It is clear that $\max_{I^{j+1}}(A|B)$ cannot be greater than $\max_{I^j}(A|B)$ for any $A, B$. Moreover, there is at least one term that strictly decreases. Suppose first that $I^{j+1}$ was obtained from $I^j$ by adding the constraint $\langle B, \text{sol}_{\text{small}}(B)\rangle$. Then

$$\log \max_{I^{j+1}}(A|B) \leq \log \sqrt{N^\varepsilon}\frac{|\text{sol}_{I^j}(A)|}{|\text{sol}_{I^j}(B)|} \leq \log(\max_{I^j}(A|B)/\sqrt{N^\varepsilon}) = \log \max_{I^j}(A|B) - (\varepsilon/2)\log N,$$

where we have used (4.2) in the second inequality. On the other hand, if $I^{j+1}$ was obtained by adding the constraint $\langle B, \mathrm{sol}_{\mathrm{large}}(B)\rangle$, then

$$\log \max\nolimits_{I^{j+1}}(B|\emptyset) = \log|\mathrm{sol}_{I^{j+1}}(B)| \leq \log(|\mathrm{sol}_{I^j}(B)|/\sqrt{N^\varepsilon}) = \log\max\nolimits_{I^j}(B|\emptyset) - (\varepsilon/2)\log N,$$

where the inequality follows from (2). In both cases, we get that at least one term decreases by at least $(\varepsilon/2)\log N$. □

### 4.3 Uniform CSP instances and submodularity

Assume for a moment that we have a 1-uniform instance $I$ with hypergraph $H$. Note that by Prop 4.9(1), this means that $\max(A|B) = |\mathrm{sol}(A)|/|\mathrm{sol}(B)|$. Suppose that every constraint contains at most $N$ tuples and let us define the function $b(S) = \log_N|\mathrm{sol}(S)|$. For every edge $e \in E(H)$, there is a corresponding constraint, which has at most $N$ tuples by the definition of $N$. Thus $|\mathrm{sol}(e)| \leq N$ and hence $b(e) \leq 1$ for every $e \in E(H)$, that is, $b$ is edge dominated. The crucial observation of this section is that this function $b$ is submodular:

$$\begin{aligned}
b(X) + b(Y) &= \log_N|\mathrm{sol}(X)| + \log_N\left(|\mathrm{sol}(X \cap Y)|\frac{|\mathrm{sol}(Y)|}{|\mathrm{sol}(X \cap Y)|}\right) \\
&= \log_N|\mathrm{sol}(X)| + \log_N(|\mathrm{sol}(X \cap Y)| \cdot \max(Y|X \cap Y)) \\
&\geq \log_N|\mathrm{sol}(X)| + \log_N(|\mathrm{sol}(X \cap Y)| \cdot \max(X \cup Y|X)) \\
&= \log_N|\mathrm{sol}(X)| + \log_N\left(|\mathrm{sol}(X \cap Y)| \cdot \frac{|\mathrm{sol}(X \cup Y)|}{|\mathrm{sol}(X)|}\right) \\
&= \log_N|\mathrm{sol}(X \cap Y)| + \log_N|\mathrm{sol}(X \cup Y)| \\
&= b(X \cap Y) + b(X \cup Y)
\end{aligned}$$

(the equalities follow from 1-uniformity; the inequality uses Prop. 4.9(2) with $A = Y$, $B = X \cap Y$, $C = X$). Therefore, if the submodular width of $H$ is at most $c$, then $H$ has a tree decomposition where $b(B) \leq c$ and hence $|\mathrm{sol}(B)| \leq N^c$ for every bag $B$. Thus we can find a solution of the instance by dynamic programming in time polynomial in $N^c$.

Lemma 4.11 guarantees some uniformity for the created instances, but not perfect 1-uniformity and only for the $N^c$-small sets. Thus in Lemma 4.12, we need to define $b$ in a slightly different and more technical way: we add some small terms to correct errors arising from the weaker uniformity and we truncate the function at large values (i.e., for sets that are not $N^c$-small).

**Lemma 4.12.** *Let $I = (V,D,C)$ be a CSP instance with hypergraph $H$ such that $|\mathrm{sol}(e)| \leq N$ for every $e \in E(H)$. If $I$ is $N^c$-consistent and $(N,c,\varepsilon^3)$-uniform for some $c \geq 1$ and $\varepsilon := 1/|V|$, then the following function $b$ is an edge-dominated, monotone, submodular function on $V(H)$ with $b(\emptyset) = 0$:*

$$b(S) := \begin{cases} (1-\varepsilon)\log_N|\mathrm{sol}(S)| + 2\varepsilon^2|S| - \varepsilon^3|S|^2 & \text{if } S \text{ is } N^c\text{-small,} \\ (1-\varepsilon)c + 2\varepsilon^2|S| - \varepsilon^3|S|^2 & \text{otherwise.} \end{cases}$$

*Proof.* Let $h(S) := 2\varepsilon^2|S| - \varepsilon^3|S|^2$. It is easy to see that $h(S)$ is monotone and $0 \leq h(S) \leq \varepsilon$ for every $S \subseteq V(H)$ (as $\varepsilon|S| \leq 1$). Furthermore, $h$ is a submodular function:

$$\begin{aligned}
h(X) &+ h(Y) - h(X \cap Y) - h(X \cup Y) \\
&= 2\varepsilon^2(|X| + |Y| - |X \cap Y| - |X \cup Y|) + \varepsilon^3(-|X|^2 - |Y|^2 + |X \cap Y|^2 + |X \cup Y|^2) \\
&= \varepsilon^3\left(-(|X \cap Y| + |X \setminus Y|)^2 - (|X \cap Y| + |Y \setminus X|)^2 + |X \cap Y|^2 + (|X \cap Y| + |X \setminus Y| + |Y \setminus X|)^2\right) \\
&= 2\varepsilon^3|X \setminus Y| \cdot |Y \setminus X| \geq 0.
\end{aligned}$$

This calculation shows that if $|X \setminus Y|, |Y \setminus X| \geq 1$, then we actually have $h(X) + h(Y) \geq h(X \cap Y) + h(X \cup Y) + 2\varepsilon^3$. We will use this extra $2\varepsilon^3$ term to dominate the error terms arising from assuming only $(N,c,\varepsilon^3)$-uniformity instead of perfect uniformity.

Let us first verify the monotonicity of $b$. If $Y$ is $N^c$-small, then every $X \subseteq Y$ is $N^c$-small, which implies $|\operatorname{sol}(X)| \leq |\operatorname{sol}(Y)|$ as $I$ is $N^c$-consistent. Therefore, $b(X) \leq b(Y)$ follows from the monotonicity of $h$. If $Y$ is not $N^c$ small, then $b(Y) = (1-\varepsilon)c + h(Y)$ and $b(X) \leq b(Y)$ is clear for every $X \subseteq Y$, no matter whether $X$ is $N^c$-small or not.

To see that $b$ is edge-dominated, consider an edge $e \in E(H)$. By assumption, $\log_N |\operatorname{sol}(e)| \leq 1$ for every $e \in E(H)$ and hence (using $N^c$-consistency and $c \geq 1$) $e$ is $N^c$-small. Thus $b(e) \leq (1-\varepsilon) + h(S) \leq 1$, as required.

Finally, let us verify the submodularity of $b$ for some $X, Y \subseteq V$. If $X \subseteq Y$ or $Y \subseteq X$, then there is nothing to show. Thus we can assume that $|X \setminus Y|, |Y \setminus X| \geq 1$. We consider 3 cases depending on which of $X$ and $Y$ are $N^c$-small. Suppose first that $X$ and $Y$ are both $N^c$-small. In this case,

$$\begin{aligned}
b(X) + b(Y) &= (1-\varepsilon)\log_N |\operatorname{sol}(X)| + (1-\varepsilon)\log_N |\operatorname{sol}(Y)| + h(X) + h(Y) \\
&= (1-\varepsilon)\log_N |\operatorname{sol}(X)| + (1-\varepsilon)\log_N \left(|\operatorname{sol}(X \cap Y)| \cdot \frac{|\operatorname{sol}(Y)|}{|\operatorname{sol}(X \cap Y)|}\right) + h(X) + h(Y) \\
&\geq (1-\varepsilon)\log_N |\operatorname{sol}(X)| + (1-\varepsilon)\log_N |\operatorname{sol}(X \cap Y)| \\
&\quad + (1-\varepsilon)\log_N(\max(Y|X \cap Y)/N^{\varepsilon^3}) + h(X) + h(Y) \\
&\geq (1-\varepsilon)\log_N |\operatorname{sol}(X \cap Y)| + (1-\varepsilon)\log_N(|\operatorname{sol}(X)| \max(X \cup Y|X)) \\
&\quad - (1-\varepsilon) \cdot \varepsilon^3 + h(X \cap Y) + h(X \cup Y) + 2\varepsilon^3 \\
&\geq (1-\varepsilon)\log_N |\operatorname{sol}(X \cap Y)| + (1-\varepsilon)\log_N |\operatorname{sol}(X \cup Y)| + h(X \cap Y) + h(X \cup Y) \\
&\geq b(X \cap Y) + b(X \cup Y)
\end{aligned}$$

(in the first inequality, we used the definition of $(N, c, \varepsilon^3)$-uniformity on $X \cap Y$ and $Y$; in the second inequality, we used the submodularity of $h$ and Prop. 4.9(2) for $A = Y$, $B = X \cap Y$, and $C = X$; in the third inequality, we used Prop. 4.9(1) for $A = X \cup Y$, $B = X$; the last inequality is strict only if $X \cup Y$ is not $N^c$-small).

For the second case, suppose that, say, $X$ is $N^c$-small but $Y$ is not. In this case, $X \cap Y$ is $N^c$-small but $X \cup Y$ is not. Thus

$$\begin{aligned}
b(X) + b(Y) &= (1-\varepsilon)\log_N |\operatorname{sol}(X)| + (1-\varepsilon)c + h(X) + h(Y) \\
&\geq (1-\varepsilon)\log_N |\operatorname{sol}(X \cap Y)| + (1-\varepsilon)c + h(X \cap Y) + h(X \cup Y) \\
&= b(X \cap Y) + b(X \cup Y)
\end{aligned}$$

(in the inequality, we used the $N^c$-consistency on $X \cap Y$ and $Y$, and the submodularity of $h$).

Finally, suppose that neither $X$ nor $Y$ is $N^c$-small. In this case, $X \cup Y$ is not $N^c$-small either. Now

$$b(X) + b(Y) = 2(1-\varepsilon)c + h(X) + h(Y) \geq 2(1-\varepsilon)c + h(X \cap Y) + h(X \cup Y) \geq b(X \cap Y) + b(X \cup Y).$$

□

Having constructed the submodular function $b$ as in Lemma 4.12, we can use the argument described at the beginning of the section: if $H$ has submodular width at most $(1-\varepsilon)c$, then there is a tree decomposition where every bag is $N^c$-small, and we can use this tree decomposition to find a solution. In fact, by Lemma 4.7, in this case $N^c$-consistency implies that every nontrivial instance has a solution.

*Proof (of Theorem 4.1).* Let $I$ be an instance of CSP($\mathcal{H}$) having hypergraph $H \in \mathcal{H}$. We decide the solvability of $I$ the following way. Let $N \leq \|I\|$ be the size of the largest constraint relation in $I$, i.e., every constraint has at most $N$ satisfying assignments. Set $\varepsilon := 1/|V(H)|$, and let $c := c_0/(1-\varepsilon)$. Let us use the algorithm of Lemma 4.11 to produce the nontrivial $N^c$-consistent $(N, c, \varepsilon^3)$-uniform instances $I_1, \ldots, I_t$. The running time of this step is $2^{2^{O(|V|) \cdot c/\varepsilon}} \cdot \operatorname{poly}(\|I\|, N^c)$, which is $2^{c_0 \cdot 2^{O(|V(H)|)}} \cdot \|I\|^{O(c_0)}$.

If $t = 0$, then we can conclude that $I$ has no solution. Otherwise, we argue that $I$ has a solution. Consider any $I_i$ and let $b$ be the edge-dominated monotone submodular function defined in Lemma 4.12. By definition of submodular width, $H$ has a tree decomposition $(T, (B_t)_{t \in V(T)})$ such that $b(B_t) \leq \operatorname{subw}(H) \leq c_0 = (1-\varepsilon)c$ for every $t \in V(T)$. Since $b(S) \leq (1-\varepsilon)c$ implies $|\operatorname{sol}(S)| \leq N^c$ and $b$ is monotone, this means that $B_t$ is $N^c$-small in $I_i$ for every $t \in V(T)$. Therefore, the conditions of Lemma 4.7 hold, and $I$ has a solution. □

# 5 From submodular functions to highly connected sets

The aim of this section is to show that if a hypergraph $H$ has large submodular width, then there is a large highly connected set in $H$. Recall that we say that a set $W$ is $(\mu, \lambda)$-connected for some fractional independent set $\mu$ and $\lambda > 0$, if for every disjoint $A, B \subseteq W$, every fractional $(A, B)$-separator has weight at least $\lambda \cdot \min\{\mu(A), \mu(B)\}$ (see Section 2). Equivalently, we can say that for every disjoint $A, B \subseteq W$, there is an $(A, B)$-flow of value $\lambda \cdot \min\{\mu(A), \mu(B)\}$. Recall also that $\text{con}_\lambda(H)$ denotes the maximum value of $\mu(W)$ taken over every fractional independent set $\mu$ and $(\mu, \lambda)$-connected set $W$.

The main result of this section allows us to identify a highly connected set if submodular width is large:

**Theorem 5.1.** *For every sufficiently small constant $\lambda > 0$, the following holds. Let b be an edge-dominated monotone submodular function of $H$ with $b(\emptyset) = 0$. If the b-width of $H$ is greater than $\frac{3}{2}(w+1)$, then $\text{con}_\lambda(W) \geq w$.*

For the proof of Theorem 5.1, we need to show that if there is no tree decomposition where $b(B)$ is small for every bag $B$, then a highly connected set exists. There is a standard recursive procedure that either builds a tree decomposition or finds a highly connected set (see e.g., [21, Section 11.2]). Simplifying somewhat, the main idea is that if the graph can be decomposed into smaller graphs by splitting a certain set of vertices into two parts, then a tree decomposition for each part is constructed using the algorithm recursively, and the tree decompositions for the parts are joined in an appropriate way to obtain a tree decomposition for the original graph. On the other hand, if the set of vertices cannot be split, then we can conclude that it is highly connected. This high-level idea has been applied for various notions of tree decompositions [48, 46, 2, 47], and it turns out to be useful in our context as well. However, we need to overcome two major difficulties:

1. Highly connected set in our context is defined as not having certain *fractional separators* (i.e., weight assignments). However, if we want to build a tree decomposition in a recursive manner, we need *integer separators* (i.e., subsets of vertices) that decompose the hypergraph into smaller parts.

2. Measuring the sizes of sets with a submodular function $b$ can lead to problems, since the size of the union of two sets can be much smaller than the sum of the sizes of the two sets. We need the property that, roughly speaking, removing a "large" part from a set makes it "much smaller." For example, if $A$ and $B$ are components of $H \setminus S$, and both $b(A)$ and $b(B)$ are large, then we need the property that both of them are much smaller than $b(A \cup B)$. Adler [1, Section 4.2] investigates the relation between some notion of highly connected sets and $f$-width, but assumes that $f$ is additive: if $A$ and $B$ do not touch, then $f(A \cup B) = f(A) + f(B)$. However, for a submodular function $b$, there is no reason to assume that additivity holds: for example, it very well may be that $b(A) = b(B) = b(A \cup B)$.

To overcome the first difficulty, we have to understand what fractional separation really means. The first question is whether fractional separation is equivalent to some notion of integral separation, perhaps up to constant factors. The first, naive, question is whether a fractional $(X,Y)$-separator of weight $w$ implies that there are $O(w)$ edges whose union is an $(X,Y)$-separator, i.e., there is a $(X,Y)$-separator $S$ with $\rho_H(S) = O(w)$. There is a simple counterexample showing that this is not true. It is well-known that for every integer $k > 0$ there is a hypergraph $H$ such that $\rho^*(H) = 2$ and $\rho(H) = k$. Let $V$ be the set of vertices of $H$ and let $H'$ be obtained from $H$ by extending it with two independent sets $X, Y$, each of size $k$, and connecting every vertex of $X \cup Y$ with every vertex of $V$. It is clear that there is a fractional $(X,Y)$-separator of weight 2, but every $(X,Y)$-separator $S$ has to fully contain at least one of $X, Y$, or $V$, implying $\rho_{H'}(S) \geq k$.

A less naive question is whether a fractional $(X,Y)$-separator with weight $w$ in $H$ implies that there exists an $(X,Y)$-separator $S$ with $\rho_H^*(S) = O(w)$ (or at most $f(w)$ for some function $f$). It can be shown that this is not true either: using the hypergraph family presented in [44, Section 5], one can construct counterexamples where the minimum weight of a fractional $(X,Y)$-separator is a constant, but $\rho_H^*(S)$ has to be arbitrarily large for every $(X,Y)$-separator $S$ (we omit the details).

We will characterize fractional separation in a very different way. We show that if there is a fractional $(A, B)$-separator of weight $w$, then there is an $(A, B)$-separator $S$ with $b(S) = O(w)$ for *every* edge-dominated monotone submodular function $b$. Note that this separator $S$ can be different for different functions $b$, so we are not claiming that there is a single $(A,B)$-separator $S$ that is small in every $b$. The converse is also true, thus this gives a novel characterization of fractional separation, tight up to a constant factor. This result is the key idea that allows us to move from the domain of submodular functions to the domain of pure hypergraph properties: if there is no $(A,B)$-separator such that $b(S)$ is small, then we know that there is no small fractional $(A,B)$-separator, which is a property of the hypergraph $H$ only and has no longer anything to do with the submodular function $b$.

To overcome the second difficulty, we introduce a transformation that turns a monotone submodular function $b$ on $V(H)$ into a function $b^*$ that encodes somehow the neighborhood structure of $H$ as well. The new function $b^*$ is no longer monotone and submodular, but it has a number of remarkable properties, for example, $b^*$ remains edge dominated and $b^*(S) \geq b(S)$ for every set $S \subseteq V(H)$, implying that $b^*$-width is not smaller than $b$-width. The main idea is to prove Theorem 5.1 for $b^*$-width instead of $b$-width (note that this makes the statement stronger). Because of the way $b^*$ encodes the neighborhoods, the second difficulty will disappear: for example, it will be true that $b^*(A \cup B) = b^*(A) + b^*(B)$ if there are no edges between $A$ and $B$, that is, $b^*$ is additive on disjoint components. Lemma 5.6 formulates (in a somewhat technical way) the exact property of $b^*$ that we will need. Furthermore, luckily it turns out that the result mentioned in the previous paragraph remains true with $b$ replaced by $b^*$: if there is a fractional $(A,B)$-separator of weight $w$, then there is an $(A,B)$-separator $S$ such that not only $b(S)$, but even $b^*(S)$ is $O(w)$.

## 5.1 The function $b^*$

We define the function $b^*$ the following way. Let $H$ be a hypergraph and let $b$ be a monotone submodular function defined on $V(H)$. Let $S_{V(H)}$ be the set of all permutations of $V(H)$. For a permutation $\pi \in S_{V(H)}$, let $N_\pi^-(v)$ be the neighbors of $v$ preceding $v$ in the ordering $\pi$. For $\pi \in S_{V(H)}$ and $Z \subseteq V(H)$, we define

$$\partial b_{\pi,Z}(v) := b(v \cup (N_\pi^-(v) \cap Z)) - b(N_\pi^-(v) \cap Z).$$

In other words, $\partial b_{\pi,Z}(v)$ is the marginal value of $v$ with respect to the set of its neighbors in $Z$ preceding it. We abbreviate $\partial b_{\pi,V(H)}$ by $\partial b_\pi$. As usual, we extend the definition to subsets by letting $\partial b_{\pi,Z}(S) := \sum_{v \in S} \partial b_{\pi,Z}(v)$. Furthermore, we define

$$b_\pi(Z) := \partial b_{\pi,Z}(Z) = \sum_{v \in Z} \partial b_{\pi,Z}(v),$$
$$b^*(Z) := \min_{\pi \in S_{V(H)}} b_\pi(Z).$$

Thus $b_\pi(Z)$ is the sum of the marginal values with respect to a given ordering, while $b^*(Z)$ is the smallest possible sum taken over all possible orderings. Let us prove some simple properties of the function $b^*$. Properties (1)–(3) and their proofs show why $b^*$ was defined this way, the other properties are only technical statements that we will need later.

**Proposition 5.2.** *Let $H$ be a hypergraph and let $b$ be a monotone submodular function defined on $V(H)$ with $b(\emptyset) = 0$. For every $\pi \in S_{V(H)}$ and $Z \subseteq V(H)$ we have*

*(1) $b_\pi(Z) \geq b(Z)$,*

*(2) $b^*(Z) \geq b(Z)$,*

*(3) $b_\pi(Z) = b(Z)$ if $Z$ is a clique,*

*(4) $\partial b_{\pi,Z_1}(v) \leq \partial b_{\pi,Z_2}(v)$ if $Z_2 \subseteq Z_1$,*

*(5) $\partial b_\pi(v) \leq \partial b_{\pi,Z}(v)$,*

*(6) $b^*(X \cup Y) \leq b^*(X) + b^*(Y)$.*

*Proof.* (1) We prove the statement by induction on $|Z|$; for $Z = \emptyset$, the claim is true (as $b(\emptyset) = 0$). Otherwise, let $v$ be the last element of $Z$ according to the ordering $\pi$. As $v$ is not preceding any element of $Z$, for every $u \in Z$ we have $N_\pi^-(u) \cap Z = N_\pi^-(u) \cap (Z \setminus v)$, and hence $\partial b_{\pi,Z}(u) = \partial b_{\pi,Z\setminus v}(u)$.

$$b_\pi(Z) = \sum_{u \in Z \setminus v} \partial b_{\pi,Z}(u) + \partial b_{\pi,Z}(v) = \sum_{u \in Z \setminus v} \partial b_{\pi,Z\setminus v}(u) + \partial b_{\pi,Z}(v)$$
$$= b_\pi(Z \setminus v) + \partial b_{\pi,Z}(v) \geq b(Z \setminus v) + b(v \cup (N_\pi^-(v) \cap Z)) - b(N_\pi^-(v) \cap Z) \geq b(Z).$$

In the first inequality, we used the induction hypothesis and the definition of $\partial b_{\pi,Z}(v)$; in the second inequality, we used the submodularity of $b$: the marginal value of $v$ with respect to $Z \setminus v$ is not greater than with respect to $N_\pi^-(v) \cap Z$.

(2) Follows immediately from (1) and from the definition of $b^*$.

(3) We prove the statement by induction on $|Z|$. As in (1), let $v$ be the last vertex of $Z$ in $\pi$. Note that since $Z$ is a clique, $N_\pi^-(v) \cap Z$ is exactly $Z \setminus v$.

$$b_\pi(Z) = \sum_{u \in Z \setminus v} \partial b_{\pi,Z}(u) + \partial b_{\pi,Z}(v) = \sum_{u \in Z \setminus v} \partial b_{\pi,Z\setminus v}(u) + b(v \cup (N_\pi^-(v) \cap Z)) - b(N_\pi^-(v) \cap Z)$$
$$= b_\pi(Z \setminus v) + b(v \cup (Z \setminus v)) - b(Z \setminus v) = b(Z \setminus v) + b(Z) - b(Z \setminus v) = b(Z).$$

(4) Follows from the submodularity of $b$: $\partial b_{\pi,Z_1}(v)$ is the marginal value of $v$ with respect to $N_\pi^-(v) \cap Z_1$, while $\partial b_{\pi,Z_2}(v)$ is the marginal value of $v$ with respect to the subset $N_\pi^-(v) \cap Z_2$ of $N_\pi^-(v) \cap Z_1$.

(5) Immediate from (4).

(6) Let $\pi_X$ be an ordering such that $b_{\pi_X}(X) = b^*(X)$ and define $\pi_Y$ similarly. Let us define ordering $\pi$ such that it starts with the elements of $X$, in the order of $\pi_X$, followed by the elements of $Y \setminus X$, in the order of $\pi_Y$, and completed by an arbitrary ordering of $V(H) \setminus (X \cup Y)$. It is clear that for every $v \in X$, we have $\partial b_{\pi,X \cup Y}(v) = \partial b_{\pi_X,X}(v)$. Furthermore, for every $v \in Y \setminus X$, $N_{\pi_Y}^-(v) \cap Y \subseteq N_\pi^-(v) \cap (X \cup Y)$: if $u$ is a neighbor of $v$ in $Y$ that precedes it in $\pi_Y$, then $u$ is either in $X$ or in $Y \setminus X$; in both cases $u$ precedes $v$ in $\pi$. Thus, similarly to (4), we have $\partial b_{\pi,X \cup Y}(v) \leq \partial b_{\pi_Y,Y}(v)$ for every $v \in Y \setminus X$: $\partial b_{\pi,X \cup Y}(v)$ is the marginal value of $v$ with respect to $N_\pi^-(v) \cap (X \cup Y)$, while $\partial b_{\pi_Y,Y}(v)$ is the marginal value of $v$ with respect to $N_{\pi_Y}^-(v) \cap Y$. Now we have

$$b^*(X \cup Y) \leq b_\pi(X \cup Y) = \sum_{v \in X \cup Y} \partial b_{\pi,X \cup Y}(v) \leq \sum_{v \in X} \partial b_{\pi_X,X}(v) + \sum_{v \in Y \setminus X} \partial b_{\pi_Y,Y}(v) \leq b^*(X) + b^*(Y).$$

$\square$

Prop. 5.2(3) implies that $\partial b_{w,Z}$ can be used to define a fractional independent set:

**Lemma 5.3.** *Let $H$ be a hypergraph and let $b$ be an edge-dominated monotone submodular function defined on $V(H)$ with $b(\emptyset) = 0$. Let $W \subseteq V(H)$ and let $\pi$ be an ordering of $W$. Let us define $\mu(v) = \partial b_{\pi,W}(v)$ for $v \in W$ and $\mu(v) = 0$ otherwise. Then $\mu$ is a fractional independent set of $H$ with $\mu(W) = b_\pi(W) \geq b^*(W)$.*

*Proof.* Let $e$ be an edge of $H$ and let $Z := e \cap W$. We have

$$\mu(e) = \mu(Z) = \partial b_{\pi,W}(Z) \leq \partial b_{\pi,Z}(Z) = b_\pi(Z) = b(Z) \leq 1,$$

where the fist inequality follows from Prop. 5.2(4), the last equality follows from Prop. 5.2(3), and the second inequality follows from the fact that $b$ is edge dominated. Furthermore, we have $\mu(W) = \partial b_{\pi,W}(W) = b_\pi(W) \geq b(W)$ from Prop. 5.2(1). $\square$

We close this section by proving the main property of $b^*$ that allows us to avoid the second difficulty described at the beginning of Section 5. First, although it is not used directly, let us state that $b^*$ is additive on sets that are independent from each other:

**Lemma 5.4.** *Let $H$ be a hypergraph, let $b$ be an edge-dominated monotone submodular function defined on $V(H)$ with $b(\emptyset) = 0$, and let $A, B \subseteq V(H)$ be disjoint sets such that there is no edge intersecting both $A$ and $B$. Then $b^*(A \cup B) = b^*(A) + b^*(B)$.*

*Proof.* By Prop. 5.2(6), we have to show only $b^*(A \cup B) \geq b^*(A) + b^*(B)$. Let $\pi$ be an ordering of $V(H)$ such that $b_\pi(A \cup B) = b^*(A \cup B)$; we can assume that $\pi$ starts with the vertices of $A \cup B$. Since there is no edge that intersects both $A$ and $B$, and no vertex outside $A \cup B$ precedes a vertex $u \in A \cup B$, we have $N_\pi^-(u) \subseteq A$ for every $u \in A$ and $N_\pi^-(u) \subseteq B$ for every $u \in B$. Thus $\partial b_{\pi,A \cup B}(u) = \partial b_{\pi,A}(u)$ for every $u \in A$ and $\partial b_{\pi,A \cup B}(u) = \partial b_{\pi,B}(u)$ for every $u \in B$. Therefore, $b^*(A \cup B) = b_\pi(A \cup B) = b_\pi(A) + b_\pi(B) \geq b^*(A) + b^*(B)$, what we had to show. □

The actual statement that we use is more complicated than Lemma 5.4: there can be edges between $A$ and $B$, but we assume that there is a small $(A,B)$-separator. We want to generalize the following simple statement to our setting:

**Proposition 5.5.** *Let $G$ be a graph, $W \subseteq V(G)$ a set of vertices, $A, B \subseteq W$ two disjoint subsets, and an $(A,B)$-separator $S$. If $|S| < |A|, |B|$, then $(C \cap W) \cup S < |W|$ for every component $C$ of $G \setminus S$.*

The proof of Prop. 5.5 is easy to see: every component $C$ of $G \setminus S$ is disjoint from either $A$ or $B$, thus $|C \cap W|$ is at most $|W| - \min\{|A|,|B|\} < |W| - |S|$, implying that $|(C \cap W) \cup S|$ is less than $|W|$. In our setting, we want to measure the size of the sets using the function $b^*$, not by the number of vertices. More precisely, we measure the size of $S$ and $(C \cap W) \cup S$ using $b^*$, while the size of $W$, $A$, and $B$ are measured using the fractional independent set $\mu$ defined by Lemma 5.3. The reason for this will be apparent in the proof of Lemma 5.10: we want to claim that if such a separator $S$ does not exist for any $A, B \subseteq W$, then $W$ is a $(\mu, \lambda)$-connected set for this fractional independent set $\mu$.

**Lemma 5.6.** *Let $H$ be a hypergraph, let $b$ be an edge-dominated monotone submodular function defined on $V(H)$ with $b(\emptyset) = 0$ and let $W$ be a set of vertices. Let $\pi_W$ be an ordering of $V(H)$, and let $\mu(v) := \partial b_{\pi_W,W}(v)$ for $v \in W$ and $\mu(v) = 0$ otherwise. Let $A, B \subseteq W$ be two disjoint sets, and let $S$ be an $(A,B)$-separator. If $b^*(S) < \mu(A), \mu(B)$, then $b^*((C \cap W) \cup S) < \mu(W)$ for every component $C$ of $H \setminus S$.*

*Proof.* Let $C$ be a component of $H \setminus S$ and let $Z := (C \cap W) \cup S$. Let $\pi_S$ be the ordering reaching the minimum in the definition of $b^*(S)$. Let us define the ordering $\pi$ that starts with $S$ in the order of $\pi_S$, followed by $C \cap W$ in the order of $\pi_W$, and finished by an arbitrary ordering of the remaining vertices. It is clear that for every $v \in S$, we have $\partial b_{\pi,Z}(v) = \partial b_{\pi_S,S}(v)$. Let us consider a vertex $v \in C \cap W$ and let $u \in W$ be a neighbor of $v$ that precedes it in $\pi_W$. Since $v \in C$ and $C$ is a component of $H \setminus S$, either $u \in S$ or $u \in C \cap W$. In both cases, $u$ precedes $v$ in $\pi$. This means that $N_{\pi_W}^-(v) \cap W \subseteq N_\pi^-(v) \cap Z$, which implies that $\partial b_{\pi,Z}(v) \leq \partial b_{\pi_W,W}(v) = \mu(v)$ for every $v \in C \cap W$. As $S$ separates $A$ and $B$, component $C$ intersects at most one of $A$ and $B$; suppose, without loss of generality, that $C$ is disjoint from $A$. Thus

$$b^*(Z) \leq b_\pi(Z) = \sum_{v \in S} \partial b_{\pi,Z}(v) + \sum_{v \in C \cap W} \partial b_{\pi,Z}(v) \leq b^*(S) + \mu(C \cap W) < \mu(A) + \mu(W \setminus A) = \mu(W).$$

□

## 5.2 Submodular separation

This section is devoted to understanding what fractional separation means: we show that having a small fractional $(A,B)$-separator is essentially equivalent to the property that for every edge-dominated submodular function $b$, there is an $(A,B)$-separator $S$ such that $b(S)$ is small. The proof is based on a standard trick that is often used for rounding fractional solutions for separation problems: we define a distance function and show by an averaging argument that cutting at some distance $t$ gives a small separator. However, in our setting, we need significant new ideas to make this trick work: the main difficulty is that the cost function $b$ is defined on *subsets* of vertices and is not a modular function defined by the cost of vertices. To overcome this problem, we use the definitions in Section 5.1 (in particular, the function $\partial b_\pi(v)$) to assign a cost to every single vertex.

**Theorem 5.7.** *Let $H$ be a hypergraph, $X, Y \subseteq V(H)$ two sets of vertices, and $b : V(H) \to \mathbb{R}^+$ an edge-dominated monotone submodular function with $b(\emptyset) = 0$. Suppose that $s$ is a fractional $(X,Y)$-separator of weight at most $w$. Then there is an $(X,Y)$-separator $S \subseteq V(H)$ with $b(S) \leq b^*(S) = O(w)$.*

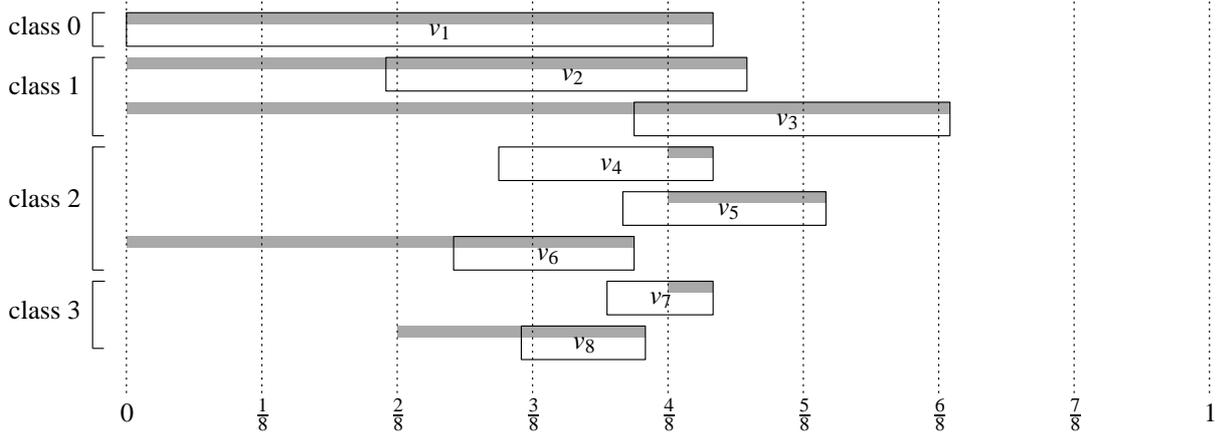

Figure 3: The intervals corresponding to a directed path $v_1, \ldots, v_8$. The shaded lines show the offsets of the vertices.

*Proof.* The total weight of the edges covering a vertex $v$ is $\sum_{e \in E(H), v \in e} s(e)$; let us define $x(v) := \min\{1, \sum_{e \in E(H), v \in e} s(e)\}$. It is clear that if $P$ is a path from $X$ to $Y$, then $\sum_{v \in P} x(v) \geq 1$. We define the distance $d(v)$ to be the minimum of $\sum_{v' \in P} x(v')$, taken over all paths from $X$ to $v$ (this means that $d(v) > 0$ is possible for some $v \in X$). It is clear that $d(v) \geq 1$ for every $v \in Y$. Let us associate the closed interval $\iota(v) = [d(v) - x(v), d(v)]$ to each vertex $v$. If $v$ is in $X$, then the left endpoint of $\iota(v)$ is 0, while if $v$ is in $Y$, then the right endpoint of $\iota(v)$ is at least 1.

Let $u$ and $v$ be two adjacent vertices in $H$ such that $d(u) \leq d(v)$. It is easy to see that $d(v) \leq d(u) + x(u)$: there is a path $P$ from $X$ to $u$ such that $\sum_{u' \in P} x(u') = d(u)$, thus the path $P'$ obtained by appending $v$ to $P$ has $\sum_{v' \in P'} x(v') = \sum_{u' \in P} x(u') + x(v) = d(u) + x(v)$. Therefore, we have:

*Claim 1.* If $u$ and $v$ are adjacent, then $\iota(u) \cap \iota(v) \neq \emptyset$.

The *class* of a vertex $v \in V(H)$ is the largest integer $\kappa(v)$ such that $x(v) \leq 2^{-\kappa(v)}$, and we define $\kappa(v) := \infty$ if $x(v) = 0$. Recall that $x(v) \leq 1$, thus $\kappa(v)$ is nonnegative. The *offset* of a vertex $v$ is the unique value $0 \leq \alpha < 2 \cdot 2^{-\kappa(v)}$ such that $d(v) = i(2 \cdot 2^{-\kappa(v)}) + \alpha$ for some integer $i$. Let us define an ordering $\pi = (v_1, \ldots, v_n)$ of $V(H)$ such that

- $\kappa(v)$ is nondecreasing,
- among vertices having the same class, the offset is nondecreasing.

Let directed graph $D$ be the orientation of the primal graph of $H$ such that if $v_i$ and $v_j$ are adjacent and $i < j$, then there is a directed edge $\overrightarrow{v_i v_j}$ in $D$. Figure 3 shows a directed path in $D$. If $P$ is a directed path in $D$, then the *width* of $P$ is the length of the interval $\bigcup_{v \in P} \iota(v)$ (note that by Claim 1, this union is indeed an interval). The following claim bounds the maximum possible width of a directed path:

*Claim 2.* If $P$ is a directed path $D$ starting at $v$, then the width of $P$ is at most $16x(v)$.

*Proof.* We first prove that if every vertex of $P$ has the same class $\kappa(v)$, then the width of $P$ is at most $4 \cdot 2^{-\kappa(v)}$. Since the class is nondecreasing along the path, we can partition the path into subpaths such that every vertex in a subpath has the same class and the classes are distinct on the different subpaths. The width of $P$ is at most the sum of the widths of the subpaths, which is at most $\sum_{i \geq \kappa(v)} 4 \cdot 2^{-i} = 8 \cdot 2^{-\kappa(v)} \leq 16x(v)$.

Suppose now that every vertex of $P$ has the same class $\kappa(v)$ as the first vertex $v$ and let $h := 2^{-\kappa(v)}$. As the offset is nondecreasing, path $P$ can be partitioned into two parts: a subpath $P_1$ containing vertices with offset less than $h$, followed by a subpath $P_2$ containing vertices with offset at least $h$ (one of $P_1$ and $P_2$ can be empty).

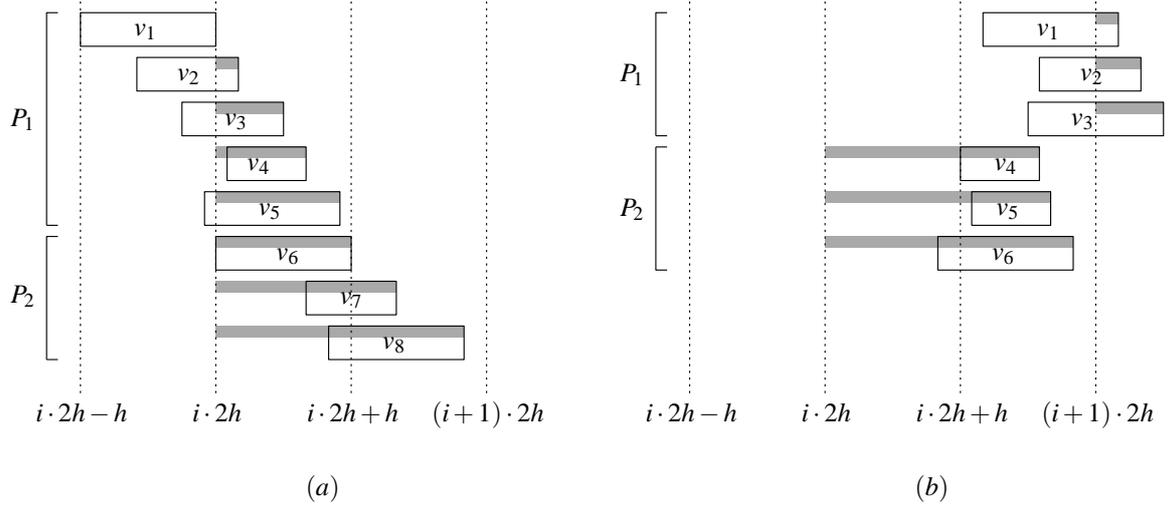

Figure 4: Proof of Claim 2: Two examples of directed paths where every vertex has the same class $\kappa$ (and $h := 2^{-\kappa}$). The shaded lines show the offsets of the vertices.

See Figure 4 for examples. We show that each of $P_1$ and $P_2$ has width at most $2h$, which implies that the width of $P$ is at most $4h$. Observe that if $u \in P_1$ and $\iota(u)$ contains a point $i \cdot 2h - h$ for some integer $i$, then, considering $x(u) \leq h$ and the bounds on the offset of $u$, this is only possible if $\iota(u) = [i \cdot 2h - h, i \cdot 2h]$, i.e., $i \cdot 2h - h$ is the left endpoint of $\iota(u)$. Thus if $I_1 = \bigcup_{u \in P_1} \iota(u)$ contains $i \cdot 2h - h$, then it is the left endpoint of $I_1$. Therefore, $I_1$ can contain $i \cdot 2h - h$ for at most one value of $i$, which immediately implies that the length of $I_1$ is at most $2h$.

We argue similarly for $P_2$. If $u \in P_2$, then $\iota(u)$ can contain the point $i \cdot 2h$ only if $\iota(u) = [i \cdot 2h, i \cdot 2h + h]$. Thus if $I_2 = \bigcup_{u \in P_2} \iota(u)$ contains $i \cdot 2h$, then it is the left endpoint of $I_2$. We get that $I_2$ can contain $i \cdot 2h$ for at most one value of $i$, which immediately implies that the width of $I_2$ is at most $2h$. This concludes the proof of Claim 2. ⌟

Let $c(v) := \partial b_\pi(v)$.

**Claim 3.** $\sum_{v \in V(H)} x(v) c(v) \leq w$.

*Proof.* Let us examine the contribution of an edge $e \in E(H)$ with value $s(e)$ to the sum. For every vertex $v \in e$, edge $e$ increases the value $x(v)$ by at most $s(e)$. Thus the total contribution of edge $e$ is at most

$$s(e) \cdot \sum_{v \in e} c(v) = s(e) \cdot \sum_{v \in e} \partial b_\pi(v) \leq s(e) \cdot \sum_{v \in e} \partial b_{\pi,e}(v) = s(e) b_\pi(e) = s(e) b(e) \leq s(e),$$

where the first inequality follows Prop. 5.2(5); the last equality follows form Prop. 5.2(3); the last inequality follows from the fact that $b$ is edge dominated. Therefore, $\sum_{v \in V(H)} x(v) c(v) \leq \sum_{e \in E(H)} s(e) \leq w$, proving Claim 3. ⌟

Let $S$ be a set of vertices. We define $\hat{S}$ to be the "inneighbor closure" of $S$, that is, the set of all vertices from which a vertex of $S$ is reachable on a directed path in $D$ (in particular, this means that $S \subseteq \hat{S}$).

**Claim 4.** For every $S \subseteq V(H)$, $\sum_{v \in \hat{S}} c(v) = b_\pi(\hat{S})$.

*Proof.* Observe that for any $v \in \hat{S}$, every inneighbor of $v$ is also in $\hat{S}$, hence $N_\pi^-(v) \subseteq \hat{S}$. Therefore, $\partial b_{\pi,\hat{S}}(v) = \partial b_\pi(v) = c(v)$ and Claim 4 follows. ⌟

Let $S(t)$ be the set of all vertices $v \in V(H)$ for which $t \in \iota(v)$. Observe that for every $0 \leq t \leq 1$, the set $S(t)$ (and hence $\hat{S}(t)$) separates $X$ from $Y$. We use an averaging argument to show that there is a $0 \leq t \leq 1$ for which $b_\pi(\hat{S}(t))$ is $O(w)$. As $b^*(\hat{S}(t)) \leq b_\pi(\hat{S}(t))$, the set $\hat{S}(t)$ satisfies the requirement of the lemma.

If we are able to show that $\int_0^1 b_\pi(\hat{S}(t))dt = O(w)$, then the existence of the required $t$ clearly follows. Let $I_v(t) = 1$ if $v \in \hat{S}(t)$ and let $I_v(t) = 0$ otherwise. If $I_v(t) = 1$, then there is a path $P$ in $D$ from $v$ to a member of $S(t)$. By Claim 2, the width of this path is at most $16x(v)$, thus $t \in [d(v) - 16x(v), d(v) + 15x(v)]$. Therefore, $\int_0^1 I_v(t)dt \leq 31x(v)$. Now we have

$$\int_0^1 b_\pi(\hat{S}(t))dt = \int_0^1 \sum_{v \in \hat{S}(t)} c(v)dt = \int_0^1 \sum_{v \in V(H)} c(v)I_v(t)dt = \sum_{v \in V(H)} c(v) \int_0^1 I_v(t)dt \leq 31 \sum_{v \in V(H)} x(v)c(v) \leq 31w$$

(we used Claim 4 in the first equality and Claim 3 in the last inequality). □

Although it is not used in this paper, we can prove the converse of Theorem 5.7 in a very simple way.

**Theorem 5.8.** *Let $H$ be a hypergraph, and let $X, Y \subseteq V(H)$ be two sets of vertices. Suppose that for every edge-dominated monotone submodular function on $H$ with $b(\emptyset) = 0$, there is an $(X,Y)$-separator $S$ with $b(S) \leq w$. Then there is a fractional $(X,Y)$-separator of weight at most $w$.*

*Proof.* If there is no fractional $(X,Y)$-separator of weight at most $w$, then by LP duality, there is an $(X,Y)$-flow $F$ of value greater than $w$. Let $b(S)$ be defined as the total weight of the paths in $F$ intersecting $S$; it is easy to see that $f$ is a monotone submodular function, and since $F$ is a flow, $b(e) \leq 1$ for every $e \in E(H)$. Thus by assumption, there is an $(X,Y)$-separator $S$ with $b(S) \leq w$. However, every $X-Y$ path of $F$ intersects $(X,Y)$-separator $S$, which implies $b(S) > w$, a contradiction. □

The problem of finding a small separator in the sense of Theorem 5.7 might seem related to submodular function minimization at a first look. We close this section by pointing out that finding an $(A,B)$-separator $S$ with $b(S)$ small for a given submodular function $b$ is *not* an instance of submodular function minimization, and hence the well-known algorithms (see [36, 37, 52]) cannot be used for this problem. If a submodular function $g(X)$ describes the weight of the *boundary* of $X$, then finding a small $(A,B)$-separator is equivalent to minimizing $g(X)$ subject to $A \subseteq X$, $X \cap B = \emptyset$, which can be expressed as an instance of submodular function minimization (and hence solvable in polynomial time). In our case, however, $b(S)$ is the weight of $S$ itself, which means that we have to minimize $g(S)$ subject to $S$ being an $(A,B)$-separator and this latter constraint cannot be expressed in the framework of submodular function minimization. A possible workaround is to define $\delta(X)$ as the neighborhood of $X$ (the set of vertices outside $X$ adjacent to $X$) and $b'(X) := b(\delta(S))$; now minimizing $b'(X)$ subject to $A \subseteq X \cup \delta(X)$, $X \cap B = \emptyset$ is the same as finding an $(X,Y)$-separator $S$ minimizing $b(S)$. However, the function $b'$ is not necessarily a submodular function in general. Therefore, transforming $b$ to $b'$ this way does not lead to a polynomial-time algorithm using submodular function minimization. In fact, it is quite easy to show that finding an $(A,B)$-separator $S$ with $b(S)$ minimum possible can be an NP-hard problem even if $b$ is a submodular function of very simple form.

**Theorem 5.9.** *Given a graph $G$, subsets of vertices $X, Y$, and collection $\mathcal{S}$ of subsets of vertices, it is NP-hard to find an $(X,Y)$-separator that intersects the minimum number of members of $\mathcal{S}$.*

*Proof.* The proof is by reduction from 3-COLORING. Let $H$ be a graph with $n$ vertices and $m$ edges; we identify the vertices of $H$ with the integers from 1 to $n$. We construct a graph $G$ consisting of $3n+2$ vertices, vertex sets $X, Y$, and a collection $\mathcal{S}$ of $6m$ sets such that there is an $(X,Y)$-separator $S$ intersecting at most $3m$ members of $\mathcal{S}$ if and only if $G$ is 3-colorable.

The graph $G$ consists of two vertices $x, y$, and for every $1 \leq i \leq n$, a path $xv_{i,1}v_{i,2}v_{i,3}y$ of length 4 connecting $x$ and $y$. The collection $\mathcal{S}$ is constructed such that for every edge $ij \in E(H)$ and $1 \leq a, b \leq 3$, $a \neq b$, there is a corresponding set $\{v_{i,a}, v_{j,b}, x, y\}$. Let $X := \{x\}$ and $Y := \{y\}$. Observe that the set $\{v_{i,a}, v_{j,b}\}$ intersects exactly 3 sets of $\mathcal{S}$ if $a \neq b$ and exactly 4 sets of $\mathcal{S}$ if $a = b$.

Let $c : V(G) \to \{1,2,3\}$ be a 3-coloring of $G$. The set $S = \{v_{i,c(i)} \mid 1 \leq i \leq n\}$ is clearly an $(X,Y)$-separator. For every $ij \in E(G)$, separator $S$ intersects only 3 of the 6 sets $\{v_{i,a}, v_{j,b}, x, y\}$. Therefore, $S$ intersects exactly $3m$ members of $\mathcal{S}$.

Consider now an $(X,Y)$-separator $S$ intersecting at most $3m$ members of $\mathcal{S}$. Since every member of $\mathcal{S}$ contains both $x$ and $y$, it follows that $x, y \notin S$. Thus $S$ has to contain at least one internal vertex of every path $xv_{i,1}v_{i,2}v_{i,3}y$. For every $1 \le i \le n$, let us fix a vertex $v_{i,c(i)} \in S$. We claim that $c$ is a 3-coloring of $G$. For every $ij \in E(G)$, $S$ intersects at least 3 of the sets $\{v_{i,a}, v_{i,b}, x, y\}$, and intersects 4 of them if $c(i) = c(j)$. Thus the assumption that $S$ intersects at most $3m$ members of $\mathcal{S}$ immediately implies that $c$ is a proper 3-coloring. □

### 5.3 Obtaining a highly connected set

The following lemma is the same as the main result of Section 5 (Theorem 5.1) we are trying to prove, with the exception that $b$-width is replaced by $b^*$-width. By Prop 5.2(2), $b^*(S) \ge b(S)$ for every set $S \subseteq V(H)$, thus $b^*$-width is not less than $b$-width. Therefore, the following is actually a stronger statement and immediately implies Theorem 5.1.

**Lemma 5.10.** *For every sufficiently small constant $\lambda > 0$, the following holds. Let $b$ be an edge-dominated monotone submodular function of $H$ with $b(\emptyset) = 0$. If the $b^*$-width of $H$ is greater than $\frac{3}{2}(w+1)$, then $\mathrm{con}_\lambda(W) \ge w$.*

*Proof.* Suppose that $\lambda < 1/c$, where $c$ is the universal constant of Lemma 5.7 hidden by the big-O notation. Suppose that $\mathrm{con}_\lambda(W_0) < w$, that is, there is no fractional independent set $\mu$ and $(\mu, \lambda)$-connected set $W_0$ with $\mu(W_0) \ge w$. We show that $H$ has a tree decomposition of $b^*$-width at most $\frac{3}{2}(w+1)$, or more precisely, we show the following stronger statement:

> For every subhypergraph $H'$ of $H$ and every $W_0 \subseteq V(H')$ with $b^*(W_0) \le w+1$, there is a tree decomposition of $H'$ having $b^*$-width at most $\frac{3}{2}(w+1)$ such that $W_0$ is contained in one of the bags.

We prove this statement by induction on $|V(H')|$. If $b^*(V(H')) \le \frac{3}{2}(w+1)$, then a decomposition consisting of a single bag proves the statement. Otherwise, let $W$ be a superset of $W_0$ such that $w \le b^*(W) \le w+1$; let us choose a $W$ that is inclusionwise maximal with respect to this property. Observe that there has to be at least one such set: from the fact that $b^*(v) \le 1$ for every vertex $v$ and from Prop. 5.2(6), we know that adding a vertex increases $b^*(W)$ by at most 1. Since $b^*(V(H')) \ge \frac{3}{2}(w+1)$, by adding vertices to $W_0$ in an arbitrary order, we eventually find a set $W$ with $b^*(W) \ge w$, and the first such set satisfies $b^*(W) \le w+1$ as well.

Let $\pi$ be an ordering of $V(H')$ such that $b_\pi(W) = b^*(W)$. As in Lemma 5.3, let us define the fractional independent set $\mu$ by $\mu(v) := \partial b_{\pi,W}(v)$ if $v \in W$ and $\mu(v) = 0$ otherwise. Clearly, we have $\mu(W) = b^*(W) \ge w$.

By assumption, $W$ is not $(\mu, \lambda)$-connected, hence there are disjoint sets $A, B \subseteq W$ and a fractional $(A, B)$-separator of weight less than $\lambda \cdot \min\{\mu(A), \mu(B)\}$. Thus by Lemma 5.7, there is an $(A, B)$-separator $S \subseteq V(H')$ with $b^*(S) < \min\{\mu(A), \mu(B)\} \le \mu(W)/2 \le (w+1)/2$ (the second inequality follows from the fact that $A$ and $B$ are disjoint subsets of $W$). Let $C_1, \ldots, C_r$ be the connected components of $H' \setminus S$; by Lemma 5.6, $b^*((C_i \cap W) \cup S) < b_\pi(W) = b^*(W) \le w+1$ for every $1 \le i \le r$. As $b^*(V(H')) \ge \frac{3}{2}(w+1)$ and $b^*(S) \le (w+1)/2$, it is not possible that $S = V(H')$, hence $r > 0$. It is not possible that $r = 1$ either: $(C_1 \cap W) \cup S$ would be a superset of $W$ with $b^*$-value less than $w+1$, and (as $b^*(V(H')) \ge \frac{3}{2}(w+1)$) we could find a set between $(C_1 \cap W) \cup S$ and $V(H')$ contradicting the maximality of the choice of $W$. Thus $r \ge 2$, which means that each hypergraph $H'_i := H'[C_i \cup S]$ has strictly fewer vertices than $H'$ for every $1 \le i \le r$.

By the induction hypothesis, each $H'_i$ has a tree decomposition $\mathcal{T}_i$ having $b^*$-width at most $\frac{3}{2}(w+1)$ such that $W_i := (C_i \cap W) \cup S$ is contained in one of the bags. Let $B_i$ be the bag of $\mathcal{T}_i$ containing $W_i$. We build a tree decomposition $\mathcal{T}$ of $H$ by joining together the tree decompositions $\mathcal{T}_1, \ldots, \mathcal{T}_r$: let $B_0 := W_0 \cup S$ be a new bag that is adjacent to bags $B_1, \ldots, B_r$. It can be easily verified that $\mathcal{T}$ is indeed a tree decomposition of $H'$. Furthermore, by Prop. 5.2(6), $b^*(B_0) \le b^*(W_0) + b^*(S) < w+1 + (w+1)/2 = \frac{3}{2}(w+1)$ and by the assumptions on $\mathcal{T}_1, \ldots, \mathcal{T}_r$, every other bag has $b^*$ value at most $\frac{3}{2}(w+1)$. □

## 6 From highly connected sets to embeddings

The main result of this section is showing that the existence of highly connected sets imply that the hypergraph has large embedding power. Recall from Section 2 that $W$ is a $(\mu, \lambda)$-connected set for some $\lambda > 0$ and fractional independent set $\mu$ if for every disjoint $X, Y \subseteq W$, the minimum weight of a fractional $(X, Y)$-separator is at least $\lambda \cdot \{\mu(X), \mu(Y)\}$. We denote by $\mathrm{con}_\lambda(H)$ the maximum value of $\mu(W)$ taken over every fractional independent set $\mu$ and $(\mu, \lambda)$-connected set $W$. Recall also that the edge depth of an embedding $\phi$ of $G$ into $H$ is the maximum of $\sum_{v \in V(G)} |\phi(v) \cap e|$, taken over every $e \in E(H)$.

**Theorem 6.1.** *For every sufficiently small $\lambda > 0$ and hypergraph $H$, there is a constant $m_{H,\lambda}$ such that every graph $G$ with $m \geq m_{H,\lambda}$ edges has an embedding into $H$ with edge depth $O(m/(\lambda^{\frac{3}{2}} \mathrm{con}_\lambda^{\frac{1}{4}}(H)))$. Furthermore, there is an algorithm that, given $G$, $H$, and $\lambda$, produces such an embedding in time $f(H, \lambda) n^{O(1)}$.*

In other words, Theorem 6.1 gives a lower bound on the embedding power of $H$:

**Corollary 6.2.** *For every sufficiently small $\lambda > 0$ and hypergraph $H$, $\mathrm{emb}(H) = \Omega(\lambda^{\frac{3}{2}} \mathrm{con}_\lambda^{\frac{1}{4}}(H))$.*

Theorem 6.1 is stated in algorithmic form, since the reduction in the hardness result of Section 7 needs to find such embeddings. For the proof, our strategy is similar to the embedding result of [43]: we show that a highly connected set implies that a uniform concurrent flow exists, the paths appearing in the uniform concurrent flow can be used to embed (a blowup of) the line graph of a complete graph, and every graph has an appropriate embedding in the line graph of a complete graph. To make this strategy work, we need generalizations of concurrent flows, multicuts, and multicommodity flows in our hypergraph setting and we need to obtain results that connect these concepts to highly connected sets. Some of these results are similar in spirit to the $O(\sqrt{n})$-approximation algorithms appearing in the combinatorial optimization literature [30, 31, 3]. However, those approximation algorithms are mostly based on clever rounding of fractional solutions, while in our setting rounding is not an option: as discussed in Section 5, the existence of a fractional $(X, Y)$-separator of small weight does not imply the existence of a small integer separator. Thus we have to work directly with the fractional solution and use the properties of the highly connected set.

It turns out that the right notion of uniform concurrent flow for our purposes is a collection of flows that connect cliques: that is, a collection $F_{i,j}$ ($1 \leq i < j \leq k$) of compatible flows, each of value $\varepsilon$, such that $F_{i,j}$ is a $(K_i, K_j)$-flow, where $K_1, \ldots, K_k$ are disjoint cliques. Thus our first goal is to find a highly connected set that can be partitioned into $k$ cliques in an appropriate way.

## 6.1 Highly connected sets with cliques

Let $(X_1, Y_1), \ldots, (X_k, Y_k)$ be pairs of vertex sets such that the minimum weight of a fractional $(X_i, Y_i)$-separator is $s_i$. Analogously to multicut problems in combinatorial optimization, we investigate weight assignments that *simultaneously* separate all these pairs. Clearly, the minimum weight of such an assignment is at least the minimum of the $s_i$'s and at most the sum of the $s_i$'s. The following lemma shows that in a highly connected set, such a simultaneous separator cannot be very efficient: roughly speaking, its weight is at least the square root of the sum of the $s_i$'s.

**Lemma 6.3.** *Let $\mu$ be a fractional independent set in hypergraph $H$ and let $W$ be a $(\mu, \lambda)$-connected set for some $0 < \lambda \leq 1$. Let $(X_1, \ldots, X_k, Y_1, \ldots, Y_k)$ be a partition of $W$, let $w_i := \min\{\mu(X_i), \mu(Y_i)\} \geq 1/2$, and let $w := \sum_{i=1}^{k} w_i$. Let $s : E(H) \to \mathbb{R}^+$ be a weight assignment of total weight $p$ such that $s$ is a fractional $(X_i, Y_i)$-separator for every $1 \leq i \leq k$. Then $p \geq (\lambda/7) \cdot \sqrt{w}$.*

*Proof.* Let us define the function $s'$ by $s'(e) = 6s(e)$ and let $x(v) := \sum_{e \in E(H), v \in e} s'(e)$. We define the distance $d(u, v)$ to be the minimum of $\sum_{v' \in P} x(v')$, taken over all paths $P$ from $u$ to $v$. It is clear that the triangle inequality holds, i.e., $d(u, v) \leq d(u, z) + d(z, v)$ for every $u, v, z \in V(H)$. If $s$ covers every $u-v$ path, then $d(u, v) \geq 6$: every edge $e$ intersecting a $u-v$ path $P$ contributes at least $s'(e)$ to the sum $\sum_{v' \in P} x(v')$ (as $e$ can intersect $P$ in more than one vertices, $e$ can increase the sum by more than $s'(e)$). On the other hand, if $d(u, v) \geq 2$, then $s'$ covers every $u-v$ path. Clearly, it is sufficient to verify this for minimal paths. Such a path $P$ can intersect an edge $e$ at most twice, hence $e$ contributes at most $2s'(e)$ to the sum $\sum_{v' \in P} x(v') \geq 2$, implying that the edges intersecting $P$ have total weight at least 1 in $s'$.

Suppose for contradiction that $p < (\lambda/7) \cdot \sqrt{w}$, that is, $w > 49p^2/\lambda^2$. As $s$ is an $(X_i, Y_i)$-separator, we have that $p \geq 1$. Let $A := \emptyset$ and $B := \bigcup_{i=1}^{k}(X_i \cup Y_i)$. Note that $\mu(B) \geq 2\sum_{i=1}^{k} w_i = 2w$. We will increase $A$ and decrease $B$ while maintaining the invariant condition that the distance of $A$ and $B$ is at least 2 in $d$. Let $T$ be the smallest integer such that $\sum_{i=1}^{T} w_i > 6p/\lambda$; if there is no such $T$, then $w \le 6p/\lambda$, a contradiction. As $w_i \ge 1/2$ for every $i$, it follows that $T \le 12p/\lambda + 1 \le 13p/\lambda$ (since $p \ge 1$ and $\lambda \le 1$).

For $i = 1, 2, \ldots, T$, we perform the following step. Let $X_i'$ (resp., $Y_i'$) be the set of all vertices of $W$ that are at distance at most 2 from $X_i$ (resp., $Y_i$). As the distance of $X_i$ and $Y_i$ is at least 6, by the triangle inequality the distance of $X_i'$ and $Y_i'$ is at least 2, hence $s'$ is a fractional $(X_i', Y_i')$-separator. Since $W$ is $(\mu, \lambda)$-connected and $s'$ is an assignment of weight $6p$, we have $\min\{\mu(X_i'), \mu(Y_i')\} \le 6p/\lambda$. If $\mu(X_i') \le 6p/\lambda$, then let us put $X_i$ (note: not $X_i'$) into $A$ and let us remove $X_i'$ from $B$. The set $X_i'$, which we remove from $B$, contains all the vertices that are at distance at most 2 from any new vertex in $A$, hence it remains true that the distance of $A$ and $B$ is at least 2. Similarly, if $\mu(X_i') > 6p/\lambda$ and $\mu(Y_i') \le 6p/\lambda$, then let us put $Y_i$ into $A$ and let us remove $Y_i'$ from $B$. Note that we may put a vertex into $A$ even if it was removed from $B$ in an earlier step.

In the $i$-th step of the procedure, we increase $\mu(A)$ by at least $w_i$ (as $\mu(X_i), \mu(Y_i) \ge w_i$ and these sets are disjoint from the sets already contained in $A$) and $\mu(B)$ is decreased by at most $6p/\lambda$. Thus at the end of the procedure, we have $\mu(A) \ge \sum_{i=1}^{T} w_i > 6p/\lambda$ and

$$\mu(B) \ge 2w - T \cdot 6p/\lambda > 98p^2/(\lambda^2) - (13p/(\lambda))(6p/\lambda) > 6p/\lambda,$$

that is, $\min\{\mu(A), \mu(B)\} > 6p/\lambda$. By construction, the distance of $A$ and $B$ is at least 2, thus $s'$ is a fractional $(A, B)$-separator of weight exactly $6p$, contradicting the assumption that $W$ is $(\mu, \lambda)$-connected. □

In the rest of the section, we need a more constrained notion of flow, where the endpoints "respect" a particular fractional independent set. Let $\mu_1, \mu_2$ be fractional independent sets of hypergraph $H$ and let $X, Y \subseteq V(H)$ be two (not necessarily disjoint) sets of vertices. A $(\mu_1, \mu_2)$-demand $(X, Y)$-flow is an $(X, Y)$-flow $F$ such that for each $x \in X$, the total weight of the paths in $F$ having first endpoint $x$ is at most $\mu_1(x)$, and similarly, the total weight of the paths in $F$ having second endpoint $y \in Y$ is at most $\mu_2(y)$. Note that there is no bound on the weight of the paths going through an $x \in X$, we only bound the paths whose first/second endpoint is $x$. The definition is particularly delicate if $X$ and $Y$ are not disjoint, in this case, a vertex $z \in X \cap Y$ can be the first endpoint of some paths and the second endpoint of some other paths, or it can be even both the first and second endpoint of a path of length 0. We use the abbreviation $\mu$-demand for $(\mu, \mu)$-demand.

The following lemma shows that if a flow connects a set $U$ with a highly connected set $W$, then $U$ is highly connected as well ("$W$ can be moved to $U$"). This observation will be used in the proof of Lemma 6.5, where we locate cliques and show that their union is highly connected, since there is a flow that connects the cliques to a highly connected set.

**Lemma 6.4.** *Let $H$ be a hypergraph, $\mu_1, \mu_2$ fractional independent sets, and $W \subseteq V(H)$ a $(\mu_1, \lambda)$-connected set for some $0 < \lambda \le 1$. Suppose that $U \subseteq V(H)$ is a set of vertices and $F$ is a $(\mu_1, \mu_2)$-demand $(W, U)$-flow of value $\mu_2(U)$. Then $U$ is $(\mu_2, \lambda/6)$-connected.*

*Proof.* Suppose that there are disjoint sets $A, B \subseteq U$ and a fractional $(A, B)$-separator $s$ of weight $w < (\lambda/6) \cdot \min\{\mu_2(A), \mu_2(B)\}$. (Note that this means $\mu_2(A), \mu_2(B) > 6w/\lambda \ge 6w$.) For a path $P$, let $s(P) = \sum_{e \in E(H), e \cap P \neq \emptyset} s(e)$ be the total weight of the edges intersecting $P$. Let $A' \subseteq W$ (resp., $B' \subseteq W$) contain a vertex $v \in W$ if there is a path $P$ in $F$ with first endpoint $v$ and second endpoint in $A$ (resp., $B$) and $s(P) \le 1/3$. If $A' \cap B' \neq \emptyset$, then it is clear that there is a path $P$ with $s(P) \le 2/3$ connecting a vertex of $A$ and a vertex of $B$ via a vertex of $A' \cap B'$, a contradiction. Thus we can assume that $A'$ and $B'$ are disjoint.

Since $F$ is a flow and $s$ has weight $w$, the total weight of the paths in $F$ with $s(P) \ge 1/3$ is at most $3w$. As the value of $F$ is exactly $\mu_2(U)$, the total weight of the paths in $F$ with second endpoint in $A$ is exactly $\mu_2(A)$. If $s(P) \le 1/3$ for such a path, then its first endpoint is in $A'$ by definition. Therefore, the total weight of the paths in $F$ with first endpoint in $A'$ is at least $\mu_2(A) - 3w$, which means that $\mu_1(A') \ge \mu_2(A) - 3w \ge \mu_2(A)/2$. Similarly, we have $\mu_1(B') \ge \mu_2(B)/2$. Since $W$ is $(\mu_1, \lambda)$-connected and $s$ is an assignment with weight less than $(\lambda/6) \cdot \min\{\mu_2(A), \mu_2(B)\} \le (\lambda/3) \cdot \min\{\mu_1(A'), \mu_1(B')\}$, there is an $A' - B'$ path $P$ with $s(P) < 1/3$. Now the concatenation of an $A' - A$ path $P_A$ having $s(P_A) \le 1/3$, the path $P$, and a $B' - B$ path $P_B$ having $s(P_B) \le 1/3$ forms an $A - B$ path that is not covered by $s$, a contradiction. □

| Primal LP | Dual LP |
|---|---|
| $$\text{maximize} \sum_{i=1}^{r} \sum_{\substack{u \in A_i, v \in B_i \\ P \in \mathcal{P}_{uv}}} x(P)$$ s. t. $$\sum_{i=1}^{r} \sum_{\substack{u \in A_i, v \in B_i \\ P \in \mathcal{P}_{uv}, P \cap e \neq \emptyset}} x(P) \leq 1 \quad \forall e \in E(H)$$ $$\sum_{v \in B_i, P \in \mathcal{P}_{uv}} x(P) \leq \mu(u) \quad \forall 1 \leq i \leq r, u \in A_i$$ $$\sum_{u \in A_i, P \in \mathcal{P}_{uv}} x(P) \leq \mu(v) \quad \forall 1 \leq i \leq r, v \in B_i$$ $$x(P) \geq 0 \quad \begin{array}{l}\forall 1 \leq i \leq r, u \in A_i, v \in B_i, \\ P \in \mathcal{P}_{uv}\end{array}$$ | $$\text{minimize} \sum_{e \in e(H)} y(e) + \sum_{u \in A} \mu(u) y(u) + \sum_{v \in B} \mu(v) y(v)$$ s. t. $$\sum_{\substack{e \in E(H), \\ e \cap P \neq \emptyset}} y(e) + y(u) + y(v) \geq 1 \quad \begin{array}{l}\forall 1 \leq i \leq r, u \in A_i, v \in B_i, \\ P \in \mathcal{P}_{uv}\end{array}$$ $$y(e) \geq 0 \quad \forall e \in E(H)$$ $$y(u) \geq 0 \quad \forall u \in A$$ $$y(v) \geq 0 \quad \forall v \in B$$ |

Figure 5: Primal and dual linear programs for $\mu$-demand multicommodity flow between pairs $(A_1, B_1)$, ..., $(A_r, B_r)$. We denote by $\mathcal{P}_{uv}$ the set of all $u - v$ paths.

A $\mu$-demand *multicommodity flow* between pairs $(A_1, B_1)$, ..., $(A_r, B_r)$ is a set $F_1$, ..., $F_r$ of compatible flows such that $F_i$ is a $\mu$-demand $(A_i, B_i)$-flow (recall that a set of flows is compatible if their sum is also a flow, that is, do not violate the edge constraints). The *value* of a multicommodity flow is the sum of the values of the $r$ flows. Let $A = \bigcup_{i=1}^{r} A_i$, $B = \bigcup_{i=1}^{r} B_i$, and let us restrict our attention to the case when $(A_1, \ldots, A_r, B_1, \ldots, B_r)$ is a partition of $A \cup B$. In this case, the maximum value of a $\mu$-demand multicommodity flow between pairs $(A_1, B_1)$, ..., $(A_r, B_r)$ can be expressed as the optimum values of the primal and dual linear programs in Figure 5.

The following lemma shows that if $\text{con}_\lambda(H)$ is sufficiently large, then there is a highly connected set that has the additional property that it is the union of $k$ cliques $K_1$, ..., $K_k$ with $\mu(K_i) \geq 1/2$ for every clique. The high-level idea of the proof is the following. Take a $(\mu, \lambda)$-connected set $W$ with $\mu(W) = \text{con}_\lambda(H)$ and find a large multicommodity flow between some pairs $(A_1, B_1)$, ..., $(A_r, B_r)$ in $W$. Consider the dual solution $y$. By complementary slackness, every edge with nonzero value in $y$ covers exactly 1 unit of the multicommodity flow. If most of the weight of the dual solution is on the edge variables, then we can choose $k$ edges that cover at least $\Omega(k)$ units of flow. These edges are connected to $W$ by a flow, and therefore by Lemma 6.4 the union of these edges is also highly connected and obviously can be partitioned into a small number cliques.

There are two things that can go wrong with this argument. First, it can happen that the dual solution assigns most of the weight to the vertex variables $y(u)$, $y(v)$ ($u \in A$, $v \in B$). This case is only possible if the value of the dual (and hence the primal) solution is close to $\sum_{i=1}^{r}(\min\{\mu(A_i) + \mu(B_i)\})$. To avoid this situation, we want to select the pairs $(A_i, B_i)$ such that they are only "moderately connected": there is a fractional $(A_i, B_i)$-separator of weight $2\lambda \min\{\mu(A_i), \mu(B_i)\}$, that is, at most twice the minimum possible. This means that the weight of the dual solution is at most $2\lambda \sum_{i=1}^{r}(\min\{\mu(A_i), \mu(B_i)\})$, which is much less than $\sum_{i=1}^{r}(\min\{\mu(A_i), \mu(B_i)\})$ (if $\lambda$ is small). If we are not able to find sufficiently many such pairs, then we argue that a larger highly connected set can be obtained by scaling $\mu$ by a factor of 2. More precisely, we show that there is a large subset $W' \subseteq W$ that is $(2\mu, \lambda)$-connected and $2\mu(W') > \text{con}_\lambda(H)$, a contradiction (a technical difficulty here that we have to make sure first that $2\mu$ is also a fractional independent set).

The second problem we have to deal with is that the value of the dual solution can be so small that we find a very small set of edges that already cover a large fraction of the multicommodity flow. However, we can use Lemma 6.3 to argue that a weight assignment on the edges that covers a large multicommodity flow in a $(\mu, \lambda)$-connected set cannot have very small weight.

**Lemma 6.5.** *Let H be a hypergraph and let $0 < \lambda < 1/16$ be a constant. Then there is fractional independent set $\mu$, a $(\mu, \lambda/6)$-connected set $W$, and a partition $(K_1, \ldots, K_k)$ of $W$ such that $k = \Omega(\lambda \sqrt{\text{con}_\lambda(H)})$, and for every $1 \leq i \leq k$, $K_i$ is a clique with $\mu(K_i) \geq 1/2$.*

*Proof.* Let $k$ be the largest integer such that $\mathrm{con}_\lambda(H) \geq 3T + 2k$ holds, where $T := (56/\lambda)^2 \cdot k^2$; it is clear that $k = \Omega(\lambda\sqrt{\mathrm{con}_\lambda(H)})$. Let $\mu_0$ be a fractional independent set and $W_0$ be a $(\mu_0, \lambda)$-connected set with $\mu_0(W_0) = \mathrm{con}_\lambda(H)$. We can assume that $\mu_0(v) > 0$ if and only if $v \in W_0$. This also implies that $W_0$ is in one connected component of $H$.

**Highly loaded edges.** First, we want to modify $\mu_0$ such that there is no edge $e$ with $\mu_0(e) \geq 1/2$. The following claim shows that we can achieve this by restricting $\mu_0$ to an appropriate subset $W$ of $W_0$.

*Claim 1.* There is a subset $W \subseteq W_0$ such that $\mu_0(W) \geq \mathrm{con}_\lambda(H) - k$ and $\mu_0(e \cap W) < 1/2$ for every edge $e$.

*Proof.* Let us choose edges $g_1, g_2, \ldots$ as long as possible with the requirement $\mu_0(K_i) \geq 1/2$ for $K_i := (g_i \cap W_0) \setminus \bigcup_{j=1}^{i-1} K_j$. If we can select at least $k$ such edges, then the cliques $K_1, \ldots, K_k$ satisfy the requirements. Indeed, $W' := \bigcup_{i=1}^{k} K_i \subseteq W_0$ is a $(\mu_0, \lambda)$-connected set, $\mu_0(K_i) \geq 1/2$, and $(K_1, \ldots, K_k)$ is a partition of $W'$ into cliques.

Thus we can assume that the selection of the edges stops at edge $g_t$ for some $t < k$. Let $W := W_0 \setminus \bigcup_{i=1}^{t} K_i$. Observe that there is no edge $e \in E(H)$ with $\mu_0(e \cap W) \geq 1/2$, as in this case the selection of the edges could be continued with $g_{t+1} := e$. Furthermore, we have $\mu_0(W) = \mu_0(W_0 \setminus \bigcup_{i=1}^{t} K_i) > \mu_0(W_0) - k = \mathrm{con}_\lambda(H) - k$, as required. ⌟

**Moderately connected pairs.** Let us define $\mu$ such that $\mu(v) = 2\mu_0(v)$ if $v \in W$ and $\mu(v) = 0$ otherwise. By Claim 1, $\mu$ is a fractional independent set. The set $W$ is $(\mu_0, \lambda)$-connected, but not necessarily $(\mu, \lambda)$-connected. In the next step, we find a large collection of pairs $(A_i, B_i)$ that violate $(\mu, \lambda)$-connectivity. Informally, we can say that these pairs $(A_i, B_i)$ are "moderately connected": denoting $w_i = \min\{\mu(A_i), \mu(B_i)\}$, the minimum value of a fractional $(A_i, B_i)$-separator for such a pair is less than $\lambda w_i$, but at least $\lambda w_i/2 = \lambda \min\{\mu_0(A_i), \mu_0(B_i)\}$ (because $W$ is $(\mu_0, \lambda)$-connected).

*Claim 2.* There are disjoint sets $A_i, B_1, \ldots, A_r, B_r \subseteq W$ such that for every $1 \leq i \leq r$ there is a fractional $(A_i, B_i)$-separator with weight less than $\lambda w_i$ for $w_i := \min\{\mu(A_i), \mu(B_i)\}$ and $w := \sum_{i=1}^{r} w_i \geq T$.

*Proof.* Let us greedily select a maximal collection of such pairs $(A_1, B_1), \ldots, (A_r, B_r)$. Note that every fractional separator has value at least 1 (as $W$ is in a single component of $H$), thus $\lambda w_i > 1$ holds, implying $w_i > 1/\lambda > 1$. We can assume that $\mu(A_i), \mu(B_i) \leq w_i + 1 \leq 2w_i$: if, say, $\mu(A_i) > \mu(B_i) + 1$, then removing an arbitrary vertex of $A_i$ decreases $\mu(A_i)$ by at most one (as $\mu$ is a fractional independent set) without changing $\min\{\mu(A_i), \mu(B_i)\}$, hence there would be a smaller pair of sets with the required properties. Therefore, we have $2w_i \leq \mu(A_i \cup B_i) \leq 2w_i + 1 \leq 3w_i$ for every $1 \leq i \leq r$.

Suppose that $w := \sum_{i=1}^{r} w_i < T$. Let $W' := W \setminus \bigcup_{i=1}^{r}(A_i \cup B_i)$. As $\mu(\bigcup_{i=1}^{r}(A_i \cup B_i)) \leq 3w < 3T$, we have $\mu(W') > \mu(W) - 3T = 2\mu_0(W) - 3T \geq 2\mathrm{con}_\lambda(H) - 2k - 3T \geq \mathrm{con}_\lambda(H)$. Since the greedy selection stopped, there is no fractional $(A', B')$-separator of value less than $\lambda \cdot \min\{\mu(A'), \mu(B')\}$ for any disjoint $A', B' \subseteq W'$, that is, $W'$ is $(\mu, \lambda)$-connected with $\mu(W') > \mathrm{con}_\lambda(H)$, contradicting the definition of $\mathrm{con}_\lambda(H)$. ⌟

**Finding a multicommodity flow.** Let $(A_1, B_1), \ldots, (A_r, B_r)$ be as in Claim 2. Since there is a fractional $(A_i, B_i)$-separator of value less than $\lambda w_i$, the maximum value of a $\mu$-demand multicommodity flow between pairs $(A_1, B_1), \ldots, (A_r, B_r)$ is less than $\lambda w$. Let $y$ be an optimum dual solution; we give a lower bound on the total weight of the edge variables.

*Claim 3.* $\sum_{e \in E(H)} y(e) \geq 2k$.

*Proof.* Let $A := \bigcup_{i=1}^{r} A_i$ and $B := \bigcup_{i=1}^{r} B_i$. Let $A^* := \{u \in A \mid y(u) \leq 1/4\}$, $B^* := \{v \in B \mid y(v) \leq 1/4\}$, $A_i^* = A_i \cap A^*$, $B_i^* = B_i \cap B^*$, and $w_i^* = \min\{\mu(A_i^*), \mu(B_i^*)\}$. For each $i$, the value of $w_i^*$ is either at least $w_i/2$, or less than that. Assume without loss of generality that there is a $1 \leq r^* \leq r$ such that $w_i^* \geq w_i/2$ if and only if $i \leq r^*$. Let $w^* = \sum_{i=1}^{r^*} w_i^*$.

We claim that $w^* \geq w/4$. Note that $w_i^* < w_i/2$ means that either $\mu(A_i^*) < w_i/2$ or $\mu(B_i^*) < w_i/2$; as $\mu(A_i), \mu(B_i) \geq w_i$, this is only possible if $\mu(A_i \setminus A^*) + \mu(B_i \setminus B^*) > w_i/2$. Suppose first that $\sum_{i=r^*+1}^{r} w_i > w/2$. This would imply

$$\mu((A \setminus A^*) \cup (B \setminus B^*)) \geq \sum_{i=r^*+1}^{r} (\mu(A_i \setminus A^*) + \mu(B_i \setminus B^*)) > \sum_{i=r^*+1}^{r} w_i/2 > w/4.$$

However, $y(u) > 1/4$ for every $u \in (A \setminus A^*) \cup (B \setminus B^*)$, thus $\sum_{v \in A \cup B} \mu(v) y(v) \geq \mu((A \setminus A^*) \cup (B \setminus B^*))/4 \geq w/16 > \lambda w$ (since $\lambda < 1/16$), a contradiction with the assumption that the optimum is at most $\lambda w$. Thus we can assume that $\sum_{i=r^*+1}^{r} w_i \leq w/2$ and hence $\sum_{i=1}^{r^*} w_i \geq w/2$. Together with $w_i^* \geq w_i/2$ for every $1 \leq i \leq r^*$, this implies $w^* \geq w/4$.

As $y(a), y(b) \leq 1/4$ for every $a \in A_i^*$, $b \in B_i^*$, it is clear that for every $A_i^* - B_i^*$ path $P$, the total weight of the edges intersecting $P$ has to be at least $1/2$ in assignment $y$. Therefore, if we define $y^* : E(H) \to \mathbb{R}^+$ by $y^*(e) = 2y(e)$ for every $e \in E(H)$, then $y^*$ covers every $A_i^* - B_i^*$ path. Let $W^* = \bigcup_{i=1}^{r^*} (A_i^* \cup B_i^*)$. We use Lemma 6.3 for the $(\mu, \lambda)$-connected set $W^*$, the pairs $(A_1^*, B_1^*)$, ..., $(A_{r^*}^*, B_{r^*}^*)$, and for the weight assignment $y^*$. Note that $w_i^* \geq w_i/2 \geq 1/2$ for every $i$. It follows that the total weight of $y^*$ on the edges is at least $(\lambda/7) \cdot \sqrt{w^*} \geq (\lambda/14) \cdot \sqrt{w}$, which means that $\sum_{e \in E(H)} y(e) \geq (\lambda/28) \cdot \sqrt{w} \geq (\lambda/28) \cdot \sqrt{T} \geq 2k$. ⌟

**Locating the cliques.** Let $y$ be an optimum dual solution for the maximum multicommodity flow problem with pairs $(A_1, B_1)$, ..., $(A_r, B_r)$ and let $F$ be the sum of the flows obtained from an optimum primal solution.

*Claim 4.* There are $k$ pairwise-disjoint cliques $K_1$, ..., $K_k$ and a $k$ subflows $f_1$, ..., $f_k$ of $F$, each of them having value at least $1/2$, such that every path appearing in $f_i$ intersects $K_i$ and is disjoint from $K_j$ for every $j \neq i$.

*Proof.* Let $F^{(0)} = F$ and for $i = 1, 2, \ldots$, let $F^{(i)}$ be the flow obtained from $F^{(0)}$ by removing $f_1$, ..., $f_i$. Let $c(e, F^{(i)})$ be the total weight of the paths in $F^{(i)}$ intersecting edge $e$ and let $C_i = \sum_{e \in E(H)} y(e) c(e, F^{(i)})$. By complementary slackness, $c(e, F^{(0)}) = 1$ for each $e \in E(H)$ with $y(e) > 0$ and hence $C_0 = \sum_{e \in E(H)} y(e) \geq 2k$.

Let us select $e_i$ to be an edge such that $c(e_i, F^{(i-1)})$ is maximum possible and let $K_i := e_i \setminus \bigcup_{j=1}^{i-1} e_j$. Let the flow $f_i$ contain all the paths of $F^{(i-1)}$ intersecting $e_i$. Observe that the paths appearing in $f_i$ do not intersect $e_1$, ..., $e_{i-1}$ (otherwise they would be in one of $f_1$, ..., $f_{i-1}$ and hence they would no longer be in $F^{(i-1)}$), thus clique $K_i$ intersects every path in $f_i$.

For every $u - v$ path $P$ appearing in $F^{(0)}$, we get $\sum_{e \in E(H), e \cap P \neq \emptyset} y(e) + y(u) + y(v) = 1$ from complementary slackness: if the primal variable corresponding to $P$ is nonzero, then the corresponding dual constraint is tight. In particular, this means that the total weight of the edges intersecting such a path $P$ is at most 1 in $y$. As $F^{(i-1)}$ is a subflow of $F^{(0)}$, this is also true for every path $P$ in $F^{(i-1)}$. This means that when we remove a path of weight $\gamma$ from $F^{(i-1)}$ to obtain $F^{(i)}$, then the total weight of the edges $e$ for which $c(e, F^{(i-1)})$ decreases by $\gamma$ is at most 1, i.e., $C_{i-1}$ decreases by at most $\gamma$. As only the paths intersecting $e_i$ are removed from $F^{(i-1)}$ and the total weight of the paths intersecting $e_i$ is at most 1, we get that $C_i \geq C_{i-1} - 1$ and hence $C_i \geq C_0 - k \geq C_0/2$ for $i \leq k$. Since $C_0 = \sum_{e \in E(H)} y(e)$ and $C_i = \sum_{e \in E(H)} y(e) c(e, F^{(i)}) \geq C_0/2$, it follows that there has to be at least one edge $e$ with $c(e, F^{(i)}) \geq 1/2$. Thus in each step, we can select an edge $e_i$ such that that the total weight of the paths in $F^{(i)}$ intersecting $e_i$ is at least $1/2$, and hence the value of $f_i$ is at least $1/2$ for every $1 \leq i \leq k$. ⌟

**Moving the highly connected set.** Let $U = \bigcup_{i=1}^{k} K_i$.

*Claim 5.* There is a fractional independent set $\mu'$ such that $U$ is a $(\mu', \lambda/6)$-connected set with $\mu'(K_i) \geq 1/2$ for every $1 \leq i \leq r$.

*Proof.* Each path $P$ in $f_i$ is a path with endpoints in $W$ and intersecting $K_i$. Let us truncate each path $P$ in $f_i$ such that its first endpoint is still in $W$ and its second endpoint is in $K_i$; let $f_i'$ be the $(W, K_i)$-flow obtained by truncating every path in $f_i$. Note that $f_i'$ is still a flow and the sum $F'$ of $f_1'$, ..., $f_k'$ is a $(W, U)$-flow. Let $\mu_1 = \mu$ and let $\mu_2(v)$ be the total weight of the paths in $F'$ with second endpoint $v$. It is clear that $\mu_2$ is a fractional independent set, $\mu_2(K_i) \geq 1/2$, and $F$ is a $(\mu_1, \mu_2)$-demand $(W, U)$-flow with value $\mu_2(U)$. Thus by Lemma 6.4, $U$ is a $(\mu_2, \lambda/6)$-connected set with the required properties. ⌟

The set $U$, the partition $(K_1, \ldots, K_r)$, and the fractional independent set $\mu'$ clearly satisfy the requirements of the lemma. □

## 6.2 Concurrent flows and embedding

Let $W$ be a set of vertices and let $(X_1, \ldots, X_k)$ be a partition of $W$. A *uniform concurrent flow of value $\varepsilon$ on $(X_1, \ldots, X_k)$* is a compatible set of $\binom{k}{2}$ flows $F_{i,j}$ ($1 \leq i < j \leq k$) where $F_{i,j}$ is an $(X_i, X_j)$-flow of value $\varepsilon$. The

|Primal LP|Dual LP|
|---|---|

Primal LP:

$$\text{maximize } \varepsilon$$

s. t.

$$\sum_{1 \leq i < j \leq k} \sum_{\substack{P \in \mathcal{P}_{i,j}, \\ P \cap e \neq \emptyset}} x(P) \leq 1 \quad \forall e \in E(H)$$

$$\sum_{P \in \mathcal{P}_{i,j}} x(P) \geq \varepsilon \quad \forall 1 \leq i < j \leq k$$

$$x(P) \geq 0 \quad \forall 1 \leq i < j \leq k, P \in \mathcal{P}_{i,j}$$

Dual LP:

$$\text{minimize } \sum_{e \in e(H)} y(e)$$

$$\sum_{e \in E(H), e \cap P \neq \emptyset} y(e) \geq \ell_{i,j} \quad \forall 1 \leq i < j \leq k, P \in \mathcal{P}_{i,j}$$

$$\sum_{1 \leq i < j \leq k} \ell_{i,j} \geq 1$$

$$y(e) \geq 0 \quad \forall e \in E(H)$$

$$\ell_{i,j} \geq 0 \quad \forall 1 \leq i < j \leq k$$

Figure 6: Primal and dual linear programs for uniform concurrent flow on $W = (X_1, \ldots, X_k)$. We denote by $\mathcal{P}_{i,j}$ the set of all $X_i - X_j$ paths.

maximum value of a uniform concurrent flow on $W$ can be expressed as the optimum values of the primal and dual linear programs in Figure 6.

If $H$ is connected, then the maximum value of a uniform concurrent flow on $(X_1, \ldots, X_k)$ is at least $1/\binom{k}{2} = \Omega(k^{-2})$: if each of the $\binom{k}{2}$ flows has value $1/\binom{k}{2}$, then they are clearly compatible. The following lemma shows that in a $(\mu, \lambda)$-connected set, if the sets $X_1, \ldots, X_k$ are cliques and $\mu(X_i) \geq 1/2$ for every $i$, then we can guarantee a better bound of $\Omega(k^{-\frac{3}{2}})$.

**Lemma 6.6.** *Let $H$ be a hypergraph, $\mu$ a fractional independent set of $H$, and $W \subseteq V(H)$ a $(\mu, \lambda)$-connected set for some $0 < \lambda < 1$. Let $(K_1, \ldots, K_k)$ (with $k \geq 1$) be a partition of $W$ such that $K_i$ is a clique and $\mu(K_i) \geq 1/2$ for every $1 \leq i \leq k$. Then there is a uniform concurrent flow of value $\Omega(\lambda/k^{\frac{3}{2}})$ on $(K_1, \ldots, K_k)$.*

*Proof.* Suppose that there is no uniform concurrent flow of value $\beta \cdot \lambda / k^{\frac{3}{2}}$, where $\beta > 0$ is a sufficiently small universal constant specified later. This means that the dual linear program has a solution having value less than that. Let us fix such a solution $(y, \ell_{i,j})$ of the dual linear program. In the following, for every path $P$, we denote by $y(P) := \sum_{e \in E(H), e \cap P \neq \emptyset} y(e)$ the total weight of the edges intersecting $P$. It is clear from the dual linear program that $y(P) \geq \ell_{i,j}$ for every $P \in \mathcal{P}_{i,j}$.

We construct two graphs $G_1$ and $G_2$: the vertex set of both graphs is $\{1, \ldots, k\}$ and for every $1 \leq i < j \leq k$, vertices $i$ and $j$ are adjacent in $G_1$ (resp., $G_2$) if and only if $\ell_{i,j} > 1/(3k^2)$ (resp., $\ell_{i,j} > 1/k^2$). Note that $G_2$ is a subgraph of $G_1$. First we prove the following claim:

*Claim 1.* If the distance of $u$ and $v$ is at most 3 in the *complement* of $G_1$, then $u$ and $v$ are not adjacent in $G_2$.

*Proof.* Suppose that $uw_1w_2v$ is a path of length 3 in the complement of $G_1$ (the same argument works for paths of length less than 3). By definition of $G_1$, there is a $K_u - K_{w_1}$ path $P_1$, a $K_{w_1} - K_{w_2}$ path $P_2$, and a $K_{w_2} - K_v$ path $P_3$ such that $y(P_1), y(P_2), y(P_3) \leq 1/(3k^2)$. Since $K_{w_1}$ and $K_{w_2}$ are cliques, paths $P_1$ and $P_2$ touch, and paths $P_2$ and $P_3$ touch. Thus by concatenating the three paths, we can obtain a $K_u - K_v$ path $P$ with $y(P) \leq y(P_1) + y(P_2) + y(P_3) \leq 1/k^2$, implying that $u$ and $v$ are not adjacent in $G_2$, proving the claim. Note that the proof of this claim is the only point where we use that the $K_i$'s are cliques. ⌟

Let $y' : E(H) \to \mathbb{R}^+$ be defined by $y'(e) := 3k^2 \cdot y(e)$, thus $y'$ has total weight less than $3\beta \cdot \lambda \sqrt{k}$. Suppose first that $G_1$ has a matching $a_1b_1, \ldots, a_mb_m$ of size $m = \lceil k/4 \rceil$. This means that $y'$ covers every $K_{a_i} - K_{b_i}$ path for every $1 \leq i \leq \lceil k/4 \rceil$. Therefore, by Lemma 6.3, $y'$ has weight at least $(\lambda/7) \cdot \sqrt{\lceil k/4 \rceil \cdot (1/2)} > 3\beta \cdot \lambda \sqrt{k}$, if $\beta$ is sufficiently small, yielding a contradiction.

Thus the size of the maximum matching in $G_1$ is less than $\lceil k/4 \rceil$, which means that there is a vertex cover $S_1$ of size at most $k/2$. Let $S_2 \subseteq S_1$ contain those vertices of $S_1$ that are adjacent to every vertex outside $S_1$ in $G_1$. We claim that $S_2$ is a vertex cover of $G_2$. Suppose that there is an edge $uv$ of $G_2$ for some $u, v \notin S_2$. By the definition of $S_2$, either $u \notin S_1$, or there is a vertex $w_1 \notin S_1$ such that $u$ and $w_1$ are not adjacent in $G_1$. Similarly, either $v$ is not in $S_1$, or it is not adjacent in $G_1$ to some $w_2 \notin S_1$. Since vertices not in $S_1$ are not adjacent in $G_1$ (as $S_1$ is a vertex cover of $G_1$), we get that the distance of $u$ and $v$ is at most 3 in the complement of $G_1$. Thus by the claim, $u$ and $v$ are not adjacent in $G_2$.

Let us give an upper bound on $\sum_{1\leq i<j\leq k}\ell_{i,j}$ by bounding $\ell_{i,j}$ separately for pairs that are adjacent in $G_2$ and for pairs that are not adjacent in $G_2$. The total weight of $y$, which is less than $\beta\cdot\lambda/k^{\frac{3}{2}}$, is an upper bound on any $\ell_{i,j}$. Furthermore, if $i$ and $j$ are not adjacent in $G_2$, then we have $\ell_{i,j}\leq 1/k^2$. The number of edges in $G_2$ is at most $|S_2|k$ (as $S_2$ is vertex cover), hence we have

$$1 \leq \sum_{1\leq i<j\leq k}\ell_{i,j} \leq |S_2|k\cdot\beta\cdot\lambda/k^{\frac{3}{2}} + \binom{k}{2}(1/k^2) \leq \beta\cdot\lambda|S_2|/\sqrt{k}+1/2,$$

which implies that $|S_2|\geq \sqrt{k}/(2\beta\lambda)$. Let $A:=\bigcup_{i\in S_2}K_i$ and $B:=\bigcup_{i\notin S_1}K_i$; we have $\mu(A)\geq |S_2|\cdot(1/2)\geq \sqrt{k}/(4\beta\lambda)$ and $\mu(B)\geq (1/2)\cdot(k-|S_1|))\geq k/4$. As every vertex of $S_2$ is adjacent in $G_1$ with every vertex outside $S_1$, assignment $y'$ covers every $A-B$ path. However, $y'$ has weight less than $3\beta\cdot\lambda\sqrt{k} < \min\{\sqrt{k}/(4\beta\lambda),k/4\}$ (using that $\lambda\leq 1$ and assuming that $\beta$ is sufficiently small), contradicting the assumption that $W$ is $(\mu,\lambda)$-connected. □

Intuitively, the intersection structure of the paths appearing in a uniform concurrent flow on cliques $K_1, \ldots, K_k$ is reminiscent of the edges of the complete graph on $k$ vertices: if $\{i_1,j_1\}\cap\{i_2,j_2\}\neq\emptyset$, then every path of $F_{i_1,j_1}$ touches every path of $F_{i_2,j_2}$. We use the following result from [43], which shows that the line graph of cliques have good embedding properties. If $G$ is a graph and $q\geq 1$ is an integer, then the *blow up* $G^{(q)}$ is obtained from $G$ by replacing every vertex $v$ with a clique $K_v$ of size $q$ and for every edge $uv$ of $G$, connecting every vertex of the clique $K_u$ with every vertex of the clique $K_v$. Let $L_k$ be the line graph of the complete graph on $k$ vertices.

**Lemma 6.7** ([43]). *For every $k>1$ there is a constant $n_k>0$ such that for every $G$ with $|E(G)|>n_k$ and no isolated vertices, the graph $G$ is a minor of $L_k^{(q)}$ for $q=\lceil 130|E(G)|/k^2\rceil$. Furthermore, a minor mapping can be found in time polynomial in the size of $G$.*

Using the terminology of embeddings, a minor mapping of $G$ into $L_k^{(q)}$ can be considered as an embedding from $G$ to $L_k$ where every vertex of $L_k$ appears in the image of at most $q$ vertices, i.e., the vertex depth of the embedding is at most $q$. Thus we can restate Lemma 6.7 the following way:

**Lemma 6.8.** *For every $k>1$ there is a constant $n_k>0$ such that for every $G$ with $|E(G)|>n_k$ and no isolated vertices, the graph $G$ has an embedding into $L_k$ with vertex depth $O(|E(G)|/k^2)$. Furthermore, such an embedding can be found in time polynomial in the size of $G$.*

Now we are ready to prove Theorem 6.1, the main result of the section:

*Proof (of Theorem 6.1).* By Lemma 6.5 and Lemma 6.6, for some $k=\Omega(\lambda\sqrt{\mathrm{con}_\lambda(H)})$, there are cliques $K_1, \ldots, K_k$ and a uniform concurrent flow $F_{i,j}$ ($1\leq i<j\leq k$) of value $\varepsilon=\Omega(\lambda/k^{\frac{3}{2}})$ on $(K_1,\ldots,K_k)$. By trying all possibilities for the cliques and then solving the uniform concurrent flow linear program, we can find these flows (the time required for this step is a constant $f(H,\lambda)$ depending only on $H$ and $\lambda$). Let $w_0$ be the smallest positive weight appearing in the flows.

Let $m=|E(G)|$ and suppose that $m\geq n_k$, for the constant $n_k$ in Lemma 6.7. Thus the algorithm of Lemma 6.8 can be used to find a an embedding $\psi$ from $G$ to $L_k$ with vertex depth $q=O(m/k^2)$. Let us denote by $v_{\{i,j\}}$ ($1\leq i<j\leq k$) the vertices of $L_k$ with the meaning that distinct vertices $v_{\{i_1,j_1\}}$ and $v_{\{i_2,j_2\}}$ are adjacent if and only if $\{i_1,j_1\}\cap\{i_2,j_2\}\neq\emptyset$.

We construct an embedding $\phi$ from $G$ to $H$ the following way. The set $\phi(u)$ is obtained by replacing each vertex of $v_{\{i,j\}}\in\psi(u)$ by a path from the flow $F_{i,j}$ (thus $\phi(u)$ is the union of $|\psi(u)|$ paths). We select the paths in such a way that the following requirement is satisfied: a path $P$ of $F_{i,j}$ having weight $w$ is selected into the images of at most $\lceil(q/\varepsilon)\cdot w\rceil$ vertices of $G$. We set $m_{H,\lambda}$ sufficiently large that $(q/\varepsilon)\cdot w_0\geq 1$ (note that $q$ depends on $m$, but $\varepsilon$ and $w_0$ depends only on $H$ and $\lambda$). Thus if $m\geq m_{H,\lambda}$, then $\lceil(q/\varepsilon)\cdot w\rceil\leq 2(q/\varepsilon)\cdot w$. Since the total weight of the paths in $F_{i,j}$ is $\varepsilon$, these paths can accommodate the image of at least $(q/\varepsilon)\cdot\varepsilon=q$ vertices. As each vertex $v_{\{i,j\}}$ of $L_k$ appears in the image of at most $q$ vertices of $G$ in the mapping $\psi$, we can satisfy the requirement.

It is easy to see that if $u_1$ and $u_2$ are adjacent in $G$, then $\phi(u_1)$ and $\phi(u_2)$ touch: in this case, there are vertices $v_{\{i_1,j_1\}} \in \psi(u_1)$, $v_{\{i_2,j_2\}} \in \psi(u_2)$ that are adjacent or the same in $L_k$ (that is, there is a $t \in \{i_1, j_1\} \cap \{i_2, j_2\}$), and the corresponding paths of $F_{i_1,j_1}$ and $F_{i_2,j_2}$ selected into $\phi(u_1)$ and $\phi(u_2)$ touch, as they both intersect the clique $K_t$. With a similar argument, we can show that $\phi(u)$ is connected.

To bound the edge depth of the embedding $\phi$, consider an edge $e$. The total weight of the paths intersecting $e$ is at most 1 and a path with weight $w$ is used in the image of at most $2(q/\varepsilon) \cdot w$ vertices. Each path intersects $e$ in at most 2 vertices (as we can assume that the paths appearing in the flows are minimal), thus a path with weight $w$ contributes at most $4(q/\varepsilon) \cdot w$ to the depth of $e$. Thus the edge depth of $\phi$ is at most $4(q/\varepsilon) = O(m/(\lambda\sqrt{k})) = O(m/(\lambda^{\frac{3}{2}} \operatorname{con}_\lambda(H)^{\frac{1}{4}}))$. $\square$

### 6.3 Connection with adaptive width

As an easy consequence of the embedding result Corollary 6.2, we can show that large submodular width implies large adaptive width:

**Lemma 6.9.** *For every hypergraph $H$, $\operatorname{adw}(H) = \Omega(\operatorname{emb}(H))$.*

*Proof.* Suppose that $\operatorname{emb}(H) > \alpha$. This means that there is an integer $m_\alpha$ such that every graph with $m \geq m_\alpha$ edges has an embedding into $H$ with edge depth $m/\alpha$. It is well-known that there are arbitrarily large sparse graphs whose treewidth is linear in the number of vertices (for example, bounded-degree expanders, see e.g., [29]): for some universal constant $\beta$, there is a graph $G$ with $m \geq m_\alpha$ edges and treewidth at least $\beta m$. Thus there is an embedding $\phi$ from $G$ to $H$ with edge depth at most $q \leq m/\alpha$. Let $d(v)$ be the depth of vertex $v$ in the embedding and let us define $\mu(v) := d(v)/q$. From the definition of edge depth, it is clear that $\mu$ is a fractional independent set. Suppose that there is a tree decomposition $(T, B_{v \in V(T)})$ of $H$ having $\mu$-width $w$. This tree decomposition can be turned into a tree decomposition $(T, B'_{v \in V(T)})$ of $G$: for every $B_t \subseteq V(H)$, let $B'_t := \{u \in V(G) \mid \phi(u) \cap B_t \neq \emptyset\}$ contain those vertices of $G$ whose images intersect $B_t$. Now $\mu(B_t) \leq w$ means that $\sum_{v \in B_t} d(v) \leq qw$, which implies that $|B'_t| \leq qw$. Thus the width of $(T, B'_{v \in V(T)})$ is less than $qw$, which means that $w$ has to be at least $\beta m/q = \Omega(\alpha)$, the required lower bound on the adaptive width of $H$. $\square$

Combining Theorem 5.1 and Lemma 6.9 gives:

**Corollary 6.10.** *For every hypergraph $H$, $\operatorname{subw}(H) = O(\operatorname{adw}(H)^4)$.*

## 7 From embeddings to hardness of CSP

We prove the main hardness result of the paper in this section:

**Theorem 7.1.** *Let $\mathcal{H}$ be a recursively enumerable class of hypergraphs with unbounded submodular width. If there is an algorithm $\mathbb{A}$ and a function $f$ such that $\mathbb{A}$ solves every instance $I$ of $\operatorname{CSP}(\mathcal{H})$ with hypergraph $H \in \mathcal{H}$ in time $f(H) \cdot \|I\|^{o(\operatorname{subw}(H)^{1/4})}$, then the Exponential Time Hypothesis fails.*

In particular, Theorem 7.1 implies that $\operatorname{CSP}(\mathcal{H})$ for such a $\mathcal{H}$ is not fixed-parameter tractable:

**Corollary 7.2.** *If $\mathcal{H}$ is a recursively enumerable class of hypergraphs with unbounded submodular width, then $\operatorname{CSP}(\mathcal{H})$ is not fixed-parameter tractable, unless the Exponential Time Hypothesis fails.*

The Exponential Time Hypothesis (ETH) states that there is no $2^{o(n)}$ time algorithm for $n$-variable 3SAT. The Sparsification Lemma of Impagliazzo, Paturi, and Zane [35] shows that ETH is equivalent to the assumption that there is no algorithm for 3SAT whose running time is subexponential *in the number of clauses*. This result will be crucial for our hardness proof, as our reduction from 3SAT is sensitive to the number of clauses.

**Theorem 7.3** (Impagliazzo, Paturi, and Zane [35]). *If there is a $2^{o(m)}$ time algorithm for $m$-clause 3-SAT, then there is a $2^{o(n)}$ time algorithm for $n$-variable 3-SAT.*

To prove Theorem 7.1, we show that a subexponential-time algorithm for 3SAT exists if CSP($\mathcal{H}$) is can be solved "too fast" for some $\mathcal{H}$ with unbounded submodular width. We use the characterization of submodular width from Section 5 and the embedding results of Section 6 to reduce 3SAT to CSP($\mathcal{H}$) by embedding the incidence graph of a 3SAT formula into a hypergraph $H \in \mathcal{H}$. The basic idea of the proof is that if the 3SAT formula has $m$ clauses and the edge depth of the embedding is $m/r$, then we can gain a factor $r$ in the exponent of the running time. If submodular width is unbounded in $\mathcal{H}$, then we can make this gap $r$ between the number of clauses and the edge depth arbitrary large, and hence the exponent can be arbitrarily smaller than the number of clauses, i.e., the algorithm is subexponential in the number of clauses.

The following simple lemma from [43] gives a transformation that turns a 3SAT instance into a binary CSP instance. We include the proof for completeness.

**Lemma 7.4.** *Given an instance of 3SAT with $n$ variables and $m$ clauses, it is possible to construct in polynomial time an equivalent CSP instance with $n+m$ variables, $3m$ binary constraints, and domain size* 3.

*Proof.* Let $\phi$ be a 3SAT formula with $n$ variables and $m$ clauses. We construct an instance of CSP as follows. The CSP instance contains a variable $x_i$ ($1 \le i \le n$) corresponding to the $i$-th variable of $\phi$ and a variable $y_j$ ($1 \le j \le m$) corresponding to the $j$-th clause of $\phi$. Let $D = \{1,2,3\}$ be the domain. We try to describe a satisfying assignment of $\phi$ with these $n+m$ variables. The intended meaning of the variables is the following. If the value of variable $x_i$ is 1 (resp., 2), then this represents that the $i$-th variable of $\phi$ is true (resp., false). If the value of variable $y_j$ is $\ell$, then this represents that the $j$-th clause of $\phi$ is satisfied by its $\ell$-th literal. To ensure consistency, we add $3m$ constraints. Let $1 \le j \le m$ and $1 \le \ell \le 3$, and assume that the $\ell$-th literal of the $j$-th clause is a positive occurrence of the $i$-th variable. In this case, we add the binary constraint ($x_i = 1 \vee y_j \ne \ell$): either $x_i$ is true or some other literal satisfies the clause. Similarly, if the $\ell$-th literal of the $j$-th clause is a negated occurrence of the $i$-th variable, then we add the binary constraint ($x_i = 2 \vee y_j \ne \ell$). It is easy to verify that if $\phi$ is satisfiable, then we can assign values to the variables of the CSP instance such that every constraint is satisfied, and conversely, if the CSP instance has a solution, then $\phi$ is satisfiable. □

Next we show that an embedding from graph $G$ to hypergraph $H$ can be used to simulate a binary CSP instance $I_1$ having primal graph $G$ by a CSP instance $I_2$ whose hypergraph is $H$. The domain size and the size of the constraint relations of $I_2$ can grow very large in this transformation: the edge depth of the embedding determines how large this increase is.

**Lemma 7.5.** *Let $I_1 = (V_1, D_1, C_1)$ be a binary CSP instance with primal graph $G$ and let $\phi$ be an embedding of $G$ into a hypergraph $H$ with edge depth $q$. Given $I_1$, $H$, and the embedding $\phi$, it is possible to construct (in time polynomial in the size of the* output*) an equivalent CSP instance $I_2 = (V_2, D_2, C_2)$ with hypergraph $H$ where the size of every constraint relation is at most $|D_1|^q$.*

*Proof.* For every $v \in V(H)$, let $U_v := \{u \in V(G) \mid v \in \phi(u)\}$ be the set of vertices in $G$ whose images contain $v$, and for every $e \in E(H)$, let $U_e := \bigcup_{v \in e} U_v$. Observe that for every $e \in E(H)$, we have $|U_e| \le \sum_{v \in e} |U_v| \le q$, since the edge depth of $\phi$ is $q$. Let $D_2$ be the set of integers between 1 and $|D_1|^q$. For every $v \in V(H)$, the number of assignments from $U_v$ to $D_1$ is clearly $|D_1|^{|U_v|} \le |D_1|^q$. Let us fix a bijection $h_v$ between these assignments on $U_v$ and the set $\{1, \dots, |D_1|^{|U_v|}\} \subseteq D_2$.

The set $C_2$ of constraints of $I_2$ are constructed as follows. For each $e \in E(H)$, there is a constraint $\langle s_e, R_e \rangle$ in $C_2$, where $s_e$ is an $|e|$-tuple containing an arbitrary ordering of the elements of $e$. The relation $R_e$ is defined the following way. Suppose that $v_i$ is the $i$-th coordinate of $s_e$ and consider a tuple $t = (d_1, \dots, d_{|e|}) \in D_2^{|e|}$ of integers where $1 \le d_i \le |D_1|^{|U_{v_i}|}$ for every $1 \le i \le |e|$. This means that $d_i$ is in the image of $h_{v_i}$ and hence $f_i := h_{v_i}^{-1}(d_i)$ is an assignment from $U_{v_i}$ to $D_1$. We define relation $R_e$ such that it contains tuple $t$ if the following two conditions hold. First, we require that the assignments $f_1, \dots, f_{|e|}$ are *consistent* in the sense that $f_i(u) = f_j(u)$ for any $i, j$ and $u \in U_{v_i} \cap U_{v_j}$. In this case, $f_1, \dots, f_{|e|}$ together define an assignment $f$ on $\bigcup_{i=1}^{|e|} U_{v_i} = U_e$. The second requirement is that this assignment $f$ satisfies every constraint of $I_1$ whose scope is contained in $U_e$, that is, for every constraint $\langle (u_1, u_2), R \rangle \in C_1$ with $\{u_1, u_2\} \subseteq U_e$, we have $(f(u_1), f(u_2)) \in R$. This completes the description of the instance $I_2$.

Let us bound the maximum size of a relation of $I_2$. Consider the relation $R_e$ constructed in the previous paragraph. It contains tuples $(d_1, \ldots, d_{|e|}) \in D_2^{|e|}$ where $1 \le d_i \le |D_1|^{|U_{v_i}|}$ for every $1 \le i \le |e|$. This means that

$$|R_e| \le \prod_{i=1}^{|e|} |D_1|^{|U_{v_i}|} = |D_1|^{\sum_{i=1}^{|e|} |U_{v_i}|} \le |D_1|^q, \tag{3}$$

where the last inequality follows from the fact that $\phi$ has edge depth at most $q$.

To prove that $I_1$ and $I_2$ are equivalent, assume first that $I_1$ has a solution $f_1 : V_1 \to D_1$. For every $v \in V_2$, let us define $f_2(v) := h_v(\mathrm{pr}_{U_v} f_2)$, that is, the integer between 1 and $|D_1|^{|U_v|}$ corresponding to the projection of assignment $f_2$ to $U_v$. It is easy to see that $f_2$ is a solution of $I_2$.

Assume now that $I_2$ has a solution $f_2 : V_2 \to D_2$. For every $v \in V(H)$, let $f_z := h_v^{-1}(f_2(v))$ be the assignment from $U_v$ to $D_1$ that corresponds to $f_2(v)$ (note that by construction, $f_2(v)$ is at most $|D_1|^{|U_v|}$, hence $h_v^{-1}(f_2(v))$ is well-defined). We claim that these assignments are compatible: if $u \in U_{v'} \cap U_{v''}$ for some $u \in V(G)$ and $v', v'' \in V(H)$, then $f_{v'}(u) = f_{v''}(u)$. Recall that $\phi(u)$ is a connected set in $H$, hence there is a path between $v'$ and $v''$ in $\phi(u)$. We prove the claim by induction on the distance between $v'$ and $v''$ in $\phi(u)$. If the distance is 0, that is, $v' = v''$, then the statement is trivial. Suppose now that the distance of $v'$ and $v''$ is $d > 0$. This means that $v'$ has a neighbor $z \in \phi(u)$ such that the distance of $z$ and $v''$ is $d - 1$. Therefore, $f_z(u) = f_{v''}(u)$ by the induction hypothesis. Since $v'$ and $z$ are adjacent in $H$, there is an edge $E \in E(H)$ containing both $v'$ and $z$. From the way $I_2$ is defined, this means that $f_{v'}$ and $f_z$ are compatible and $f_{v'}(u) = f_z(u) = f_{v''}(u)$ follows, proving the claim. Thus the assignments $f_v$, $v \in V(H)$ are compatible and these assignments together define an assignment $f_1 : V(G) \to D$. We claim that $f_1$ is a solution of $I_1$. Let $c = \langle (u_1, u_2), R \rangle$ be an arbitrary constraint of $I_1$. Since $u_1 u_2' \in E(G)$, sets $\phi(u_1)$ and $\phi(u_2)$ touch, thus there is an edge $e \in E(H_k)$ that contains a vertex $v_1 \in \phi(u_1)$ and a vertex $v_2 \in \phi(u_2)$ (or, in other words, $u_1 \in U_{v_1}$ and $u_2 \in U_{v_2}$). The definition of $c_e$ in $I_2$ ensures that $f_1$ restricted to $U_{v_1} \cup U_{v_2}$ satisfies every constraint of $I_1$ whose scope is contained in $U_{v_1} \cup U_{v_2}$; in particular, $f_1$ satisfies constraint $c$. □

Now we are ready to prove Theorem 7.1, the main result of the section. We show that if there is a class $\mathcal{H}$ of hypergraphs with unbounded submodular width such that $\mathrm{CSP}(\mathcal{H})$ is FPT, then this algorithm can be used to solve 3SAT in subexponential time. The main ingredients are the embedding result of Theorem 6.1, and Lemmas 7.4 and 7.5 above on reduction to CSP. Furthermore, we need a way of choosing an appropriate hypergraph from the set $\mathcal{H}$. As discussed earlier, the larger the submodular width of the hypergraph is, the more we gain in the running time. However, we should not spend too much time on constructing the hypergraph and on finding an embedding. Therefore, we use the same technique as in [43]: we enumerate a certain number of hypergraphs and we try all of them simultaneously. The number of hypergraphs enumerated depends on the size of the 3SAT instance. This will be done in such a way that guarantees that we do not spend too much time on the enumeration, but eventually every hypergraph in $\mathcal{H}$ is considered for sufficiently large input sizes.

*Proof (of Theorem 7.1).* Let us fix a $\lambda > 0$ that is sufficiently small for Theorems 5.1 and 6.1. Suppose that there is an $f_1(H)n^{o(\mathrm{subw}(H)^{1/4})}$ time algorithm $\mathbb{A}$ for $\mathrm{CSP}(\mathcal{H})$. We can express the running time as $f_1(H)n^{\mathrm{subw}(H)^{1/4}/\iota(\mathrm{subw}(H))}$ for some unbounded nondecreasing function $\iota$ with $\iota(1) \ge 1$. We construct an algorithm $\mathbb{B}$ that solves 3SAT in subexponential time by using algorithm $\mathbb{A}$ as subroutine.

Given an instance $I$ of 3SAT with $n$ variables and $m$ clauses and a hypergraph $H \in \mathcal{H}$, we can solve $I$ the following way. First we use Lemma 7.4 to transform $I$ into a CSP instance $I_1 = (V_1, D_1, C_1)$ with $|V_1| = n + m$, $|D_1| = 3$, and $|C_1| = 3m$. Let $G$ be the primal graph of $I_1$, which is a graph having $3m$ edges. It can be assumed that $m$ is greater than some constant $m_{H,\lambda}$ of Theorem 6.1, otherwise the instance can be solved in constant time. Therefore, the algorithm of Theorem 6.1 can be used to find an embedding $\phi$ of $G$ into $H$ with edge depth $q = O(m/(\lambda^{\frac{3}{2}} \mathrm{con}_\lambda(H)^{1/4}))$; by Theorem 5.1, we have that $\mathrm{con}_\lambda(H) = \Omega(\mathrm{subw}(H))$ and hence $q \le c_\lambda m / \mathrm{subw}(H)^{1/4}$ for some constant $c_\lambda$ depending only on $\lambda$. By Lemma 7.5, we can construct an equivalent instance $I_2 = (V_2, D_2, C_2)$ whose hypergraph is $H$. By solving $I_2$ using the assumed algorithm $\mathbb{A}$ for $\mathrm{CSP}(\mathcal{H})$, we can answer if $I_1$ has a solution, or equivalently, if the 3SAT instance $I$ has a solution.

We will call "running algorithm $\mathbb{A}[I, H]$" this way of solving the 3SAT instance $I$. Let us determine the running time of $\mathbb{A}[I, H]$. The two dominating terms are the time required to find embedding $\phi$ using the $f(H,\lambda)m^{O(1)}$ time algorithm of Theorem 7.1 and the time required to run $\mathbb{A}$ on $I_2$. The size of every constraint relation in $I_2$ is at most $|D_1|^q = 3^q$, hence $\|I_2\| = O((|E(H)|+|V(H)|)3^q)$. Let $k = \mathrm{subw}(H)$. The total running time of $\mathbb{A}[I,H]$ can be bounded by

$$f(H,\lambda)m^{O(1)} + f_1(H)\|I_2\|^{k^{1/4}/\iota(k)} = f(H,\lambda)m^{O(1)} + f_1(H)(|E(H)|+|V(H)|)^{k^{1/4}/\iota(k)} \cdot 3^{q \cdot k^{1/4}/\iota(k)}$$
$$= f_2(H,\lambda) \cdot m^{O(1)} \cdot 3^{c_\lambda m/\iota(k)}$$

for an appropriate function $f_2(H,\lambda)$ depending only on $H$ and $\lambda$.

Algorithm $\mathbb{B}$ for 3SAT proceeds as follows. Let us fix an arbitrary computable enumeration $H_1, H_2, \ldots$ of the hypergraphs in $\mathcal{H}$. Given an $m$-clause 3SAT formula $I$, algorithm $\mathbb{B}$ spends the first $m$ steps on enumerating these hypergraphs; let $H_\ell$ be the last hypergraph produced by this enumeration (we assume that $m$ is sufficiently large that $\ell \geq 1$). Next we start simulating the algorithms $\mathbb{A}[I,H_1]$, $\mathbb{A}[I,H_2]$, ..., $\mathbb{A}[I,H_\ell]$ in *parallel*. When one of the simulations stops and returns an answer, then we stop all the simulations and return the answer. It is clear that algorithm $\mathbb{B}$ will correctly decide the satisfiability of $I$.

We claim that there is a universal constant $d$ such that for every $s$, there is an $m_s$ such that for every $m > m_s$, the running time of $\mathbb{B}$ is at most $(m \cdot 2^{m/s})^d$ on an $m$-clause formula. Clearly, this means that the running time of $\mathbb{B}$ is $2^{o(m)}$.

Let $k_s$ be the smallest positive integer such that $\iota(k_s) \geq s$ (as $\iota$ is unbounded, this is well defined). Let $i_s$ be the smallest positive integer such that $\mathrm{subw}(H_{i_s}) \geq k_s$ (as $\mathcal{H}$ has unbounded submodular width, this is also well defined). Set $m_s$ sufficiently large that $m_s \geq f_2(H_{i_s},\lambda)$ and the fixed enumeration of $\mathcal{H}$ reaches $H_{i_s}$ in less then $m_s$ steps. This means that if we run $\mathbb{B}$ on a 3SAT formula $I$ with $m \geq m_s$ clauses, then $\ell \geq i_s$ and hence $\mathbb{A}[I,H_{i_s}]$ will be one of the $\ell$ simulations started by $\mathbb{B}$. The simulation of $\mathbb{A}[I,H_{i_s}]$ terminates in

$$f_2(H_{i_s},\lambda)m^{O(1)} \cdot 3^{c_\lambda m/\iota(\mathrm{subw}(H_{i_s}))} \leq m \cdot m^{O(1)} \cdot 3^{c_\lambda m/s}$$

steps. Taking into account that we simulate $\ell \leq m$ algorithms in parallel and all the simulations are stopped not later than the termination of $\mathbb{A}[I,H_{i_s}]$, the running time of $\mathbb{B}$ can be bounded polynomially by the running time of $\mathbb{A}[I,H_{i_s}]$. Therefore, there is a constant $d$ such that the running time of $\mathbb{B}$ is at most $(m \cdot 2^{m/s})^d$, as required. $\square$

*Remark* 7.6. Recall that if $\phi$ is an embedding of $G$ into $H$, then the depth of an edge $e \in E(H)$ is $d_\phi(e) = \sum_{v \in V(G)} |\phi(v) \cap e|$. A variant of this definition would be to define the depth of $e$ as $d'_\phi(e) = |\{v \in V(G) \mid \phi(v) \cap e \neq \emptyset\}|$, i.e., if $\phi(v)$ intersects $e$, then $v$ contributes only 1 to the depth of $e$, not $|\phi(v) \cap e|$ as in the original definition. Let us call this variant *weak edge depth*, it is clear that the weak edge depth of an embedding is at most the edge depth of the embedding.

Lemma 7.5 can be made stronger by requiring only that the weak edge depth is at most $q$. Indeed, the only place where we use the bound on edge depth is in Inequality (3). However, the size of the relation $R_e$ can be bounded by the number of possible assignments on $U_e$ in instance $I_1$. If weak edge depth is at most $q$, then $|U_e| \leq q$, and the $|D_1|^q$ bound on the size of $R_e$ follows.

*Remark* 7.7. A different version of CSP was investigated in [44], where each variable has a different domain, and each constraint relation is represented by a full truth table (see the exact definition in [44]). Let us denote by $\mathrm{CSP}_{tt}(\mathcal{H})$ this variant of the problem. It is easy to see that $\mathrm{CSP}_{tt}(\mathcal{H})$ can be reduced to $\mathrm{CSP}(\mathcal{H})$ in polynomial time, but a reduction in the other direction can possibly increase the representation of a constraint by an exponential factor. Nevertheless, the hardness results of this section apply to the "easier" problem $\mathrm{CSP}_{tt}(\mathcal{H})$ as well. What we have to verify is that the proof of Lemma 7.5 works even if $I_2$ is an instance of $\mathrm{CSP}_{tt}$, i.e., the constraint relations have to be represented by truth tables. Inspection of the proof shows that it indeed works: the product in Inequality (3) is exactly the size of the truth table describing the constraint corresponding to edge $e$, thus the $|D_1|^q$ upper bound remains valid even if constraints are represented by truth tables. Therefore, the hardness results of [44] are subsumed by the following corollary:

**Corollary 7.8.** *If $\mathcal{H}$ is a recursively enumerable class of hypergraphs with unbounded submodular width, then $\mathrm{CSP}_{tt}(\mathcal{H})$ is not fixed-parameter tractable, unless the Exponential Time Hypothesis fails.*

# 8 Conclusions

The main result of the paper is introducing submodular width and proving that bounded submodular width is the property that determines the fixed-parameter tractability of CSP($\mathcal{H}$). The hardness result is proved assuming the Exponential Time Hypothesis. This conjecture was formulated relatively recently [35], but it turned out to be very useful in proving lower bounds in a variety of settings [43, 6, 41, 49].

For the hardness proof, we had to understand what large submodular width means and we had to explore the connection between submodular width and other combinatorial properties. We have obtained several equivalent characterizations of bounded submodular width, in particular, we have showed that bounded submodular width is equivalent to bounded adaptive width:

**Corollary 8.1.** *The following are equivalent for every class $\mathcal{H}$ of hypergraphs:*

1. *There is a constant $c_1$ such that $\mu$-width$(H) \le c_1$ for every $H \in \mathcal{H}$ and fractional independent set $\mu$.*

2. *There is a constant $c_2$ such that b-width$(H) \le c_2$ for every $H \in \mathcal{H}$ and edge-dominated monotone submodular function $b$ on $V(H)$ with $b(\emptyset) = 0$.*

3. *There is a constant $c_3$ such that $b^*$-width$(H) \le c_3$ for every $H \in \mathcal{H}$ and edge-dominated monotone submodular function $b$ on $V(H)$ with $b(\emptyset) = 0$.*

4. *There is a constant $c_4$ such that $\mathrm{con}_\lambda(H) \le c_4$ for every $H \in \mathcal{H}$, where $\lambda > 0$ is a universal constant.*

5. *There is a constant $c_5$ such that $\mathrm{emb}(H) \le c_5$ for every $H \in \mathcal{H}$.*

Implications (2)$\Rightarrow$(1) and (3)$\Rightarrow$(2) are trivial; (4)$\Rightarrow$(3) follows from Lemma 5.10; (5)$\Rightarrow$(4) follows from Corollary 6.2; (1)$\Rightarrow$(5) follows from Lemma 6.9.

Let us briefly review the main ideas that were necessary for proving the main result of the paper:

- Recognizing that submodular width is the right property characterizing the complexity of the problem.

- A CSP instance can be partitioned into a bounded number of uniform instances (Section 4.2).

- The number of solutions in a uniform CSP instance can be described by a submodular function (Section 4.3).

- There is a connection between fractional separation and finding a separator minimizing an edge-dominated submodular cost function (Section 5.2).

- The transformation that turns $b$ into $b^*$, the properties of $b^*$ (Section 5.1).

- Our results on fractional separation and the standard framework of finding tree decompositions show that large submodular width implies that there is highly connected set (Section 5.3).

- A highly connected set can be turned into a highly connected set that is partitioned into cliques in an appropriate way (Section 6.1).

- A highly connected set with appropriate cliques implies that there is a uniform concurrent flow of large value between the cliques (Section 6.2).

- Similarly to [43], we use the observation that a concurrent flow is analogous to a line graph of a clique, hence it has good embedding properties (Section 6.2).

- Similarly to [43], an embedding in a hypergraph gives a way of simulating 3SAT with CSP($\mathcal{H}$) (Section 7).

It is possible that the main result can be proved in a simpler way by bypassing some of the ideas above. In particular, a surprising consequence of our results is that bounded submodular width and bounded adaptive width are the same, i.e., if a class $\mathcal{H}$ has unbounded submodular width, then for every $k$ there is a $H_k \in \mathcal{H}$ and a fractional independent set $\mu_k$ such that $\mu_k$-width$(H_k) \geq k$, or in other words, large submodular width can be certified by the *modular* function $\mu_k$. To prove this, we need all the results of Sections 5 and 6. Having a better understanding and an independent proof of this fact could simplify the proofs considerably. Another possible target for simplification is Section 6.1, where a lot of effort is spent on proving that if there is a large highly connected set, then there is a large highly connected set that is partitioned into cliques in an appropriate way. It might be possible to strengthen the results of Section 5 (perhaps by better understanding the role of cliques in separators) so that they give such a highly connected set directly.

An obvious question for further research is whether it is possible to prove a similar dichotomy result with respect to polynomial-time solvability. At this point, it is hard to see what the answer could be if we investigate the same question using the more restricted notion of polynomial time solvability. We know that bounded fractional hypertree width implies polynomial-time solvability [42] and Theorem 7.1 shows that unbounded submodular width implies that the problem is not polynomial-time solvable (as it is not even fixed-parameter tractable). So only those classes of hypergraphs are in the "gray zone" that have bounded submodular width but unbounded fractional hypertree width.

What could be the truth in this gray zone? A first possibility is that CSP($\mathcal{H}$) is polynomial-time solvable for every such class, i.e., Theorem 4.1 can be improved from fixed-parameter tractability to polynomial-time solvability. However, Theorem 4.1 uses the power of fixed-parameter tractability in an essential way (splitting into a double-exponential number of uniform instances), so it is not clear how such improvement is possible. A second possibility is that unbounded fractional hypertree width implies that CSP($\mathcal{H}$) is not polynomial-time solvable. Substantially new techniques would be required for such a hardness proof. The hardness proofs of this paper and of [27, 43] are based on showing that a large problem space can be efficiently embedded into an instance with a particular hypergraph. However, the fixed-parameter tractability results show that no such embedding is possible in case of classes with bounded submodular width. Therefore, a possible hardness proof should embed a problem space that is comparable (in some sense) with the size of the hypergraph and should create instances where the domain size is bounded by a function of the size of the hypergraph. A third possibility is that the boundary of polynomial-time solvability is somewhere between bounded fractional hypertree width and bound submodular width. Currently, there is no natural candidate for a property that could correspond to this boundary and, again, the hardness part of the characterization should be substantially different than what was done before. Finally, there is a fourth possibility: the boundary of the polynomial-time cases cannot be elegantly characterized by a simple combinatorial property. In general, if we consider the restriction of a problem to all possible classes of (hyper)graphs, then there is no a priori reason why an elegant characterization should exist that that describes the easy and hard classes. For example, it is highly unlikely that there is an elegant characterization of those classes of graphs where solving the MAXIMUM INDEPENDENT SET problem is polynomial-time solvable. As discussed earlier, the fixed-parameter tractability of CSP($\mathcal{H}$) is a more robust question than its polynomial-time solvability, hence it is very well possible that only the former question has an elegant answer.